\documentclass[12pt, draftclsnofoot, onecolumn]{IEEEtran}
\usepackage{url}
\usepackage[utf8]{inputenc}
\usepackage{xcolor}
\usepackage{amsmath}
\usepackage{amssymb}
\usepackage{pifont}
\newcommand{\cmark}{\ding{51}}%
\newcommand{\xmark}{\ding{55}}%

\usepackage[acronyms,nonumberlist,nopostdot,nomain,nogroupskip]{glossaries}
\usepackage{tablefootnote}
\usepackage{booktabs}
\usepackage{tabularx}
\usepackage{epsfig}
\usepackage[outdir=img/]{epstopdf}
\usepackage{tikz}
\usepackage{pgfplots}
\pgfplotsset{compat=newest} 
\pgfplotsset{plot coordinates/math parser=false} 
\newlength\fheight
\newlength\fwidth
\usetikzlibrary{plotmarks,shapes,patterns,decorations.pathreplacing,backgrounds,calc,arrows,arrows.meta,spy,matrix,shadows,trees,positioning}
\usepgfplotslibrary{patchplots,groupplots}
\usepackage{tikzscale}
\usepackage{tikz-qtree}
\usepackage{hyperref}
\usepackage{physics}
\usepackage{siunitx}

\usepackage{multirow}
\usepackage{tkz-kiviat}

\usepackage[font=scriptsize]{subcaption}
\usepackage[font=footnotesize]{caption}

\usepackage{mathtools}

\newcommand{\normalDistrib}[2]{\ensuremath{\mathcal{N} \qty(#1, #2)}}
\newcommand{\ricianDistrib}[2]{\ensuremath{\mathcal{R} \qty(#1, #2)}} 

\newcommand{\PG}{\ensuremath{{\rm PG}}}
\newcommand{\AoD}{\ensuremath{{\rm AoD}}}
\newcommand{\AoA}{\ensuremath{{\rm AoA}}}

\usepackage{dblfloatfix}    
\usepackage{colortbl}

\newacronym{3gpp}{3GPP}{3rd Generation Partnership Project}
\newacronym{5g}{5G}{5th generation}
\newacronym{5gc}{5GC}{5G Core}
\newacronym{adc}{ADC}{Analog to Digital Converter}
\newacronym{afbw}{AFBW}{Average Fading Bandwidth}
\newacronym{aimd}{AIMD}{Additive Increase Multiplicative Decrease}
\newacronym{am}{AM}{Acknowledged Mode}
\newacronym{amc}{AMC}{Adaptive Modulation and Coding}
\newacronym{aoa}{AoA}{Angle of Arrival}
\newacronym{aod}{AoD}{Angle of Departure}
\newacronym{ap}{AP}{Access Point}
\newacronym{app}{APP}{Application Layer}
\newacronym{aqm}{AQM}{Active Queue Management}
\newacronym{awgn}{AGWN}{Additive White Gaussian Noise}
\newacronym{balia}{BALIA}{Balanced Link Adaptation}
\newacronym{bdp}{BDP}{Bandwidth-Delay Product}
\newacronym{ber}{BER}{Bit Error Rate}
\newacronym{bf}{BF}{Beamforming}
\newacronym{cad}{CAD}{Computer-Aided Design}
\newacronym{cbr}{CBR}{Constant Bit Rate}
\newacronym{cc}{CC}{Congestion Control}
\newacronym{cdf}{CDF}{Cumulative Distribution Function}
\newacronym{ci}{CI}{Confidence Interval}
\newacronym{cir}{CIR}{Channel Impulse Response}
\newacronym{cn}{CN}{Core Network}
\newacronym{cp}{CP}{Control Plane}
\newacronym{cqi}{CQI}{Channel Quality Information}
\newacronym{crs}{CRS}{Cell Reference Signal}
\newacronym{csirs}{CSI-RS}{Channel State Information - Reference Signal}
\newacronym{dc}{DC}{Dual Connectivity}
\newacronym{dce}{DCE}{Direct Code Execution}
\newacronym{dci}{DCI}{Downlink Control Information}
\newacronym{dl}{DL}{Downlink}
\newacronym{dmr}{DMR}{Deadline Miss Ratio}
\newacronym{dmrs}{DMRS}{DeModulation Reference Signal}
\newacronym{dray}{D-Ray}{Deterministic Ray}
\newacronym{e2e}{E2E}{End-to-End}
\newacronym{ecn}{ECN}{Explicit Congestion Notification}
\newacronym{ecdf}{ECDF}{Empirical Cumulative Distribution Function}
\newacronym{edf}{EDF}{Earliest Deadline First}
\newacronym{em}{EM}{electromagnetic}
\newacronym{enb}{eNB}{evolved Node Base}
\newacronym{endc}{EN-DC}{E-UTRAN-\gls{nr} \gls{dc}}
\newacronym{epc}{EPC}{Evolved Packet Core}
\newacronym{es}{ES}{Edge Server}
\newacronym{fdd}{FDD}{Frequency Division Duplexing}
\newacronym{fdma}{FDMA}{Frequency Division Multiple Access}
\newacronym{fray}{F-Ray}{Flashing Ray}
\newacronym{fs}{FS}{Fast Switching}
\newacronym{ftp}{FTP}{File Transfer Protocol}
\newacronym{gmm}{GMM}{Gaussian Mixture Model}
\newacronym{gnb}{gNB}{Next Generation Node Base}
\newacronym{harq}{HARQ}{Hybrid Automatic Repeat reQuest}
\newacronym{hetnet}{HetNet}{Heterogeneous Network}
\newacronym{hh}{HH}{Hard Handover}
\newacronym{hol}{HOL}{Head-of-Line}
\newacronym{hqf}{HQF}{Highest-quality-first}
\newacronym{ia}{IA}{Initial Access}
\newacronym{iab}{IAB}{Integrated Access and Backhaul}
\newacronym{ieee}{IEEE}{Institute of Electrical and Electronics Engineers}
\newacronym{imt}{IMT}{International Mobile Telecommunication}
\newacronym{inr}{INR}{Interference to Noise Ratio}
\newacronym{iot}{IoT}{Internet of Things}
\newacronym{ked}{KED}{Knife-Edge Diffraction}
\newacronym{kpi}{KPI}{Key Performance Indicator}
\newacronym{ks}{KS}{Kolmogorov–Smirnov}
\newacronym{lcf}{LCF}{Level Crossing Frequency}
\newacronym{lcr}{LCR}{Level Crossing Rate}
\newacronym{los}{LoS}{Line-of-Sight}
\newacronym{lte}{LTE}{Long Term Evolution}
\newacronym{m2m}{M2M}{Machine to Machine}
\newacronym{mac}{MAC}{Medium Access Control}
\newacronym{mc}{MC}{Multi-Connectivity}
\newacronym{mcs}{MCS}{Modulation and Coding Scheme}
\newacronym{mec}{MEC}{Mobile Edge Cloud}
\newacronym{mi}{MI}{Mutual Information}
\newacronym{mib}{MIB}{Master Information Block}
\newacronym{mimo}{MIMO}{Multiple Input, Multiple Output}
\newacronym{mlr}{MLR}{Maximum-local-rate}
\newacronym[plural=\gls{mme}s,firstplural=Mobility Management Entities (MMEs)]{mme}{MME}{Mobility Management Entity}
\newacronym{mmwave}{mmWave}{millimeter wave}
\newacronym{moi}{MoI}{Method of Images}
\newacronym{mpc}{MPC}{Multi Path Component}
\newacronym{mptcp}{MPTCP}{Multipath TCP}
\newacronym{mr}{MR}{Maximum Rate}
\newacronym{mrdc}{MR-DC}{Multi \gls{rat} \gls{dc}}
\newacronym{mss}{MSS}{Maximum Segment Size}
\newacronym{mtd}{MTD}{Machine-Type Device}
\newacronym{mtu}{MTU}{Maximum Transmission Unit}
\newacronym{nfv}{NFV}{Network Function Virtualization}
\newacronym{nist}{NIST}{National Institute of Standards and Technology}
\newacronym{nlos}{NLoS}{Non-Line-of-Sight}
\newacronym{nr}{NR}{New Radio}
\newacronym{nrmse}{NRMSE}{Normalized Root Mean Square Error}
\newacronym{ns3}{ns-3}{Network Simulator 3}
\newacronym{nsa}{NSA}{Non Stand Alone}
\newacronym{o2i}{O2I}{Outdoor-to-Indoor}
\newacronym{ofdm}{OFDM}{Orthogonal Frequency Division Multiplexing}
\newacronym{pa}{PA}{Position-aware}
\newacronym{pbch}{PBCH}{Physical Broadcast Channel}
\newacronym{pdcch}{PDCCH}{Physical Downlonk Control Channel}
\newacronym{pdcp}{PDCP}{Packet Data Convergence Protocol}
\newacronym{pdsch}{PDSCH}{Physical Downlink Shared Channel}
\newacronym{pdu}{PDU}{Packet Data Unit}
\newacronym{per}{PER}{Packet Error Rate}
\newacronym{pf}{PF}{Proportional Fair}
\newacronym{pgw}{PGW}{Packet Gateway}
\newacronym{phy}{PHY}{Physical}
\newacronym{pl}{PL}{Path Loss}
\newacronym{ppp}{PPP}{Poisson Point Process}
\newacronym{prb}{PRB}{Physical Resource Block}
\newacronym{pss}{PSS}{Primary Synchronization Signal}
\newacronym{pucch}{PUCCH}{Physical Uplink Control Channel}
\newacronym{pusch}{PUSCH}{Physical Uplink Shared Channel}
\newacronym{qd}{QD}{Quasi Deterministic}
\newacronym{rach}{RACH}{Random Access Channel}
\newacronym{ran}{RAN}{Radio Access Network}
\newacronym[firstplural=Radio Access Technologies (RATs)]{rat}{RAT}{Radio Access Technology}
\newacronym{red}{RED}{Random Early Detection}
\newacronym{rf}{RF}{Radio Frequency}
\newacronym{rlc}{RLC}{Radio Link Control}
\newacronym{rlf}{RLF}{Radio Link Failure}
\newacronym{rr}{RR}{Round Robin}
\newacronym{rray}{R-Ray}{Random Ray}
\newacronym{rrc}{RRC}{Radio Resource Control}
\newacronym{rrm}{RRM}{Radio Resource Management}
\newacronym{rs}{RS}{Remote Server}
\newacronym{rsrp}{RSRP}{Reference Signal Received Power}
\newacronym{rsrq}{RSRQ}{Reference Signal Received Quality}
\newacronym{rss}{RSS}{Received Signal Strength}
\newacronym{rssi}{RSSI}{Received Signal Strength Indicator}
\newacronym{rt}{RT}{Ray-Tracer}
\newacronym{rtt}{RTT}{Round Trip Time}
\newacronym{rw}{RW}{Receive Window}
\newacronym{rx}{RX}{Receiver}
\newacronym{sa}{SA}{standalone}
\newacronym{sack}{SACK}{Selective Acknowledgment}
\newacronym{sap}{SAP}{Service Access Point}
\newacronym{sch}{SCH}{Secondary Cell Handover}
\newacronym{scm}{SCM}{Spatial Channel Model}
\newacronym{scoot}{SCOOT}{Split Cycle Offset Optimization Technique}
\newacronym{sdma}{SDMA}{Spatial Division Multiple Access}
\newacronym{sf}{SF}{Shadow Fading}
\newacronym{si}{SI}{Study Item}
\newacronym{sib}{SIB}{Secondary Information Block}
\newacronym{sinr}{SINR}{Signal-to-Interference-plus-Noise Ratio}
\newacronym{sir}{SIR}{Signal-to-Interference Ratio}
\newacronym{sm}{SM}{Saturation Mode}
\newacronym{snr}{SNR}{Signal-to-Noise Ratio}
\newacronym{son}{SON}{Self-Organizing Network}
\newacronym{srs}{SRS}{Sounding Reference Signal}
\newacronym{ss}{SS}{Synchronization Signal}
\newacronym{sss}{SSS}{Secondary Synchronization Signal}
\newacronym{sta}{STA}{Station}
\newacronym{svd}{SVD}{Singular Value Decomposition}
\newacronym{tb}{TB}{Transport Block}
\newacronym{tcp}{TCP}{Transmission Control Protocol}
\newacronym{udp}{UDP}{User Datagram Protocol}
\newacronym{tdd}{TDD}{Time Division Duplexing}
\newacronym{tdma}{TDMA}{Time Division Multiple Access}
\newacronym{tfl}{TfL}{Transport for London}
\newacronym{tgad}{TGad}{Task Group ad}
\newacronym{tgay}{TGay}{Task Group ay}
\newacronym{tm}{TM}{Transparent Mode}
\newacronym{trp}{TRP}{Transmitter Receiver Pair}
\newacronym{tti}{TTI}{Transmission Time Interval}
\newacronym{ttt}{TTT}{Time-to-Trigger}
\newacronym{tx}{TX}{Transmitter}
\newacronym{ue}{UE}{User Equipment}
\newacronym{ul}{UL}{Uplink}
\newacronym{um}{UM}{Unacknowledged Mode}
\newacronym{uma}{UMa}{Urban Macro}
\newacronym{uml}{UML}{Unified Modeling Language}
\newacronym{utc}{UTC}{Urban Traffic Control}
\newacronym{vm}{VM}{Virtual Machine}
\newacronym{wbf}{WBF}{Wired Bias Function}
\newacronym{wf}{WF}{Wired-first}
\newacronym{wifi}{Wi-Fi}{Wireless Fidelity}
\newacronym{wigig}{WiGig}{Wireless Gigabit}
\newacronym{wlan}{WLAN}{Wireless Local Area Network}
\newacronym{xpr}{XPR}{Cross Polarization Ratio}
\tikzstyle{startstop} = [rectangle, rounded corners, minimum width=2cm, minimum height=0.5cm,text centered, draw=black]
\tikzstyle{io} = [trapezium, trapezium left angle=70, trapezium right angle=110, minimum width=3cm, minimum height=1cm, text centered, draw=black]
\tikzstyle{process} = [rectangle, minimum width=2cm, minimum height=0.5cm, text centered, draw=black, alignb=center]
\tikzstyle{decision} = [ellipse, minimum width=2cm, minimum height=1cm, text centered, draw=black]
\tikzstyle{arrow} = [thick,<->,>=stealth]
\tikzstyle{line} = [thick,>=stealth]
\tikzstyle{darrow} = [thick,<->,>=stealth,dashed]
\tikzstyle{sarrow} = [thick,->,>=stealth]
\tikzstyle{larrow} = [line width=0.1mm,dashdotted,->,>=stealth]

\makeatletter
\def\grd@save@target#1{%
  \def\grd@target{#1}}
\def\grd@save@start#1{%
  \def\grd@start{#1}}
\tikzset{
  grid with coordinates/.style={
    to path={%
      \pgfextra{%
        \edef\grd@@target{(\tikztotarget)}%
        \tikz@scan@one@point\grd@save@target\grd@@target\relax
        \edef\grd@@start{(\tikztostart)}%
        \tikz@scan@one@point\grd@save@start\grd@@start\relax
        \draw[minor help lines] (\tikztostart) grid (\tikztotarget);
        \draw[major help lines] (\tikztostart) grid (\tikztotarget);
        \grd@start
        \pgfmathsetmacro{\grd@xa}{\the\pgf@x/1cm}
        \pgfmathsetmacro{\grd@ya}{\the\pgf@y/1cm}
        \grd@target
        \pgfmathsetmacro{\grd@xb}{\the\pgf@x/1cm}
        \pgfmathsetmacro{\grd@yb}{\the\pgf@y/1cm}
        \pgfmathsetmacro{\grd@xc}{\grd@xa + \pgfkeysvalueof{/tikz/grid with coordinates/major step x}}
        \pgfmathsetmacro{\grd@yc}{\grd@ya + \pgfkeysvalueof{/tikz/grid with coordinates/major step y}}
        \foreach \x in {\grd@xa,\grd@xc,...,\grd@xb}
        \node[anchor=north] at (\x,\grd@ya) {\pgfmathprintnumber{\x}};
        \foreach \y in {\grd@ya,\grd@yc,...,\grd@yb}
        \node[anchor=east] at (\grd@xa,\y) {\pgfmathprintnumber{\y}};
      }
    }
  },
  minor help lines/.style={
    help lines,
    gray,
    line cap =round,
    xstep=\pgfkeysvalueof{/tikz/grid with coordinates/minor step x},
    ystep=\pgfkeysvalueof{/tikz/grid with coordinates/minor step y}
  },
  major help lines/.style={
    help lines,
    line cap =round,
    line width=\pgfkeysvalueof{/tikz/grid with coordinates/major line width},
    xstep=\pgfkeysvalueof{/tikz/grid with coordinates/major step x},
    ystep=\pgfkeysvalueof{/tikz/grid with coordinates/major step y}
  },
  grid with coordinates/.cd,
  minor step x/.initial=.5,
  minor step y/.initial=.2,
  major step x/.initial=1,
  major step y/.initial=1,
  major line width/.initial=1pt,
}
\makeatother

\usepackage[capitalize]{cleveref}
\crefname{section}{Sec.}{Secs.}

\IEEEoverridecommandlockouts
\newcommand\copyrightnotice{%
\begin{tikzpicture}[remember picture,overlay]
\node[anchor=south,yshift=10pt] at (current page.south) {\fbox{\parbox{\dimexpr\textwidth-\fboxsep-\fboxrule\relax}{
\footnotesize \textcopyright 2020 IEEE. Personal use of this material is permitted.
Permission from IEEE must be obtained for all other uses, in any current or future media,
including reprinting/republishing this material for advertising or promotional purposes,
creating new collective works, for resale or redistribution to servers or lists,
or reuse of any copyrighted component of this work in other works.}}};
\end{tikzpicture}
}

\makeglossaries

\begin{document}

\title{Accuracy vs. Complexity for mmWave Ray-Tracing: A Full Stack Perspective}

\author{{{Mattia Lecci},~\IEEEmembership{Student Member, IEEE},
        {Paolo Testolina},~\IEEEmembership{Student Member, IEEE},
        {Michele Polese},~\IEEEmembership{Member, IEEE},
        {Marco Giordani},~\IEEEmembership{Member, IEEE},\\
        {Michele Zorzi},~\IEEEmembership{Fellow, IEEE}}
        \thanks{Mattia Lecci, Paolo Testolina, Marco Giordani, and Michele Zorzi are with the Department of Information Engineering, University of Padova, Padova, Italy (email: \{name.surname\}@dei.unipd.it). Michele Polese is with Institute for the Wireless Internet of Things, Northeastern University, Boston, MA, 02120 USA (email: m.polese@northeastern.edu).}
        \thanks{This work was partially supported by NIST through Award No. 70NANB18H273.
                Mattia Lecci's and Paolo Testolina's activities were also supported by \textit{Fondazione CaRiPaRo} under the grants ``Dottorati di Ricerca'' 2018 and 2019, respectively.}
        \thanks{The identification of any commercial product or trade name does not imply endorsement or recommendation by the National Institute of Standards and Technology, nor is it intended to imply that the materials or equipment identified are necessarily the best available for the purpose.}
        \thanks{A preliminary version of this paper that did not consider end-to-end performance evaluations was presented at the \emph{IEEE Information Theory and Applications Workshop (ITA)}, February 2020~\cite{testolina2020simplified}.}
}

\maketitle
\copyrightnotice

\glsunset{nr}

\begin{abstract}

The millimeter wave (mmWave) band will provide multi-gigabits-per-second connectivity in the radio access of future wireless systems.
The high propagation loss in this portion of the spectrum calls for the deployment of large antenna arrays to compensate for the loss through high directional gain, thus introducing a spatial dimension in the channel model to accurately represent the performance of a mmWave network.
In this perspective, ray-tracing can characterize the channel in terms of \glspl{mpc} to provide a highly accurate model, at the price of extreme computational complexity (e.g., for processing detailed environment information about the propagation), which limits the scalability of the simulations.
In this paper, we present possible simplifications to improve the trade-off between accuracy and complexity in ray-tracing simulations at mmWaves by reducing the total number of \glspl{mpc}.
The effect of such simplifications is evaluated from a full-stack perspective through end-to-end simulations, testing different configuration parameters, propagation scenarios, and higher-layer protocol implementations.
We then provide guidelines on the optimal degree of simplification, for which it is possible to reduce the complexity of simulations with a minimal reduction in accuracy for different deployment scenarios.
\end{abstract}

\begin{tikzpicture}[remember picture,overlay]
\node[anchor=north,yshift=-10pt] at (current page.north) {\fbox{\parbox{\dimexpr\textwidth-\fboxsep-\fboxrule\relax}{
\centering\footnotesize This paper has been submitted to IEEE for publication. Copyright may change without notice.}}};
\end{tikzpicture}

\section{Introduction} 
\label{sec:introduction}
\glsresetall

Recent developments have paved the way towards \gls{5g} cellular networks and enhanced \gls{wlan} designs, to address the traffic demands of the 2020 digital society~\cite{itu-r-2083}.
In particular, 5G systems will support very high data rates (with a peak of 20 Gbps in ideal conditions),  ultra-low latency (around 1 ms for ultra-reliable communications), and a 100x increase in energy efficiency with respect to previous wireless generations.
To meet those requirements, the \gls{3gpp} has released a set of specifications for NR, the new 5G \gls{ran}, which include (i) a flexible frame structure, with adaptive numerologies, (ii) network slicing and virtualization in a new core network design, and (iii) communications in the \gls{mmwave} bands~\cite{38300}.
Similarly, the \gls{ieee} has developed amendments to 802.11 networks, namely 802.11ad and 802.11ay~\cite{zhou2018ieee}, which operate at \glspl{mmwave}.
3GPP NR carrier frequency can be as high as 52.6~GHz for Release 15 (even though future Releases will not exclude extensions up to 71~GHz~\cite{qualcomm201971}), while IEEE 802.11ad and 802.11ay exploit the unlicensed spectrum at 60~GHz~\cite{zhou2018ieee}.

The vast amount of available spectrum at \gls{mmwave} frequencies can enable multi-Gbps transmission rates~\cite{rappaport2013millimeter}.
Moreover, the very short wavelength makes it practical to build large antenna arrays (e.g., with hundreds of elements) and establish highly directional communications, thus boosting the network performance through beamforming and spatial diversity~\cite{sun2014mimo}.

Despite these promising characteristics, propagation at \glspl{mmwave} raises several challenges for the design and performance of the whole protocol stack~\cite{zhang2019will}.
First, the communication suffers from severe path loss (which is inversely  proportional to the square of the wavelength), thereby preventing long-range omni-directional transmissions.
Second, mmWave links are highly sensitive to blockage from common materials (e.g., brick and mortar), which may result in more than 40 dB of attenuation at 28~GHz when losing \gls{los}~\cite{lu2012modeling}.
Third, the delay and the Doppler spread (which determine the temporal and frequency selectivity of the channels) are particularly strong at these frequencies and may lead to network disconnections~\cite{giordani2016channel}.
Finally, directional communications require the precise alignment of the transmitter and receiver beams, hence implying an increased control overhead for channel estimation and mobility management~\cite{giordani2018tutorial,giordani2016initial}.

The combination of these phenomena makes the mmWave channel extremely volatile to mobile users. Although some early performance evaluations have suggested that \gls{mmwave} networks can offer orders of magnitude greater capacity than legacy systems (e.g.,~\cite{sun2018propagation}), a deeper understanding of the propagation channel is required to reliably characterize such networks.
In this sense, experimental testbeds make it possible to examine the network performance in real-world environments with extreme accuracy~\cite{saha2019x60}.
However, the prohibitive cost and limited flexibility of these platforms make this approach impractical for most of the research community~\cite{polese2019millimetera}.

Therefore, theoretical analyses and computer-aided simulations have emerged as an important tool in evaluating the performance of novel solutions and the interplay between the \gls{mmwave} channel and the deployment and protocol design.
Both analysis and simulation, however, require proper modeling of signal propagation to accurately reproduce the behavior of \gls{mmwave} systems~\cite{ferrand2016trends,polese2018impact}.
On one side, analytical studies model the channel using a Nakagami-m or Rayleigh distribution~\cite{bai2015coverage,park2016tractable,andrews2017modeling}.
This approach, while simplifying the analysis significantly, assumes a rich multipath channel when in fact it is sparse at \glspl{mmwave}~\cite{hemadeh2018millimeter}.
Similarly, stochastic \glspl{scm}, e.g.,~\cite{3gpp.38.901} for 3GPP NR, characterize the channel as a combination of random variables fitted from real-world measurements, providing a more realistic assessment of the mmWave network performance compared to their analytical counterparts~\cite{akdeniz2014millimeter}, however for measurements at sub-6~GHz.
Still, the stochastic nature of these models may prevent researchers from evaluating the impact of the channel dynamics in specific environments, and may respond poorly to the need of accurately characterizing the spatio-temporal evolution of the channel \glspl{mpc}~\cite{testolina2019scalable}.

Conversely, \glspl{rt} can be used to precisely model the propagation of \gls{mmwave} signals in specific scenarios~\cite{degliesposti2014rt,larew2013air}.
Unlike analytical or stochastic models, \glspl{rt} are based on the geometry of the scenario and characterize the different propagation properties of each \gls{mpc}, including time delay, Doppler shift, polarization, \gls{aod} at the \gls{tx}, and \gls{aoa} at the \gls{rx}, thus providing higher accuracy~\cite{lai2019methodology}.
Moreover, simulators can use ray-tracing to model the temporal and spatial evolution of the channel, a necessary feature for a proper planning of wireless systems.
However,  the generation  of the \glspl{mpc} can be computationally expensive, limiting the scalability of simulations.
It is thus fundamental to find a compromise between accuracy and reliability, a research challenge that, to date, has not yet been exhaustively addressed in the literature.

This paper represents a first, comprehensive study on how simplifications to a \gls{mmwave} \glspl{rt}-based channel modeling can reduce the computational complexity of system-level simulations without compromising their accuracy.
Specifically, we target the following objectives:
\begin{itemize}
  \item We propose simplification techniques for ray-tracing based on the \gls{moi} to speed up network simulations and the \gls{rt} itself.
  Specifically, along the lines of our previous investigations~\cite{testolina2020simplified,testolina2019scalable}, we consider a simplified \gls{rt} implementation that processes only MPCs whose received power is above a certain threshold (relative to the strongest MPC) and another that limits the maximum number of reflections for each MPC.
  \item We showcase a publicly available and open-source \gls{rt}\footnote{\url{https://github.com/wigig-tools/qd-realization}} supporting a \gls{qd} model (which combines deterministic channel components  with random rays representing the diffusion due to the roughness of the surfaces on which the rays reflect), the aforementioned simplifications strategies, and all the scenarios shown in our results.
  \item As our main contribution, we carry out an extensive system-level simulation campaign to quantify the impact of the \gls{rt} simplifications on several network metrics.
  Unlike in our previous contribution~\cite{testolina2020simplified}:
  \begin{itemize}
    \item We consider a full-stack performance evaluation and the impact of the simplifications on different simulators.
    To do so, we integrate the \gls{rt} implementation in ns-3~\cite{henderson2008network,mezzavilla2018end}, a network simulator with a complete TCP/IP protocol stack that makes it possible to simulate the end-to-end network performance, and in a custom MATLAB simulator (as in~\cite{testolina2020simplified}), for link-level metrics.
    Specifically, we characterize the \gls{sinr}, throughput, and latency for different traffic regimes at the application layer and different antenna architectures, as a function of the degree of simplifications that are introduced on the \gls{rt}.
    We also consider both indoor and outdoor scenarios, to characterize different mobility and propagation characteristics, as well as different antenna array configurations.
    Additionally, with ns-3, we simulate the interaction of the simplifications with different transport layer protocols, namely \gls{udp} and \gls{tcp}, and with different applications.
    Our results show that there exist some configurations for which the impact of the \gls{rt} simplifications is limited, especially when higher-layer performance metrics are considered, with a reduction in simulation complexity of up to 4 times.
    \item We assess the effect of the \gls{qd} model on the network performance.
    Notably, we show that the additional stochastic diffuse components make the throughput and latency fluctuate more over time, while the \gls{rt} simplifications generally result in more stable channels.
  \end{itemize}
  \item We provide guidelines on the optimal working points, corresponding to the best combination of simplifications and system configurations, for which it is possible to decrease the computation time and complexity of simulations  with a minimal reduction in accuracy with respect to the baseline \gls{rt} implementation (i.e., without simplifications).
\end{itemize}

The rest of this article is structured as follows.
In \cref{sec:channel_modeling_at_mmwaves} we review the most relevant analytical, stochastic, and deterministic models for the \gls{mmwave} channel, while in \cref{sec:ray_tracing_at_mmwaves} we present the \gls{qd} model we adopted for ray-tracing operations at \glspl{mmwave}.
In \cref{sec:mmwave_channel_simplifications} we describe the proposed simplifications to be applied to the \gls{rt}, while performance results are provided in \cref{sec:performance_results}.
Finally, \cref{sec:conclusions_and_future_work} concludes the paper with suggestions for future work.

\section{Channel Modeling at MmWaves} 
\label{sec:channel_modeling_at_mmwaves}

The modeling of the \gls{mmwave} channel, in terms of propagation and fading, has been a key research topic in recent years~\cite{ferrand2016trends}.
Multiple measurement campaigns have tried to characterize the properties of different portions of the \gls{mmwave} spectrum, in a variety of different scenarios and environments, e.g., urban~\cite{rappaport2013millimeter}, rural~\cite{maccartney2017rural}, and indoor~\cite{gentile2018millimeter,lai2019methodology}.

These campaigns have led to the definition of different channel models.
Comprehensive reviews, discussing propagation, fading, and beamforming models can be found in~\cite{hemadeh2018millimeter,wang2018survey}, while~\cite{rappaport2017overview} focuses on the various results on propagation loss.
These efforts have identified the key factors for an accurate modeling of the \gls{mmwave} channel.
First, multipath components are sparse in the angular domain, thus an accurate model should explicitly characterize the angle of arrival and departure of the different taps.
Moreover, as discussed in \cref{sec:introduction}, blockage at the considered frequencies has a more remarkable impact on the link dynamics than at sub-6~GHz, which should be accounted for.
Finally, rough surfaces could generate more diffuse scatterers than at longer wavelengths.

The aforementioned measurement campaigns have led to different modeling approaches for the \gls{mmwave} channel, which have various degrees of complexity and accuracy, and can be applied to different contexts and evaluations.
In the next paragraphs, we will review three broad families, i.e., channel models used for mathematical analysis, stochastic, and map-based models.

\paragraph*{Analytical Channel Models} 
\label{sub:analytical_channel_models}
Analytical studies for the coverage and capacity evaluation of \gls{mmwave} networks generally considered simplified channel models, based on propagation and a single random variable for fading.
Rayleigh and Nakagami-m fading models have been widely used in stochastic geometry analysis of \gls{mmwave} systems, such as, for example, in~\cite{andrews2017modeling,bai2015coverage,park2016tractable}.
Nakagami-m fading, introduced in~\cite{NAKAGAMI19603}, controls the amplitude of fading through the parameter $m$, so that it is possible to model differently the \gls{los} and \gls{nlos} fading phenomena.
Similarly, Rayleigh fading is widely used, as in a stochastic geometry context it provides an easily tractable exponential form of the Nakagami fading for $m=1$~\cite{park2016tractable}.
Moreover, the Nakagami-m or Rayleigh fading is generally employed with a sectored beamforming model for directional transmissions.
This accounts for the beamforming gain $G$ by assigning a maximum gain $G_M$ to a main lobe, of angular width $\theta_b$, and a lower gain $G_m$ for the simplified side lobes in the complementary angular space~\cite{andrews2017modeling}.
This simplified model limits the accuracy in the representation of the interaction between the mmWave propagation, the realistic antenna arrays, and the beamforming strategies~\cite{ferrand2016trends}.
While only being relevant for omnidirectional, multipath-rich channels at sub-6~GHz, its tractability is desirable for analytical derivations even if it may not accurately model a realistic mmWave channel.

\paragraph*{Stochastic Channel Models} 
\label{sub:stochastic_channel_models}

An improved characterization can be achieved using \glspl{scm}~\cite{saleh1987statistical}.
These are based on a channel matrix $\mathbf{H} \in \mathcal{C}^{U \times S}$, with $S$ ($U$) being the number of antenna elements at the transmitter (receiver) array.
Each entry $(i,j)$ in the matrix $\mathbf{H}$ models the channel between two specific antenna elements, and represents the joint effect of different \glspl{mpc}.
Each \gls{mpc} is characterized by angles of departure and arrival, power, and delay.
The interaction with the antenna arrays can be modeled by pre- and post-multiplying to $\mathbf{H}$ the beamforming vectors of the transmitter and receiver, respectively~\cite{akdeniz2014millimeter}.

A popular class of \glspl{scm} is that in which the \glspl{mpc} are generated from a set of random distributions, whose parameters are determined by statistical fits on channel measurements.
The channel matrix thus has a stochastic nature, with the advantage that multiple instances can be randomly generated for generic, large scale scenarios.
The random variables of these models are used to characterize two phenomena, i.e., large scale fading, which depends on the scenario and the user mobility, and fast fading, which is given by small scale variations in power resulting from the interference among the \glspl{mpc}.

Notable examples of stochastic channel models are those derived from WINNER and WINNER-II models~\cite{winnerII}, e.g., the 3GPP channel model for the system-level evaluation of 5G deployments~\cite{3gpp.38.901}.
These models have been extensively used in the performance evaluations of mmWave networks~\cite{gapeyenko2018analytical}, and are also integrated with popular open source network simulators~\cite{zugno2020implementation}.
The NYU channel model for 28~GHz and 73~GHz is also based on a stochastic \gls{scm}~\cite{akdeniz2014millimeter}.

\paragraph*{Quasi-Deterministic Channel Models} 
\label{sub:quasi_deterministic_channel_models}

The stochastic nature of the aforementioned channel models makes them generic: they can model a common rural or an urban scenario, but not a specific scenario (e.g., Times Square in NYC).
Therefore, they do not provide an accurate model for the interactions of the \gls{mmwave} propagation with a peculiar deployment, making them unfit for detailed planning and capacity studies in real-world contexts.

As discussed in \cref{sec:introduction}, \glspl{rt} can, instead, provide extremely accurate propagation results in a given environment, provided that its characterization in the simulation is accurate enough.
With respect to stochastic channels, an \gls{rt} generates the exact \glspl{mpc} that can arise from a direct or reflected propagation path in the scenario~\cite{degliesposti2014rt}.
\acrlongpl{rt} have thus been the basis for several performance evaluation studies at \glspl{mmwave}~\cite{larew2013air,palacios2017tracking,khawaja2017uav}.
\gls{rt}-based channel models have also been adopted as candidates for the evaluation of IEEE 802.11ay networks, with a \gls{qd} extension~\cite{tgay_evaluation_methodology,lai2019methodology} that combines the geometry-based \glspl{mpc} and a number of random diffuse components that model the interaction of the mmWave signal with rough surfaces.

However, while being extremely precise, \glspl{rt} and \gls{qd} models are also more computationally intensive than stochastic models for the generation of a single channel instance, especially if the number of scattering and reflecting surfaces in the scenario is large.
While a number of optimizations have been studied for \glspl{rt} in general~\cite{fuschini2015rayTracingReview}, in the remainder of this paper we analyze the complexity of the considered open source \gls{mmwave} \gls{rt} and \gls{qd} model, and propose ready-to-use recipes to reduce it, with specific focus on preserving the accuracy of system-level simulation results.

\section{Ray-Tracing at Millimeter Waves} 
\label{sec:ray_tracing_at_mmwaves}

As previously discussed, ray-tracing is widely used to accurately simulate \gls{em} propagation in an environment, whose geometry is described by a \gls{cad} model~\cite{raytraceingSurveyYun2015}.
The fundamentals of this technique are derived by solving Maxwell's equations for the far field in the high-frequency regime, where the \gls{em} wave exhibits ray-like properties, i.e., the flow of power propagates in a straight line and reflects specularly on locally flat surfaces.
In practice, the high-frequency regime is assumed whenever the wavelength of the signal is much shorter than the typical size of obstacles in the scenario.
In this case, secondary effects, i.e., diffraction, diffuse scattering, polarization, and refraction, should also be accounted for.

Some of these effects can be particularly significant in the propagation of \gls{mmwave} signals.
In this frequency band, the shorter wavelength leads to a higher effective roughness of the surfaces, thus increasing the amount of scattered power and, consequently, the reflection loss.
The effect is twofold: on one side, higher-order reflections are expected to be weaker and thus affect the communication less than at lower frequency but, on the other side, proper modeling of the scattered rays should be taken into consideration~\cite{pascualGarcia2016diffuseScatteringMmwave}.
Conversely, higher penetration loss will reduce received power in the cluttered areas, improving frequency reuse and reducing the cross-interference between close-by radiators.
Finally, diffraction shadows are deeper at higher frequency, making diffraction a less prominent means of propagation in the mmWave band.

In this section, we describe the tools we use in this work to  simulate the mmWave channel.
Specifically, in \cref{sub:the_mmwave_ray_tracer} we describe the architecture and the assumptions of the \gls{rt} used in this paper, while in \cref{sub:qd_model} we briefly describe the diffuse scattering model.

\subsection{The Millimeter Wave Ray-Tracer} 
\label{sub:the_mmwave_ray_tracer}
\glsunset{moi}

For the results of this paper, we use an open-source \gls{rt} developed jointly by the SIGNET group at the University of Padova and the U.S. \gls{nist}.
It uses triangles described in \gls{cad} files as the basic 3D surface element unit, which can then be combined to define complex shapes.

The \gls{rt} only supports specular reflections and, optionally, diffuse scattering, ignoring effects such as diffraction, penetration, and polarization.
The latter is not considered to further simplify the software from the Fresnel equations, which dictate the laws of reflection.
The reflected rays thus experience a $180^\circ$ phase rotation and a random reflection loss $\vb*{RL}$ between 7~dB and 25~dB, depending on the material, but irrespective of the angle of incidence.

Multiple network nodes, playing the role of \glspl{tx} and \glspl{rx}, are modeled as points, and can be deployed simultaneously, allowing the calculation of interfering channels.
Furthermore, trace-based mobility is supported, making it possible to create complex scenarios with multiple base stations and mobile users.
Given $N$ nodes and $t$ time-steps, the simulator computes a channel instance for each time-step and for each node pair.
Considering a symmetric channel for a given node pair, $t N (N-1) /2 $ channel instances have to be calculated.

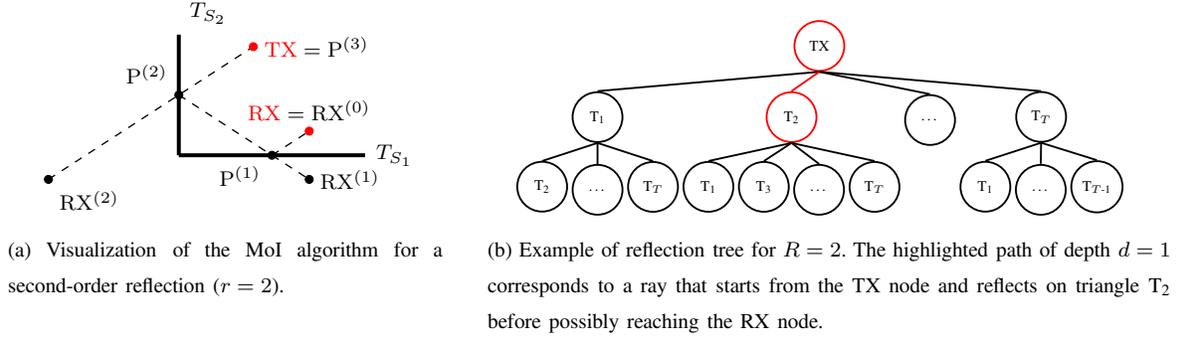
\begin{figure*}[tbp]
  \hfill
  \begin{subfigure}[t]{0.35\textwidth}
    \centering
    \setlength\fwidth{0.9\textwidth} 
    \setlength\fheight{0.5\textwidth}
%
%
\begin{tikzpicture}

\tikzset{font=\scriptsize}

\begin{axis}[%
width=\fwidth,
height=\fheight,
at={(0\fwidth,0\fheight)},
scale only axis,
xmin=-8,
xmax=13,
xtick={\empty},
ytick={\empty},
ymin=-5,
ymax=13,
axis background/.style={fill=white},
hide axis,
legend style={legend cell align=left, align=left, draw=white!15!black}
]
\addplot [color=black, line width=1.5pt, forget plot]
  table[row sep=crcr]{%
0	0\\
10	0\\
};
\node[anchor=west] at (10,0) {$T_{S_1}$};

\addplot [color=black, line width=1.5pt, forget plot]
  table[row sep=crcr]{%
0	0\\
0	10\\
};
\node[anchor=south west] at (0,10) {$T_{S_2}$};

\addplot[only marks, mark=*, mark options={}, mark size=1.5000pt, color=red, fill=red] table[row sep=crcr, forget plot]{%
x	y\\
7	2\\
};
\node[anchor=south] at (7,2) {$\rm {\color{red}RX} = RX^{(0)}$};

\addplot[only marks, mark=*, mark options={}, mark size=1.5000pt, color=red, fill=red] table[row sep=crcr, forget plot]{%
x	y\\
4	9\\
};
\node[anchor=west] at (4,9) {$\rm {\color{red}TX} = P^{(3)}$};

\addplot[only marks, mark=*, mark options={}, mark size=1.5000pt, color=black, fill=black] table[row sep=crcr, forget plot]{%
x	y\\
7	-2\\
};
\node[anchor=west] at (7,-2) {$\rm RX^{(1)}$};

\addplot[only marks, mark=*, mark options={}, mark size=1.5000pt, color=black, fill=black] table[row sep=crcr, forget plot]{%
x	y\\
-7	-2\\
};
\node[anchor=north west] at (-7,-2) {$\rm RX^{(2)}$};

\addplot [color=black, dashed, line width=.5pt, forget plot]
  table[row sep=crcr]{%
-7	-2\\
4	9\\
};

\addplot[only marks, mark=*, mark options={}, mark size=1.5000pt, color=black, fill=black] table[row sep=crcr, forget plot]{%
x	y\\
0	5\\
};
\node[anchor=south east] at (0,5) {$\rm P^{(2)}$};

\addplot [color=black, dashed, line width=.5pt, forget plot]
  table[row sep=crcr]{%
0	5\\
7	-2\\
};

\addplot[only marks, mark=*, mark options={}, mark size=1.5000pt, color=black, fill=black] table[row sep=crcr, forget plot]{%
x	y\\
5	0\\
};
\node[anchor=north east] at (5,0) {$\rm P^{(1)}$};

\addplot [color=black, dashed, line width=.5pt, forget plot]
  table[row sep=crcr]{%
5	0\\
7	2\\
};

\end{axis}
\end{tikzpicture}%
    \caption{Visualization of the MoI algorithm for a second-order reflection ($r = 2$).}
    \label{fig:moi_visualization}
  \end{subfigure}
  \hfill
  \begin{subfigure}[t]{0.55\textwidth}
    \setlength\belowcaptionskip{-0cm}
    \centering
    \setlength\fwidth{0.9\textwidth}
    \resizebox{\fwidth}{!}{%
  \begin{tikzpicture}
  \tikzset{font=\scriptsize,
    level distance=7ex,
    every node/.style = {
      shape=circle,
      draw,
      line width=1pt,
      minimum height=5ex,
      align=center},
    edge from parent/.append style = {
      line width=1pt
      }
  }

  \Tree [.\node[draw=red]{TX};
    [.T\textsubscript{1}
      [.T\textsubscript{2} ]
      [.{\ldots} ]
      [.T\textsubscript{$T$} ]
    ]
    \edge[red];  
    [.\node[draw=red]{T\textsubscript{2}};
      [.T\textsubscript{1} ]
      [.T\textsubscript{3} ]
      [.{\ldots} ]
      [.T\textsubscript{$T$} ]
    ]
    [.{\ldots} ]
    [.T\textsubscript{$T$}
      [.T\textsubscript{1} ]
      [.{\ldots} ]
      [.T\textsubscript{$T$-1} ]
    ]
  ]
  \end{tikzpicture}
}
    \caption{Example of reflection tree for $R = 2$.
    The highlighted path of depth $d=1$ corresponds to a ray that starts from the \gls{tx} node and reflects on triangle T\textsubscript{2} before possibly reaching the \gls{rx} node.}
    \label{fig:reflectionTree}
  \end{subfigure}
  \hfill
  \setlength\belowcaptionskip{-.6cm}
  \caption{Visualization of the basic principles behind the \gls{rt} software used.}
  \label{fig:rt_visualizations}
\end{figure*}%

The \gls{rt} uses the \gls{moi}~\cite{raytraceingSurveyYun2015} to compute specular reflections, assumed to be independent across time and node pairs.
For the simplest case, i.e., first order reflections, the \gls{moi} defines the virtual image of a node, for example the \gls{rx}, to be a specular image with respect to a surface.
Formally, $\rm RX^{(1)}$ is the specular image of the \gls{rx}, defined as $\rm RX^{(0)}$, across the surface $S$.
The specular reflection point $\rm P^{(1)}$ between the \gls{rx} and the \gls{tx} coincides with the intersection of the segment $\rm \qty( RX^{(1)}, TX)$ with the surface $S$, as shown in \cref{fig:moi_visualization}.

Since triangles are used as the basic surface units of the \gls{cad} environment, $S_i$ is the plane generated by a given triangle $T_i, \; i=1, \ldots, T$, where $T$ is the total number of triangles of the CAD environment, and it must be verified that $\rm P^{(1)}$ is a point within the area shaped by $T_i$, otherwise the reflection will not be valid and will thus be discarded.
Finally, every segment of the ray, namely $\rm \qty( RX^{(0)}, P^{(1)})$ and $\rm \qty(P^{(1)}, TX)$, has to be checked against the remaining triangles of the environment $T_j, \; j= 1, \ldots, T, \; j \neq i$ for obstruction.
If any segment of the ray is obstructed, the whole ray is considered obstructed and thus discarded.

The \gls{moi} applies recursively when multiple reflections are considered, computing the $n$-th virtual image of the \gls{rx}, $\rm RX^{(n)}$, and the respective specular reflection point $\rm P^{(n)}$ as shown in \cref{fig:moi_visualization}.
Thus, for each ray of reflection order $r>0$, $r$ geometrical operations must be done to compute the ray path and, if the ray is valid, each of the $r+1$ segments needs to be checked for obstruction over the $T-1$ triangles of the \gls{cad} environment.
This totals to about $r + (r+1)T$ operations per ray, for $T \gg 1$.

To compute all possible reflections between a given node pair, all possible paths have to be computed and tested for obstruction.
Considering increasing reflection orders, the first one to be tested is the direct ray, meaning the segment $\rm \qty(RX, TX)$.
Subsequently, all first order reflections are computed, meaning those rays starting from the \gls{tx}, reflecting off a triangle $T_i, \; i=1, \ldots, T$ and reaching the \gls{rx}.
Then, second order reflections starting from the \gls{tx}, reflecting first on triangle $T_{i_1}, \; i_1=1, \ldots, T$ and then on triangle $T_{i_2}, \; i_2=1, \ldots, T, \; i_2 \neq i_1$ to finally reach the \gls{rx}, and so on up to a maximum reflection order $R$.

All the reflections can be encoded in a \textit{reflection tree}, as represented  in \cref{fig:reflectionTree}, where each node of the tree corresponds to a possible ray, the node depth corresponds to the reflection order $r$ starting from 0 from the root of the tree, and the path starting from the root describes the ordered tuple of reflecting triangles to be tested.

The complexity of a single channel instance, then, is determined by the total number of operations required for all the nodes of the reflection tree.
At depth $r=0$ we consider only the direct ray, at depth $r=1$ we consider the $T$ possible first-order reflections, then, in general, at depth $r \geq 2$ we consider $T(T-1)^{r-1} < T^r$ possible $r$-order reflections.
The total number of operations per channel instance is thus upper bounded as 
\begin{equation}\label{eq:single_channel_instance_complexity}
\begin{aligned}
  & \sum_{r=0}^{R} (r + (r+1)T) T^r = T \sum_{r=0}^{R} T^r + (T+1) \sum_{r=0}^{R} r T^r\\
  &= T \frac{T^{R+1} - 1}{T - 1} + (T+1) \frac{T(R T^{R+1} - (R+1)T^R + 1}{(T-1)^2}\:\xrightarrow{T \to \infty} \: T^{R+1} + R T^{R+1},
\end{aligned}
\end{equation}
thus denoting a complexity per channel instance equal to $\mathcal{O}\qty(R T^{R+1})$, and a total complexity equal to $\mathcal{O}\qty(t N^2 R T^{R+1})$.
The last step in \cref{eq:single_channel_instance_complexity} is justified by considering that typical values for $R$ and $T$ are in the order of $1$--$4$ and $100$--$10\,000$, respectively, thus making $T$ the dominating term of the formula.


\subsection{Quasi-Deterministic Model Integration} 
\label{sub:qd_model}
\glsunset{qd}

Besides the deterministic \gls{rt}, in \cref{sec:performance_results} we also evaluate the performance with and without a stochastic model for diffuse components, which alone can account for up to 40\% of the total received power according to measurements campaigns~\cite{lai2019methodology}.
It is based on the specifications proposed for IEEE 802.11ay channel modeling~\cite{tgay_channel_model} with parameters obtained from accurate measurement campaigns~\cite{lai2019methodology}.
Further details are given in~\cite{lecci2020qd}.
Please note that in this section, boldface letters denote random variables while non-boldface letters denote deterministic variables or realizations of random variables.

The \gls{qd} model is built upon the deterministic channel, as described in \cref{sub:the_mmwave_ray_tracer}.
For first-order reflections, the path gain of the \gls{dray} is equal to
\begin{equation}\label{eq:pg_db}
  \vb*{PG}_{0, \mathrm{dB}} = 20 \log_{10}\qty(\frac{\lambda_c}{4\pi \ell_{\rm ray}}) - \vb*{RL}_{\rm dB},
\end{equation}
where $\lambda_c$ is the wavelength of the carrier frequency, $\ell_{\rm ray}$ is the total ray length, and $\vb*{RL} \sim \ricianDistrib{s_{RL}}{\sigma_{RL}}$ is the Rician-distributed random reflection loss factor given by the reflecting surface's material, whose parameters have been fitted from measurements.
The computed \glspl{dray} will then be the baseline for the multipath components randomly generated by the \gls{qd} model.
If present, the direct ray is treated separately as it does not generate any diffuse component.
For first-order reflections, a cluster can then be defined as the set of rays including a \gls{dray} and its diffuse components.
The total number of \glspl{mpc} of a given cluster will be $N_{\rm MPC} = N_{\rm pre} + 1 + N_{\rm post}$, including pre-cursors (i.e., diffuse components that are received before the \gls{dray}), main cursor (i.e., the \gls{dray}), and post-cursors (i.e., received after the \gls{dray}).

\begin{figure}[t]
  \centering
  \setlength\belowcaptionskip{-.6cm}
  \setlength\fwidth{0.6\columnwidth} 
  \setlength\fheight{0.2\columnwidth}
%
%
\begin{tikzpicture}
\pgfplotsset{every tick label/.append style={font=\scriptsize}}

\begin{axis}[%
width=0.951\fwidth,
height=\fheight,
at={(0\fwidth,0\fheight)},
scale only axis,
xmin=-1,
xmax=2,
xtick={-0.3, 0, 0.4},
xticklabels={{$\tau_{1, pre}$}, {$\tau_0$}, {$\tau_{1, post}$}},
ymin=0,
ymax=12,
ytick={10},
yticklabels={{$PG_0$}},
axis background/.style={fill=white},
axis lines=left, 
xmajorgrids,
ymajorgrids,
legend style={legend cell align=left, align=left, draw=white!15!black}
]
\addplot[ycomb, color=black, mark=*, mark options={solid, fill=black, black}, forget plot] table[row sep=crcr] {%
0 10\\
};

\draw [|-|] (-0.05, 10) -- (-0.05, 8.5);
\node[left, align=right] at (axis cs:-0.05,9.25) {$K_{pre}$};

\addplot [color=black, line width=1pt]
  table[row sep=crcr]{%
0 8.5\\
-1  5.5\\
};

\addplot [color=black, line width=0.5pt]
  table[row sep=crcr]{%
-0.4 7.3\\
-0.6 7.3\\
-0.6 6.7\\
};
\node[right, align=left] at (axis cs:-0.7,7.9) {$\gamma_{pre}$};

\draw [|-|] (-0.9, 5.4) -- (-0.9, 6.2);
\node[right, align=left] at (axis cs:-0.9, 5.2) {$S_{pre}$};

\draw [|-|] (0.05, 10) -- (0.05, 8);
\node[right, align=left] at (axis cs:0.05,9) {$K_{post}$};

\addplot [color=black, line width=1pt]
  table[row sep=crcr]{%
0 8\\
2 6\\
};

\addplot [color=black, line width=0.5pt]
  table[row sep=crcr]{%
1 7\\
1.4 7\\
1.4 6.6\\
};
\node[right, align=left] at (axis cs:1,7.5) {$\gamma_{post}$};

\draw [|-|] (1.9, 5.4) -- (1.9, 6.8);
\node[left, align=right] at (axis cs:1.9, 5.5) {$S_{post}$};

\addplot[scatter, only marks, color=blue, mark=*, mark options={solid, fill=blue, blue}, forget plot] table[row sep=crcr] {%
0.4 8\\
-0.3 7\\
};

\draw [<->] (0, 0.4) -- (0.4, 0.4);
\node [above right, align=left] at (axis cs:0.2, 0.4) {\scriptsize $\Delta_{1, post}$};

\draw [<->] (0, 0.4) -- (-0.3, 0.4);
\node [above left, align=right] at (axis cs: -0.15, 0.4) {\scriptsize $\Delta_{1, pre}$};

\end{axis}
\end{tikzpicture}%
\setlength\belowcaptionskip{-.6cm}
  \caption{Graphical representation of the \gls{qd} parameters.}
  \label{fig:qdFig}
\end{figure}
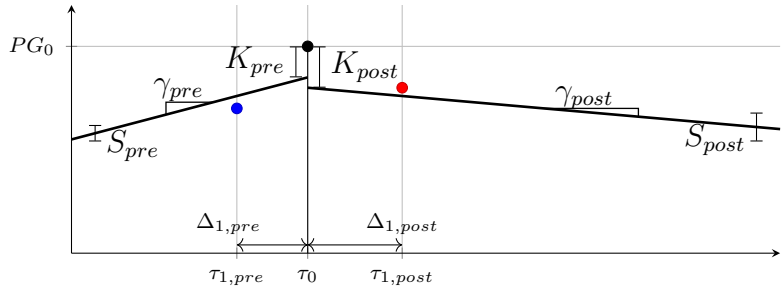

The arrival time of the $i$-th \gls{mpc} $\vb*{\tau}_i$ is modeled as a Poisson process, meaning that their inter-arrival times are independent and exponentially distributed, with an exponential power law, i.e.,
\begin{equation}\label{eq:pg_lin}
  \vb*{PG}_{i} = \frac{\vb*{PG}_0}{\vb*{K}} \cdot
                            \exp \qty(-\frac{|\vb*{\tau}_{i} - \tau_0|}{\vb*{\gamma}} + \vb*{S} ),
\end{equation}
where $\vb*{K}_{\mathrm{dB}} \sim \ricianDistrib{s_K}{\sigma_K}$ is a loss factor, $\tau_0$ is the \gls{dray} arrival delay, $\vb*{\gamma} \sim \ricianDistrib{s_\gamma}{\sigma_\gamma}$ is the power-delay decay constant, $\vb*{S} \sim \normalDistrib{0}{\vb*{\sigma}_s^2}$ is the power-delay decay standard deviation, and $\vb*{\sigma}_s \sim \ricianDistrib{s_{\sigma_s}}{\sigma_{\sigma_{s}}}$.
A graphic representation of these parameters is shown in \cref{fig:qdFig}.

Finally, the departure and arrival angles follow a Laplacian distribution around the \gls{dray}, while the phase shift $\vb*{\phi}_i$ due to both diffusion and Doppler shift is assumed to be $\mathcal{U}[0,2\pi)$ independently for each diffuse \gls{mpc}.

In general, for the $r$-th reflection order, with $r >1$, the definitions presented above are heuristically adapted, considering independent statistics for the different reflectors.
To reduce the computational complexity of the complete model, the multi-bounce model neglects diffuse rays beyond the first order, given their fast increasing attenuation.
Instead, only diffuse rays generated directly by the deterministic ray are taken into account, each generated with the \gls{qd} parameters relative to the impinging reflecting surface.
Moreover, we assume that every diffuse component closely follows the main cursor, thus reflecting on the same reflectors.
Consequently, every reflector produces $N_{\rm pre} + N_{\rm post}$ diffuse components, thus yielding  $N_{\rm MPC} \sim r (N_{\rm pre} + N_{\rm post})+1$.
Finally, diffuse components are independently extracted at each timestep.

\section{MmWave Channel Simplifications} 
\label{sec:mmwave_channel_simplifications}
The accuracy of the \gls{moi}-based ray-traced channels --- especially when considering the \gls{qd} model --- comes at a high computational cost, which may limit the scalability of the simulations, especially when considering a very large number of devices.
In this perspective, the main objective of this work is to evaluate how channel simplifications affect the results of link-level and network-level simulations while speeding-up the overall simulation runtime.
In this section, we present two techniques that were designed with this objective in mind~\cite{testolina2019scalable}, starting from the analysis of the complexity discussed in~\cref{sub:the_mmwave_ray_tracer}.

The overall computational complexity of a simulation of $N$ TX/RX nodes lasting $t$ time steps in a scenario composed of $T$ triangles when considering up to $R$ reflections is $\mathcal{O} \qty(tN^2RT^{R+1})$.
In this approach, $t$ and $N$ are simulation parameters set by the user, $T$ is determined by the CAD model of the simulated environment, while the maximum reflection order $R$ for the ray-tracing depends on the channel model.
Understanding how different values of $R$ affect the low- and high-layer performance metrics of the network with respect to the model complexity is the core of the first simplification strategy proposed in this work, referred to as \emph{Maximum Reflection Order Reduction} (see~\cref{sec:max_ref}).

Differently, the second technique aims at reducing the number of rays between a pair of nodes.
Specifically, given a set of $M$ rays connecting two nodes, we propose and evaluate a selection criteria that aims to reduce the number of \glspl{mpc} to $M'<M$,  to reduce the overall simulation time.
This operation, named \emph{\gls{mpc} Thresholding}, is applied on a time-step basis (see~\cref{sec:mpc_thresh}).

The rationale behind both strategies is to decrease the number of \glspl{mpc} by removing the least significant ones, i.e., those with the lowest power, as they are expected to provide a limited contribution to the overall received signal strength.
More details will be given in the following subsections.

\subsection{Maximum Reflection Order Reduction}
\label{sec:max_ref}
Each reflection of the \glspl{mpc} on a surface is associated to a partial power loss and an increased path length, translating into a higher path loss.
Namely, from \cref{eq:pg_db}, the path gain for a ray reflected on $r$ surfaces is
\begin{equation}\label{eq:pg_multiple_reflections}
  \vb*{PG}_{\mathrm{dB}} = 20 \log_{10}\qty(\frac{\lambda_c}{4\pi \sum_{i=1}^{r}\ell_{i}}) - \sum_{r=1}^{r}\vb*{RL}_{i,\rm dB},
\end{equation}
where $\ell_{i}$ is the length of the segment associated with the $i$-th reflection.
The summation is decomposed into two terms to underline the different contributions: both the path length and the reflection losses degrade the path gain when the reflection order increases.
Therefore, it is reasonable to assume that \glspl{mpc} that bounce across multiple scattering surfaces have a low contribution to the overall received power, and can be omitted from the \gls{rt} computations.
Setting the maximum reflection order to $R'<R$, the \gls{rt} complexity is decreased to $\mathcal{O} \qty(tN^2R'T^{R'+1})<\mathcal{O} \qty(tN^2RT^{R+1})$ with significant savings in terms of computation time, given the super-exponential dependency of the complexity on $R$.

\subsection{MPC Thresholding}
\label{sec:mpc_thresh}

Beside the reflection order, there are other elements that contribute to reducing the \gls{mpc} path gain.
For example, even if $R$ is small, in large scenarios surfaces that are located far from the TX and RX nodes are associated to \glspl{mpc} with a large path length (i.e., the first term in \cref{eq:pg_multiple_reflections}).
As the path gain from these scatterers is much smaller than that from close-by reflecting surfaces, it is possible to prune them from the list of \glspl{mpc} to compute.

For these \glspl{mpc}, the path gain plays a key role and can thus be used as an indicator to perform the selection of the most significant rays.
Specifically, the selection is performed at each time step considering a threshold $\gamma_{\rm th}$, according to the following rules:
\begin{itemize}
  \item the strongest ray is identified and the corresponding path gain $PG_{\text{strong}}$ is computed;
  \item all the other \glspl{mpc} are identified.
  Note that $PG_i<PG_{\text{strong}}$ for every \gls{mpc} $i$.
  \item the selection is carried out discarding every \gls{mpc} $i$ such that $PG_i-PG_{\text{strong}}<\gamma_{\rm th}$.
\end{itemize}

While being more general than the previous approach, as it tackles directly the weakest, least significant rays, this strategy requires the computation of all the geometric paths in order to obtain the path gain list.
However, this method, which removes $(M-M')$ rays, eases the load on the subsequent \gls{rt} operations, i.e., the obstruction check, reducing by a factor of $\frac{M-M'}{M}$ the complexity of each time step.
In fact, following the same logic as in \cref{sub:the_mmwave_ray_tracer} for every ray of reflection order $r$, after the $r$ geometrical operations required to compute the path of the ray, none of the $(r+1)T$ obstruction checks are performed if the ray is discarded.
Updating \cref{eq:single_channel_instance_complexity} with $r\ge 0$ instead of $r+(r+1)T$ operations per ray, the complexity can be reduced to $\mathcal{O} \qty(tN^2RT^{R})$ and thus by a factor up to $T$.
Note that, whereas this approach achieves a constant factor improvement, $T$ can be in the order of tens to thousands, depending on the details included in the \gls{cad} file and on the adopted triangulation, thus being one of the dominant terms in the complexity expression.

Absolute thresholding can also be used to limit the number of extremely weak rays similarly to the previous technique.
This approach can be useful when considering high values for $R$, low values for $\gamma_{\rm th}$, and especially when using the \gls{qd} model.
In this case, setting a threshold $\Gamma_{\rm th}$, every \gls{mpc} $i$ such that $PG_i<\Gamma_{\rm th}$ is discarded.

The complexity of the \gls{rt} can be significantly reduced thanks to the removal of \glspl{mpc} and to the reduction of the maximum reflection order $R$.
On the other hand, these simplifications degrade the accuracy of the simulation results at the different levels of the network stack.
In the remainder of this paper, we will quantify this trade-off for three realistic propagation scenarios.
The overall end-to-end network performance and the runtime of the simplified \gls{rt} settings will be compared with those of the complete, non-simplified channel traces. 


\section{Performance Evaluation} 
\label{sec:performance_results}

This section reports the details of an extensive performance evaluation aimed at understanding the impact that the simplifications introduced in \cref{sec:mmwave_channel_simplifications} have at different layers of the protocol stack.
We first describe the scenarios and tools used for the performance evaluation (\cref{sub:simulation_scenarios}), then the link and higher layer performance (\cref{sub:link_level_performance_results,sub:end_to_end_performance_results}, respectively), and conclude with the computational performance given by the simplifications (\cref{sub:performance}) and guidelines for the more efficient design configurations (\cref{sub:design_guidelines}).

\subsection{Simulation Scenarios} 
\label{sub:simulation_scenarios}

Three representative scenarios with distinctive features have been selected to make the performance evaluation as general as possible.
Their main characteristics are hereby described, and are summarized in \Cref{tab:scenarios_summary}.
Without loss of generality, only downlink channels are considered.

\begin{table}[t!]
  \caption{Characteristics of the simulated scenarios.
  Other important simulation parameters are bandwidth $B=400$~MHz, and noise figure $NF=9$~dB.
  All nodes of a given scenario transmit with the same power $P_{\rm tx}$.}
  \label{tab:scenarios_summary}
  \centering
  \begin{tabular}{lcccccccc}
    \toprule
    \textbf{Scenario} & \textbf{Time steps} $t$ & \textbf{\acrshort{los}} & \textbf{\acrshort{nlos}} & \textbf{Environment} & \textbf{RX velocity} & \textbf{Interferer} & $T$ & $P_{\rm TX}$\\
    \midrule
    Indoor1     & 3133  & \cmark  & \xmark  & Indoor   & 1.2 [m/s]  & \xmark  & 12  & 20~dBm\\
    L-Room      & 3831  & \cmark  & \cmark  & Indoor   & 1.2 [m/s]  & \cmark  & 16  & 20~dBm\\
    ParkingLot  & 3971  & \cmark  & \xmark  & Outdoor  & 4.17 [m/s] & \cmark  & 284 & 30~dBm\\
    \bottomrule
  \end{tabular}
\end{table}

\begin{figure*}[tbp]
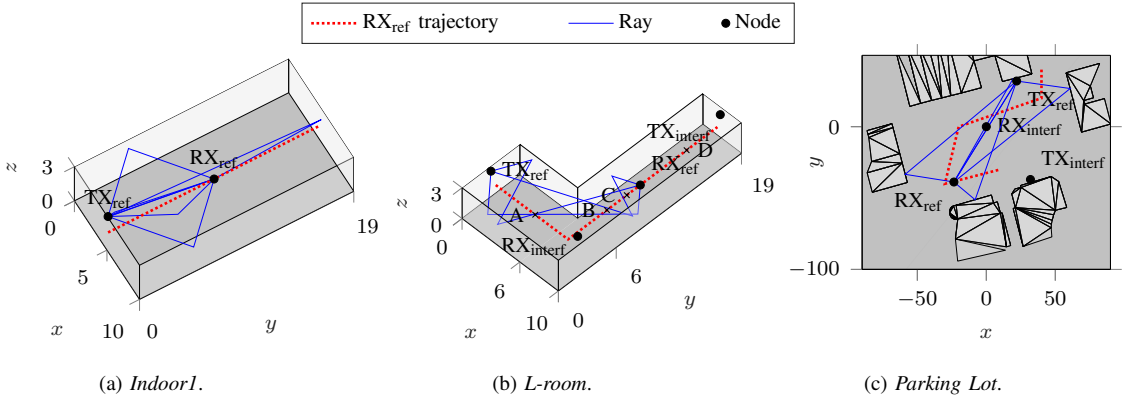

  \centering
  \begin{subfigure}[t]{\textwidth}
    \centering
%

\begin{tikzpicture}
\pgfplotsset{every tick label/.append style={font=\scriptsize}}

\begin{axis}[%
width=0,
height=0,
at={(0,0)},
scale only axis,
xmin=0,
xmax=0,
xtick={},
ymin=0,
ymax=0,
ytick={},
zmin=0,
zmax=0,
ztick={},
axis background/.style={fill=white},
legend style={legend cell align=center, align=center, draw=white!15!black, font=\scriptsize, at={(0, 0)}, anchor=center, /tikz/every even column/.append style={column sep=2em}},
legend columns=3,
]

\addplot3 [color=red, densely dotted, line width=1pt]
 table[row sep=crcr] {%
-1 -1 -1\\
};
\addlegendentry{RX\textsubscript{ref} trajectory}

\addplot3 [color=blue, line width=.1pt]
 table[row sep=crcr] {%
-1 -1 -1\\
};
\addlegendentry{Ray}

\node[right, align=left, font=\bfseries]
at (axis cs:0,3,2.5) {TX};
\addplot3[only marks, mark=*, mark options={}, mark size=1.5000pt, color=black, fill=black] table[row sep=crcr]{%
x y z\\
-1 -1  -1\\
};
\addlegendentry{Node}

\end{axis}
\end{tikzpicture}%
  \end{subfigure}
  \\
  \hfill
  \begin{subfigure}[t]{0.25\textwidth}
      \centering
      \setlength\fwidth{0.9\columnwidth}
      \input{img/Indoor1.tex}
      \caption{\textit{Indoor1}.}
      \label{fig:indoor1}
  \end{subfigure}
  \hfill
  \begin{subfigure}[t]{0.25\textwidth}
      \centering
      \setlength\fwidth{0.9\columnwidth}
      \input{img/L-room.tex}
      \caption{\textit{L-room}.}
      \label{fig:l-room}
  \end{subfigure}
  \hfill
  \begin{subfigure}[t]{0.25\textwidth}
      \centering
      \setlength\fwidth{0.8\columnwidth}
      \input{img/ParkingLot.tex}
        \caption{\textit{Parking Lot}.}
        \label{fig:parking-lot}
  \end{subfigure}
  \hfill
  \setlength\belowcaptionskip{-.6cm}
  \caption{Visual representations of our simulation scenarios.
  Distances are measured in meters.}
  \label{fig:scenarios}
\end{figure*}

\begin{enumerate}
  \item \textit{Indoor1}: The most basic scenario, with a rectangular room (see \cref{fig:indoor1}) of size $10 \mbox{ m} \times 19 \mbox{ m} \times 3 \mbox{ m}$.
  The \gls{tx} is positioned close to the ceiling at $(5, 0.1, 2.9)$~m. The \gls{rx}, at height $1.5$~m, moves away from it at a speed of $1.2$~m/s along a straight line.
  This scenario was deliberately designed to be simple, to analyze the propagation characteristics simulated by the \gls{rt} focusing on the received power pattern when different simplifications are used;
  \item \textit{L-Room}: An L-shaped hallway (see~\cref{fig:l-room}).
  A static \gls{tx}, placed at $(0.2,3,2.5)$~m, transmits to the reference \gls{rx} that moves away from it at a speed of $1.2$~m/s across the corridor.
  The shape of the room is such that the \gls{rx} is in \gls{nlos} condition for a significant portion of the path.
  Moreover, in order to analyze the impact of interference on the network performance, a second \gls{tx} placed at $(8,18.8,2.5)$~m and acting as interferer, communicates with an \gls{rx} at $(9,3,1.5)$~m.
  Furthermore, the shape of the room plays an important role when comparing the proposed simplification techniques, as it may create blind spots where no signal is received;
  \item \textit{Parking-Lot}: The only outdoor scenario, representing a parking area of about $120 \mbox{ m} \times 70 \mbox{ m}$ enclosed by buildings (see \cref{fig:parking-lot}).
  The reference \gls{tx} transmits from an access point placed at $3$~m height on a building to the \gls{rx}, that moves at a speed of $4.17$~m/s ($15$~km/h) around the parking lot.
  By far the largest scenario, it makes it possible to analyze the effect of the simplifications in terms of time savings when the CAD file contains a large number $T$ of triangles.
  Moreover, the reference \gls{rx} moves at a much higher speed than in the previous scenarios, as in a basic vehicular scenario.
\end{enumerate}
The ray-tracing parameters for the \textit{Indoor1}~\cite{lai2019methodology} and for the \textit{Parking Lot} scenarios are obtained from detailed measurement campaigns.

For each scenario, the \gls{rt} and \gls{qd} model software described in \cref{sec:ray_tracing_at_mmwaves} have been used to generate the channel instances for the specified devices and mobility patterns sampled every 5~ms.
These traces were then integrated in a custom MATLAB simulator~\cite{testolina2019scalable,testolina2020simplified} to evaluate metrics at the link layer (e.g., the \gls{sinr}), and with the mmWave module~\cite{mezzavilla2018end} of the ns-3 network simulator~\cite{henderson2008network} to investigate the performance of the full protocol~stack.

Both simulators rely on the computation of the channel matrix $\vb*{H}$ to describe the channel obtained from the \glspl{mpc} provided by the \gls{rt} and \gls{qd}.
Given $M$ rays, with path power gain $\PG_m$, phase shift $\Phi_m$, delay $\tau_m$, and angles of departure $\vb*{\AoD}_m$ and angles of arrival $\vb*{\AoA}_m$, the matrix for the carrier frequency $f_c$ is computed as
\begin{equation}\label{eq:matrix}
   \vb*{H} = \sum_{m = 1}^{M} \sqrt{PG_m} \, e^{j(-2\pi\tau_m f_c + \Phi_m)} \, \vb*{a}_{\rm rx}^*(\vb*{\AoA}_m) \, \vb*{a}^{H}_{\rm tx}(\vb*{\AoD}_m),
 \end{equation}
where $\vb*{a}_{\rm rx}(\vb*{\theta})$ and $\vb*{a}_{\rm tx}(\vb*{\theta})$ are the receiver and transmitter array responses in the 3D angle $\vb*{\theta}$, $(\cdot)^*$ is the conjugate operator, and $(\cdot)^H$ is the Hermitian operator.
The \gls{sinr} for the link between the transmitter $t$ and the receiver $r$ is
\begin{equation}
    \Gamma_{t,r} = \frac{P_{{\rm tx}, t} \, \vb*{w}_{t,r}^T \vb*{H}_{t,r} \vb*{w}_{r,t}}
    {\sum_{m \ne t}P_{{\rm tx}, m} \, \vb*{w}_{m,*}^T \vb*{H}_{m,r} \vb*{w}_{r,t} + N_0 B F},
\end{equation} 
where $P_{{\rm tx}, t}$ is the transmit power of device $t$, $\vb*{w}_{i,j}$ is the beamforming vector used by device $i$ to communicate with device $j$ (and $\vb*{w}_{i,*}$ is used with abuse of notation to indicate the beamforming vector used by device $i$ to transmit towards a connected device, or $\vb*{0}$ if $i$ is not transmitting), $N_0$ is the noise power spectral density, $B$ is the bandwidth, and $F=10^{\frac{NF}{10}}$ is the noise factor of the receiver.

For ns-3, we extended the channel model implementation described in~\cite{zugno2020implementation} to account for a generic channel matrix computed, in this case, as expressed in \cref{eq:matrix}\footnote{The implementation can be found at \url{https://github.com/signetlabdei/qd-channel}.}.
In the performance evaluation, this channel model has been combined with the 3GPP-like protocol stack of the 5G mmWave module for ns-3~\cite{mezzavilla2018end}, which features physical and \gls{mac} layers with an \gls{ofdm}-based frame structure, dynamic \gls{tdd}, \gls{amc}, and several scheduler implementations.
Besides, the \glspl{ue} and base stations protocol stacks are completed by 3GPP \gls{rlc} and \gls{pdcp} layers, together with a realistic control plane based on the \gls{rrc} layer which supports mobility-related procedures~\cite{polese2017jsac}.
We consider two configurations for the uniform planar antenna arrays: large arrays, comprising 8$\times$8 elements for the \glspl{tx} and 4$\times$4 elements for the \glspl{rx}, and small arrays, comprising 2$\times$2 arrays for both \glspl{tx} and \glspl{rx}, all of them with omni-directional elements spaced apart by $\frac{\lambda}{2}$.
The planes on which all planar arrays lie are parallel to the $y$-$z$ plane with a fixed orientation throughout the simulation.
Finally, thanks to the integration with ns-3, it is possible to equip the \glspl{ue} with the TCP/IP stack and applications which connect to remote servers in the Internet.

The results shown in the following sections will assume as default parameters, unless stated differently, a maximum reflection order $R=3$ for the \textit{Parking Lot} scenario, and $R=4$ for the others, a relative threshold $\gamma_{\rm th}=-\infty$~dB, a conservative absolute threshold $\Gamma_{\rm th}=-200$~dB, 
a large antenna array configuration, only deterministic rays (i.e., no \gls{qd} model), and a \gls{udp} stream with an offered traffic equal to 800~Mbps.
The beamforming is based on the \gls{svd} of the channel matrix $\vb*{H}$.

\subsection{Link-Level Performance Results} 
\label{sub:link_level_performance_results}

The first step towards proper protocol design is gaining a deep understanding of how the proposed ray-tracing simplifications impact the link-level performance of the network, neglecting, at this stage, the effects at the upper layers.
In this perspective, we are interested in investigating how the strategies described in \cref{sec:mmwave_channel_simplifications} result in different \gls{sinr} regimes.

\begin{figure*}[tbp]
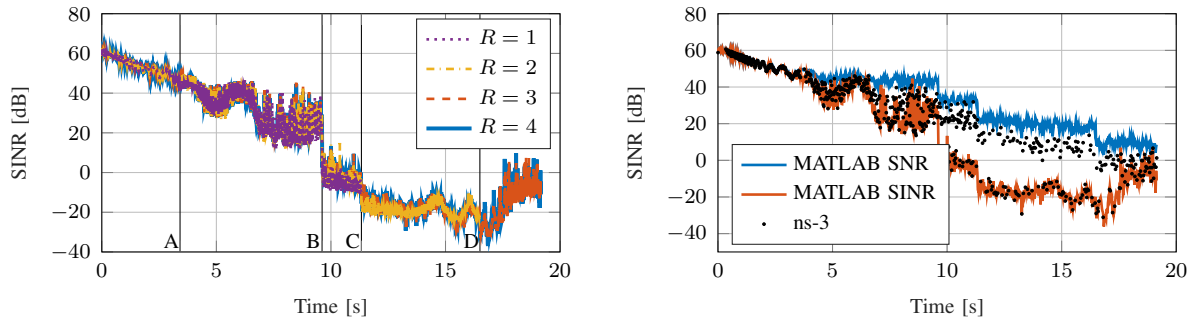
 
  \hfill
  \begin{subfigure}[t]{0.48\textwidth}
      \centering
      \setlength\fwidth{0.8\columnwidth}
      \setlength\fheight{0.4\columnwidth}
      \input{img/lroom_sinr_time.tex}
      \caption{SINR vs.
time for different values of $R$, with $\gamma_{\rm th} = -\infty$~dB.}
      \label{fig:lroom_sinr_time}
  \end{subfigure}
  \hfill
  \begin{subfigure}[t]{0.48\textwidth}
      \centering
      \setlength\fwidth{0.8\columnwidth}
      \setlength\fheight{0.4\columnwidth}
      \input{img/lroom_sinr_matlabVsNs3.tex}
      \caption{SINR vs. time considering MATLAB and ns-3 simulations.}
      \label{fig:lroom_sinr_matlabVsNs3}
  \end{subfigure}
  \hfill
  \setlength\belowcaptionskip{-.6cm}
  \caption{Evolution of the SINR experienced when the test RX moves in the \emph{L-room} scenario along the path described in \cref{fig:l-room}.}
  \label{fig:lroom_sinr}
\end{figure*}%

From~\cref{fig:lroom_sinr_time}, which plots the temporal evolution of the \gls{sinr} experienced when the \gls{rx} moves in the \emph{L-Room} scenario along the path described in~\cref{fig:l-room}, it is possible to see that the impact of $R$ is certainly non-negligible: the trend of the SINR visibly changes when progressively reducing the number of reflections per MPC.
Moreover, we see that the SINR evolves consistently with the mobility of the RX.
The SINR indeed drops by more than 30 dB when the RX loses its \gls{los} condition (position B in~\cref{fig:l-room}), while the SINR degradation that is experienced at time $t=3.4$ s (position A in~\cref{fig:l-room}) is due to the interference from the $\text{RX}_{\rm interf}$.
Rapid fluctuations within the SINR trace are then due to the fact that different \glspl{mpc} travel different paths.
At 60~GHz, where the wavelength is as short as $\lambda = 5$ mm, even small variations of the path length between the direct ray and the reflected ones from the back wall (behind the TX), side walls, ceiling, and floor of the room, may result in strong fading.
\cref{fig:lroom_sinr_time} also shows that the impact of the \gls{rt} simplifications is particularly evident when the RX operates in \gls{nlos}: in this region, in fact, the received power drops to zero when first-order and second-order reflections  are removed (positions C and D in \cref{fig:l-room}, respectively).

For completeness, in \cref{fig:lroom_sinr_matlabVsNs3} we compare the metrics from the MATLAB (straight lines) and the ns-3 (dots) simulations.
The MATLAB \gls{sinr} assumes an always-on interferer, representing a lower bound for the \gls{sinr} metric, while the MATLAB \gls{snr} metric assumes an interference-free channel.
On the other hand, ns-3 models a realistic transmission pattern for the primary and interferer links, which could occupy the channel in overlapping, partially overlapping, or non-overlapping time instants.
Therefore, for each time instant, the \gls{sinr} generated by the ns-3 simulations is lower and upper bounded by the MATLAB \gls{sinr} and \gls{snr}, respectively.
On the other hand, if the reference TX and the second TX/RX pair communicate in non-overlapping time slots, interference is minimized and the ns-3 SINR curve closely approximates the MATLAB configuration without interferer.

The same conclusions can be drawn from \cref{fig:scenarios_sinr_cdf}, which illustrates the \gls{cdf} of the SINR for the three scenarios described in \cref{sub:simulation_scenarios} as a function of $R$.
Specifically, the \emph{L-Room} scenario, due to the presence of \gls{nlos} conditions, is again the only one for which a reduction of the \glspl{mpc} of the channel may result in a significant reshape of the CDF of the SINR with respect to the baseline configuration with no simplifications.
\begin{figure*}[t!] 
  \hfill
\setlength\belowcaptionskip{-.6cm}
  \begin{minipage}[t]{.45\textwidth}
    \centering
    \setlength\fwidth{0.8\columnwidth}
    \setlength\fheight{0.4\columnwidth}
    \input{img/scenarios_sinr_cdf.tex}
    \caption{Cumulative Distribution Function of the SINR in different scenarios vs.
$R$, with $\gamma_{\rm th} = -\infty$~dB.}
    \label{fig:scenarios_sinr_cdf}
  \end{minipage}
  \hfill
  \begin{minipage}[t]{.45\textwidth} 
    \centering
    \setlength\fwidth{0.8\columnwidth}
    \setlength\fheight{0.4\columnwidth}
%
%
\definecolor{mycolor1}{rgb}{0.00000,0.44700,0.74100}%
\definecolor{mycolor2}{rgb}{0.85000,0.32500,0.09800}%
\begin{tikzpicture}

\pgfplotsset{every tick label/.append style={font=\scriptsize}}

\begin{axis}[%
width=0.961\fwidth,
height=\fheight,
at={(0\fwidth,0\fheight)},
scale only axis,
bar shift auto,
xmin=0.5,
xmax=4.5,
xtick={1,2,3,4},
xlabel style={font=\scriptsize\color{white!15!black}},
xticklabels={{$-\infty$},{$-40$},{$-25$},{$-15$}},
xlabel={$\gamma_{\rm th}$},
ymin=-2,
ymax=18,
ylabel style={font=\scriptsize\color{white!15!black}},
ylabel={Average SINR [dB]},
axis background/.style={fill=white},
xmajorgrids,
ymajorgrids,
legend style={legend cell align=left, align=left, draw=white!15!black, at={(0.5,0.97)},/tikz/every even column/.append style={column sep=0.35cm}, font=\scriptsize,
  anchor=north ,legend columns=-1},
]
\addplot[ybar, bar width=0.2, fill=mycolor1, draw=black, area legend] table[row sep=crcr] {%
1	-0.328777706024901\\
2	-0.360429062654857\\
3	-0.292059019732808\\
4	-0.302207167877643\\
};
\addlegendentry{Small Arrays}

\addplot [color=black, draw=none, xshift=-0.20cm, forget plot]
 plot [error bars/.cd, y dir = both, y explicit]
 table[row sep=crcr, y error plus index=2, y error minus index=3]{%
1	-0.328777706024901	0	0\\
2	-0.360429062654857	0	0\\
3	-0.292059019732808	0	0\\
4	-0.302207167877643	0	0\\
};
\addplot[ybar, bar width=0.2, fill=mycolor2, draw=black, area legend] table[row sep=crcr] {%
1	13.0395509075006\\
2	12.8601163247711\\
3	13.6410035455084\\
4	13.3663852603162\\
};
\addlegendentry{Large Arrays}

\addplot [color=black, draw=none, xshift=0.20cm, forget plot]
 plot [error bars/.cd, y dir = both, y explicit]
 table[row sep=crcr, y error plus index=2, y error minus index=3]{%
1	13.0395509075006	0	0\\
2	12.8601163247711	0	0\\
3	13.6410035455084	0	0\\
4	13.3663852603162	0	0\\
};
\end{axis}
\end{tikzpicture}%
    \caption{Average SINR vs.
$\gamma_{\rm th}$ as a function of the antenna architecture in the \emph{L-Room} scenario, with $R=4$.}
    \label{fig:lroom_avg_sinr_bars}
  \end{minipage}
  \hfill
\end{figure*}%
On the other hand, both the \emph{Indoor1} and the \emph{Parking Lot} scenarios are able to preserve the LoS for the whole duration of the simulation, thereby making it possible for the signal to propagate with a minor impact on the received power even when limiting the number of reflections $R$ per MPC.
Notice that the 20 dB gap of SINR between the  \emph{Parking Lot} and the \emph{Indoor1}  configurations is due to the larger distance between the TX and the RX, and to the reflecting surfaces (e.g., buildings) in the outdoor scenario.

Finally, in Fig.~\ref{fig:lroom_avg_sinr_bars} we plot the SINR vs.
the relative threshold $\gamma_{\rm th}$ as a function of the antenna size at the TX and the RX.
As expected,  the SINR increases when increasing the number of antenna elements, which increases the beamforming gain.
Also, while the impact of $R$ severely affects the SINR in the \emph{L-Room} scenario, increasing the relative threshold $\gamma_{\rm th}$ to reduce the number of MPCs to be processed by the \gls{rt} has negligible deterioration at the link-level, while speeding up the simulation, as will be discuss in \cref{sub:performance}.



\subsection{End-to-End Performance Results} 
\label{sub:end_to_end_performance_results}

Many of the conclusions we derived from the link-level performance metrics in \cref{sub:link_level_performance_results} can be extended to the end-to-end ones, namely throughput and delay at the \gls{pdcp} layer.
Statistics have been collected at this layer since they can easily profile both \gls{udp} and \gls{tcp} traffic indistinctly and are very close to the application layer performance, without the addition of extra delays due to the specific architecture of the simulation scenario.

In this section we study three types of traffic, namely full-buffer \gls{tcp} traffic and \gls{udp} traffic with a \gls{cbr}  of 100~Mbps and 800~Mbps.
The latter was chosen to be the default for the results shown in this section, unless stated differently.
Both throughput and delay are averaged over 100~ms windows.

\begin{figure*}[tbp] 
  \centering
  \begin{subfigure}[t]{\textwidth}
      \centering
%
%
\definecolor{mycolor1}{rgb}{0.00000,0.44700,0.74100}%
\definecolor{mycolor2}{rgb}{0.85000,0.32500,0.09800}%
\definecolor{mycolor3}{rgb}{0.92900,0.69400,0.12500}%
\definecolor{mycolor4}{rgb}{0.49400,0.18400,0.55600}%

\begin{tikzpicture}
\pgfplotsset{every tick label/.append style={font=\scriptsize}}

\begin{axis}[%
width=0,
height=0,
at={(0,0)},
scale only axis,
xmin=0,
xmax=0,
xtick={},
ymin=0,
ymax=0,
ytick={},
zmin=0,
zmax=0,
ztick={},
axis background/.style={fill=white},
legend style={legend cell align=center, align=center, draw=white!15!black, font=\scriptsize, at={(0, 0)}, anchor=center, /tikz/every even column/.append style={column sep=2em}},
legend columns=4,
]

\addplot [color=mycolor4, dotted, line width=1pt]
  table[row sep=crcr]{%
0 0\\
};
\addlegendentry{$R = 1$}

\addplot [color=mycolor3, dashdotted, line width=1pt]
  table[row sep=crcr]{%
0 0\\
};
\addlegendentry{$R = 2$}

\addplot [color=mycolor2, dashed, line width=1pt]
  table[row sep=crcr]{%
0 0\\
};
\addlegendentry{$R = 3$}

\addplot [color=mycolor1, line width=1pt]
  table[row sep=crcr]{%
0 0\\
};
\addlegendentry{$R = 4$}

\end{axis}
\end{tikzpicture}%
  \end{subfigure}
  \\
  \hfill
  \begin{subfigure}[t]{0.45\textwidth}
      \centering
      \setlength\fwidth{0.8\columnwidth}
      \setlength\fheight{0.4\columnwidth}
      \input{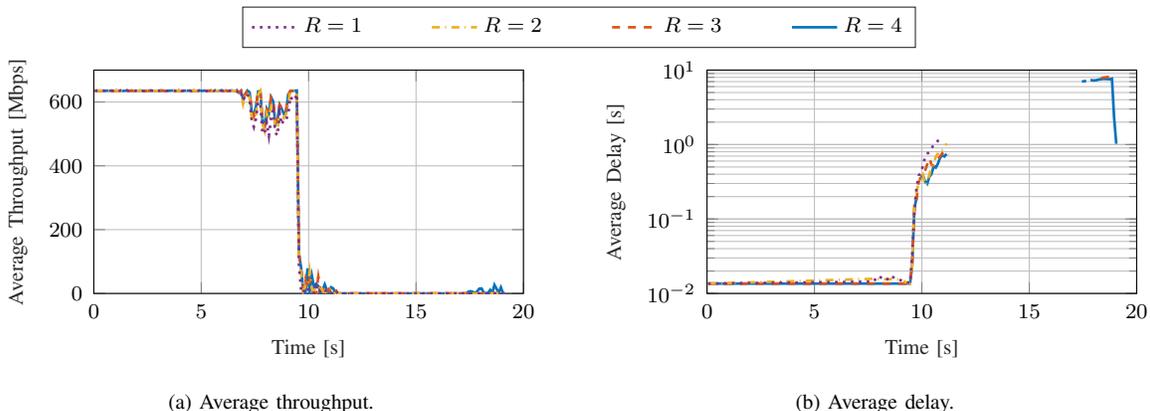}
      \caption{Average throughput.}
      \label{fig:lroom_thr_udp12_time}
  \end{subfigure}
  \hfill
  \begin{subfigure}[t]{0.45\textwidth}
      \centering
      \setlength\fwidth{0.8\columnwidth}
      \setlength\fheight{0.4\columnwidth}
%
%
\definecolor{mycolor1}{rgb}{0.00000,0.44700,0.74100}%
\definecolor{mycolor2}{rgb}{0.85000,0.32500,0.09800}%
\definecolor{mycolor3}{rgb}{0.92900,0.69400,0.12500}%
\definecolor{mycolor4}{rgb}{0.49400,0.18400,0.55600}%
\begin{tikzpicture}

\pgfplotsset{every tick label/.append style={font=\scriptsize}}

\begin{axis}[%
width=0.961\fwidth,
height=\fheight,
at={(0\fwidth,0\fheight)},
scale only axis,
unbounded coords=jump,
xmin=0,
xmax=20,
xlabel style={font=\scriptsize\color{white!15!black}},
xlabel={Time [s]},
ymode=log,
ymin=0.01,
ymax=10,
yminorticks=true,
ylabel style={font=\scriptsize\color{white!15!black}},
ylabel={Average Delay [s]},
axis background/.style={fill=white},
xmajorgrids,
ymajorgrids,
yminorgrids,
legend style={legend cell align=left, align=left, draw=white!15!black, font=\scriptsize}
]
\addplot [color=mycolor1, line width=1pt]
  table[row sep=crcr]{%
0.0500000000000007	0.0135618522118968\\
9.45	0.0135621621278226\\
9.55	0.0235966415714302\\
9.65	0.137184550436892\\
9.75	0.191526333115942\\
nan	nan\\
9.95	0.367158243599999\\
10.05	0.385399472216495\\
10.15	0.319896224574467\\
10.25	0.297047410360825\\
10.35	0.368171486052631\\
10.45	0.404313029135022\\
10.55	0.508220034387755\\
10.65	0.490317292272728\\
10.75	0.555502872692308\\
10.85	0.668621139257425\\
10.95	0.712714094411766\\
11.05	0.691519072614213\\
11.15	0.749772590316457\\
nan	nan\\
17.45	7.05739456762295\\
17.55	7.06693073401409\\
17.65	7.200550815\\
nan	nan\\
17.85	7.3780982910274\\
17.95	7.27964604887097\\
18.05	7.272734765\\
18.15	7.35426063907407\\
18.25	7.482765565\\
18.35	7.55686102653846\\
18.45	7.620609115\\
18.55	7.60653238318182\\
18.65	7.57160635658672\\
18.75	7.55259169395105\\
18.85	7.66062741348485\\
18.95	2.29757491733161\\
19.05	1.023540565\\
};

\addplot [color=mycolor2, dashed, line width=1pt]
  table[row sep=crcr]{%
0.0500000000000007	0.0135618522118968\\
9.45	0.0135621893887325\\
9.55	0.0252699988461522\\
9.65	0.125544435422537\\
9.75	0.181999953429752\\
9.85	0.308762235995673\\
9.95	0.368382486194029\\
10.05	0.387493021621005\\
10.15	0.447416792848102\\
10.25	0.541676492710842\\
10.35	0.657754565000001\\
10.45	0.535529133181818\\
10.55	0.517237516456312\\
10.65	0.600990308119265\\
10.75	0.647840655909089\\
10.85	0.690786070617978\\
10.95	0.762130292891158\\
11.05	0.720940564999999\\
11.15	0.786823052804877\\
11.25	0.726861898333333\\
nan	nan\\
17.65	7.13909304775862\\
nan	nan\\
18.35	7.82559823846939\\
18.45	7.92814903558823\\
18.55	7.99998289833333\\
18.65	8.077650565\\
18.75	8.180666565\\
18.85	8.26275323166667\\
18.95	8.38062727386076\\
19.05	8.459317765\\
};

\addplot [color=mycolor3, dashdotted, line width=1pt]
  table[row sep=crcr]{%
0.0500000000000007	0.0135618522118968\\
7.85	0.0157614237770183\\
8.05	0.0155082609663868\\
8.85	0.0155140174825732\\
9.45	0.0135622515119156\\
9.55	0.0233698539056597\\
9.65	0.172120472046476\\
9.75	0.230983913837209\\
9.85	0.30423511542735\\
9.95	0.368769280596331\\
10.05	0.397242336754637\\
10.15	0.285984165\\
10.25	0.303533987222222\\
10.35	0.390852294729729\\
10.45	0.503565539358974\\
10.55	0.554604179545455\\
10.65	0.688596633965517\\
10.75	0.758098905425532\\
10.85	0.852141949615385\\
10.95	0.852346077195122\\
11.05	0.888886000121952\\
11.15	1.03215068807692\\
};

\addplot [color=mycolor4, dotted, line width=1pt]
  table[row sep=crcr]{%
0.0500000000000007	0.0135618522118968\\
7.75	0.0144074258996536\\
7.95	0.0171849480739148\\
8.35	0.0150414252683184\\
8.55	0.0173919110433651\\
8.75	0.0159212067877093\\
9.45	0.013763907163522\\
9.55	0.0339663221364646\\
9.65	0.121238085\\
9.75	0.257574565000001\\
9.85	0.383807620837564\\
9.95	0.413277537067039\\
10.05	0.518687965\\
10.15	0.681428565000001\\
10.25	0.751630564999999\\
10.35	0.791734725\\
10.45	0.880127231666666\\
10.55	0.997892605816327\\
10.65	1.05469777430233\\
10.75	1.12560147676471\\
10.85	1.162340565\\
nan	nan\\
11.05	1.33744020136364\\
nan	nan\\
11.25	1.527440565\\
};

\end{axis}
\end{tikzpicture}%
      \caption{Average delay.}
      \label{fig:lroom_delay_udp12_time}
  \end{subfigure}
  \hfill
\setlength\belowcaptionskip{-.6cm}
  \caption{End-to-end performance vs.
$R$ for the \textit{L-Room} scenario with a \gls{udp} CBR traffic of 800~Mbps averaged over 100~ms windows.}
  \label{fig:lroom_udp12_time}
\end{figure*}%

\cref{fig:lroom_udp12_time} reports end-to-end metrics over time for the \textit{L-Room} scenario as a function of the maximum number of reflections $R$, making it possible to analyze the behavior at specific time instants.
First of all, we notice that the 800~Mbps data rate for \gls{udp} was chosen to saturate the channel capacity, as the physical layer only supports 630~Mbps at peak performance.
The interference starts to impact the \gls{udp} performance from 7~s, followed by a rapid performance degradation when the direct ray is lost in point B (see \cref{fig:scenarios}), at about 9.65~s.
Between point B and point C (11.33~s), the signal is still strong enough to allow for some transmissions, resulting, however, in a rapid increase of the delay of received packets due to buffering and retransmissions.
When RX\textsubscript{ref} gets closer to TX\textsubscript{interf}  (i.e., around 17~s), the interference decreases enough to still support data communication, although with extremely high delay, but only for $R$ large enough to reach the end of the corridor, i.e., $R \geq 3$.
As a matter of fact, TX\textsubscript{interf} points its 64 antenna elements towards RX\textsubscript{interf}, in \gls{los} at the end of the corridor.
Thus, when RX\textsubscript{ref} gets closer to the source of the interference, the angular separation with respect to RX\textsubscript{interf} is large enough to reduce the received interfering power due to the highly directional communication of the interfering \gls{tx}, resulting in an increased \gls{sinr}.
In this case, the second-order reflections between points C and D cannot alone guarantee a sufficiently high \gls{sinr} for the transmission, given the significant amount of interference.

\begin{figure*}[tbp]
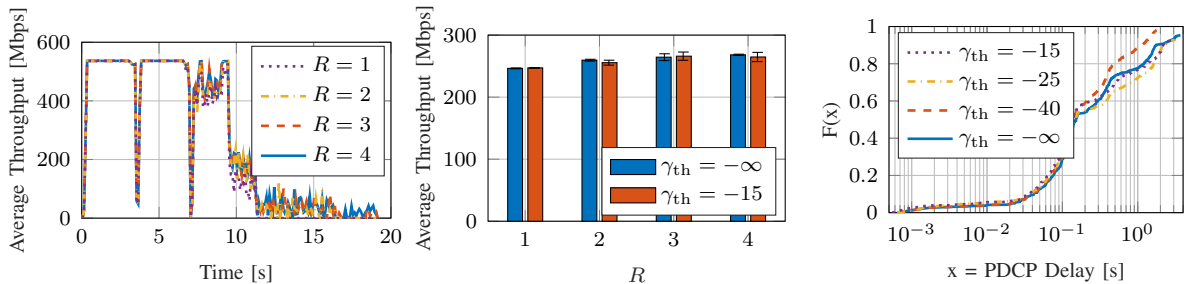
 
  \hfill
  \begin{subfigure}[t]{0.3\textwidth}
    \centering
    \setlength\fwidth{0.83\columnwidth}
    \setlength\fheight{0.5\columnwidth}
    \input{img/lroom_thr_tcp_time.tex}
    \caption{Throughput over time, $\gamma_{\rm th}=-\infty$~dB.}
    \label{fig:lroom_thr_tcp_time}
  \end{subfigure}
  \hfill
  \begin{subfigure}[t]{0.3\textwidth}
      \centering
      \setlength\fwidth{0.83\columnwidth}
      \setlength\fheight{0.5\columnwidth}
%
%
\definecolor{mycolor1}{rgb}{0.00000,0.44700,0.74100}%
\definecolor{mycolor2}{rgb}{0.85000,0.32500,0.09800}%
\begin{tikzpicture}

\pgfplotsset{every tick label/.append style={font=\scriptsize}}

\begin{axis}[%
width=0.961\fwidth,
height=\fheight,
at={(0\fwidth,0\fheight)},
scale only axis,
bar shift auto,
xmin=0.5,
xmax=4.5,
xtick={1,2,3,4},
xlabel style={font=\scriptsize\color{white!15!black}},
xlabel={$R$},
ymin=0,
ymax=300,
ylabel style={font=\scriptsize\color{white!15!black}},
ylabel={Average Throughput [Mbps]},
axis background/.style={fill=white},
xmajorgrids,
ymajorgrids,
legend style={legend cell align=left, align=left, draw=white!15!black, font=\scriptsize, legend pos=south east},
ylabel shift=-5pt
]
\addplot[ybar, bar width=0.2, fill=mycolor1, draw=black, area legend] table[row sep=crcr] {%
1	246.369779085051\\
2	259.603451687408\\
3	264.422240428048\\
4	268.1851554901\\
};
\addplot [color=black, draw=none, xshift=-0.125cm, forget plot]
 plot [error bars/.cd, y dir = both, y explicit]
 table[row sep=crcr, y error plus index=2, y error minus index=3]{%
1 246.369779085051  0.687707790179273 0.687707790179273\\
2 259.603451687408  1.45023903506841  1.45023903506841\\
3 264.422240428048  5.49896882270133  5.49896882270133\\
4 268.1851554901  1.05653815845405  1.05653815845405\\
};
\addlegendentry{$\gamma_{\rm th} = -\infty$}

\addplot[ybar, bar width=0.2, fill=mycolor2, draw=black, area legend] table[row sep=crcr] {%
1	247.159338129599\\
2	255.550421163736\\
3	266.031035437883\\
4	264.709965617389\\
};
\addplot [color=black, draw=none, xshift=0.125cm, forget plot]
 plot [error bars/.cd, y dir = both, y explicit]
 table[row sep=crcr, y error plus index=2, y error minus index=3]{%
1 247.159338129599  0.37672441151631  0.37672441151631\\
2 255.550421163736  3.98352405141304  3.98352405141304\\
3 266.031035437883  6.65693246202084  6.65693246202084\\
4 264.709965617389  7.45534070657864  7.45534070657864\\
};
\addlegendentry{$\gamma_{\rm th} = -15$}

\end{axis}
\end{tikzpicture}%
      \caption{Average throughput.}
      \label{fig:lroom_avg_thr_tcp}
  \end{subfigure}
  \hfill
  \begin{subfigure}[t]{0.3\textwidth}
      \centering
      \setlength\fwidth{0.83\columnwidth}
      \setlength\fheight{0.5\columnwidth}
      \input{img/lroom_delay_tcp_cdf.tex}
      \caption{\gls{pdcp} delay, $R=4$.}
      \label{fig:lroom_delay_tcp_cdf}
  \end{subfigure}
  \hfill
   \setlength\belowcaptionskip{-.6cm}
  \caption{End-to-end performance vs.
$R$ and $\gamma_{\rm th}$ for the \textit{L-Room} scenario with full-buffer \gls{tcp} traffic.}
  \label{fig:lroom_tcp}
\end{figure*}

\cref{fig:lroom_thr_tcp_time} shows the corresponding simulations using a full-buffer \gls{tcp} traffic stream, which reaches 536~Mbps at peak.
As expected, sudden jumps in the channel quality lead to sudden performance drops in \gls{tcp}.
This is the case in point A at 3.415~s (see \cref{fig:scenarios}), where the strong first-order rays of TX\textsubscript{interf} are received by RX\textsubscript{ref}, at the beginning of the interfering regime at about 7~s, when the direct ray is lost in point B at 9.65~s, and finally when the strong first-order reflections are lost in point C at 11.33~s.

For these reasons, \cref{fig:lroom_avg_thr_tcp} highlights a positive correlation between the average throughput and the maximum reflection order $R$, although only a 9\% increase is observed with $\gamma_{\rm th}=-\infty$~dB, going from 246~Mbps for $R=1$ to 268~Mbps for $R=4$.
In general, instead, delay statistics do not show a clear trend in the reflection order, nor in the relative threshold, probably due to the extra complexity created by retransmissions and queues.
An example is shown in \cref{fig:lroom_delay_tcp_cdf} where  most statistics follow a very similar trend, separating only towards extreme values of delay, corresponding to the portion of the scenario after point B, i.e., when the direct ray is lost.
Notice that the \glspl{cdf} for the delay do not reach 1 since windows where no packets were received were considered to have infinite average delay.

Unlike for the \gls{udp} case at 800~Mbps, \gls{tcp} decreases the congestion window when strong interference affects the communication for both the reference and the interfering streams.
For this reason, packets sent during the interfering regime do not always collide with each other.
The reduced interference greatly increases the perceived \gls{sinr}, as explained in~\cref{sub:link_level_performance_results} for \cref{fig:lroom_sinr_matlabVsNs3}, thus triggering transmissions even after point B for $R\geq 2$.
Although not shown here, similar conclusions can be drawn for  \gls{udp} traffic at 100~Mbps, which is able to transmit after point B as well, sending data at a rate that depends on the small scale fading affecting the communication.

\begin{figure*}[t!] 
  \hfill
  \begin{subfigure}[t]{0.48\textwidth}
      \centering
      \setlength\fwidth{0.8\columnwidth}
      \setlength\fheight{0.4\columnwidth}
%
%
\definecolor{mycolor1}{rgb}{0.00000,0.44700,0.74100}%
\definecolor{mycolor2}{rgb}{0.85000,0.32500,0.09800}%
\begin{tikzpicture}

\pgfplotsset{every tick label/.append style={font=\scriptsize}}

\begin{axis}[%
width=0.961\fwidth,
height=\fheight,
at={(0\fwidth,0\fheight)},
scale only axis,
xmin=0,
xmax=20,
xlabel style={font=\scriptsize\color{white!15!black}},
xlabel={Time [s]},
ymin=0,
ymax=700,
ylabel style={font=\scriptsize\color{white!15!black}},
ylabel={Average Throughput [Mbps]},
axis background/.style={fill=white},
xmajorgrids,
ymajorgrids,
legend style={legend cell align=left, align=left, draw=white!15!black, font=\scriptsize}
]
\addplot [color=mycolor1, line width=1pt]
  table[row sep=crcr]{%
0.05	635.2704\\
0.15	635.3688\\
0.25	635.2704\\
0.35	635.172\\
0.45	635.4672\\
0.55	635.172\\
0.65	635.2704\\
0.75	635.172\\
0.85	635.4672\\
0.95	634.5816\\
1.05	635.8608\\
1.15	634.7784\\
1.25	635.8608\\
1.35	634.5816\\
1.45	635.172\\
1.55	636.0576\\
1.65	634.5816\\
1.75	635.8608\\
1.85	634.7784\\
1.95	635.2704\\
2.05	635.172\\
2.15	635.8608\\
2.25	635.3688\\
2.35	634.68\\
2.45	635.172\\
2.55	635.3688\\
2.65	635.8608\\
2.75	635.2704\\
2.85	634.7784\\
2.95	635.172\\
3.05	635.2704\\
3.15	635.8608\\
3.25	635.3688\\
3.35	634.5816\\
3.45	635.2704\\
3.55	635.3688\\
3.65	635.8608\\
3.75	635.2704\\
3.85	634.5816\\
3.95	636.0576\\
4.05	634.5816\\
4.15	635.172\\
4.25	635.4672\\
4.35	635.172\\
4.45	635.8608\\
4.55	634.5816\\
4.65	636.0576\\
4.75	634.5816\\
4.85	635.2704\\
4.95	635.9592\\
5.05	634.5816\\
5.15	635.8608\\
5.25	634.7784\\
5.35	635.2704\\
5.45	635.8608\\
5.55	634.5816\\
5.65	636.0576\\
5.75	634.5816\\
5.85	635.172\\
5.95	636.0576\\
6.05	634.5816\\
6.15	635.8608\\
6.25	634.5816\\
6.35	635.4672\\
6.45	635.8608\\
6.55	634.5816\\
6.65	636.0576\\
6.75	629.3664\\
6.85	635.172\\
6.95	607.9152\\
7.05	624.0528\\
7.15	635.8608\\
7.25	630.4488\\
7.35	570.1296\\
7.45	559.1088\\
7.55	585.972\\
7.65	633.0072\\
7.75	634.5816\\
7.85	539.4288\\
7.95	524.8656\\
8.05	578.8872\\
8.15	565.0128\\
8.25	615.0984\\
8.35	601.8144\\
8.45	530.2776\\
8.55	543.4632\\
8.65	591.4824\\
8.75	578.1\\
8.85	562.2576\\
8.95	569.6376\\
9.05	622.872\\
9.15	634.5816\\
9.25	635.2704\\
9.35	635.3688\\
9.45	635.8608\\
9.55	123.984\\
9.65	81.0816\\
9.75	6.7896\\
9.85	0\\
9.95	78.72\\
10.05	9.5448\\
10.15	55.4976\\
10.25	19.0896\\
10.35	29.9136\\
10.45	23.3208\\
10.55	4.8216\\
10.65	10.824\\
10.75	28.1424\\
10.85	9.9384\\
10.95	5.0184\\
11.05	19.3848\\
11.15	15.5472\\
11.25	0\\
11.35	0\\
11.45	0\\
11.55	0\\
11.65	0\\
11.75	0\\
11.85	0\\
11.95	0\\
12.05	0\\
12.15	0\\
12.25	0\\
12.35	0\\
12.45	0\\
12.55	0\\
12.65	0\\
12.75	0\\
12.85	0\\
12.95	0\\
13.05	0\\
13.15	0\\
13.25	0\\
13.35	0\\
13.45	0\\
13.55	0\\
13.65	0\\
13.75	0\\
13.85	0\\
13.95	0\\
14.05	0\\
14.15	0\\
14.25	0\\
14.35	0\\
14.45	0\\
14.55	0\\
14.65	0\\
14.75	0\\
14.85	0\\
14.95	0\\
15.05	0\\
15.15	0\\
15.25	0\\
15.35	0\\
15.45	0\\
15.55	0\\
15.65	0\\
15.75	0\\
15.85	0\\
15.95	0\\
16.05	0\\
16.15	0\\
16.25	0\\
16.35	0\\
16.45	0\\
16.55	0\\
16.65	0\\
16.75	0\\
16.85	0\\
16.95	0\\
17.05	0\\
17.15	0\\
17.25	0\\
17.35	0\\
17.45	6.0024\\
17.55	6.9864\\
17.65	4.7232\\
17.75	0\\
17.85	7.1832\\
17.95	15.252\\
18.05	15.744\\
18.15	2.6568\\
18.25	0.3936\\
18.35	1.2792\\
18.45	11.0208\\
18.55	2.1648\\
18.65	26.6664\\
18.75	14.0712\\
18.85	3.2472\\
18.95	18.9912\\
19.05	0.540659340659387\\
19.15	0\\
};
\addlegendentry{Purely deterministic}

\addplot [color=mycolor2, dashed, line width=1pt]
  table[row sep=crcr]{%
0.05	635.2704\\
0.15	635.3688\\
0.25	635.2704\\
0.35	635.172\\
0.45	635.4672\\
0.55	635.172\\
0.65	635.2704\\
0.75	635.172\\
0.85	635.4672\\
0.95	634.5816\\
1.05	635.8608\\
1.15	634.7784\\
1.25	635.8608\\
1.35	634.5816\\
1.45	635.172\\
1.55	636.0576\\
1.65	634.5816\\
1.75	635.8608\\
1.85	634.7784\\
1.95	635.2704\\
2.05	635.172\\
2.15	635.8608\\
2.25	559.0104\\
2.35	634.5816\\
2.45	635.9592\\
2.55	634.68\\
2.65	635.8608\\
2.75	635.2704\\
2.85	634.7784\\
2.95	635.172\\
3.05	635.2704\\
3.15	635.9592\\
3.25	635.2704\\
3.35	634.5816\\
3.45	635.2704\\
3.55	635.3688\\
3.65	635.8608\\
3.75	635.172\\
3.85	558.2232\\
3.95	635.8608\\
4.05	634.5816\\
4.15	635.4672\\
4.25	635.172\\
4.35	635.2704\\
4.45	636.0576\\
4.55	634.5816\\
4.65	635.172\\
4.75	610.9656\\
4.85	634.5816\\
4.95	635.8608\\
5.05	631.0392\\
5.15	635.9592\\
5.25	630.4488\\
5.35	627.792\\
5.45	636.0576\\
5.55	634.5816\\
5.65	635.8608\\
5.75	634.5816\\
5.85	635.3688\\
5.95	635.8608\\
6.05	634.5816\\
6.15	636.0576\\
6.25	634.5816\\
6.35	635.2704\\
6.45	635.8608\\
6.55	634.68\\
6.65	635.8608\\
6.75	632.3184\\
6.85	629.9568\\
6.95	610.9656\\
7.05	621.0024\\
7.15	627.6936\\
7.25	625.5288\\
7.35	614.016\\
7.45	592.4664\\
7.55	593.0568\\
7.65	617.1648\\
7.75	627.0048\\
7.85	580.8552\\
7.95	533.9184\\
8.05	566.6856\\
8.15	401.472\\
8.25	647.1768\\
8.35	587.5464\\
8.45	480.8808\\
8.55	522.012\\
8.65	586.8576\\
8.75	594.0408\\
8.85	569.244\\
8.95	583.02\\
9.05	618.2472\\
9.15	608.5056\\
9.25	604.668\\
9.35	529.6872\\
9.45	645.7008\\
9.55	168.756\\
9.65	38.5728\\
9.75	136.0872\\
9.85	172.9872\\
9.95	198.0792\\
10.05	102.828\\
10.15	85.4112\\
10.25	21.156\\
10.35	229.9608\\
10.45	84.0336\\
10.55	58.2528\\
10.65	60.024\\
10.75	163.344\\
10.85	112.2744\\
10.95	86.1984\\
11.05	85.8048\\
11.15	48.5112\\
11.25	42.2136\\
11.35	1.1808\\
11.45	128.8056\\
11.55	5.5104\\
11.65	0\\
11.75	2.8536\\
11.85	8.5608\\
11.95	0\\
12.05	9.6432\\
12.15	33.0624\\
12.25	8.5608\\
12.35	2.8536\\
12.45	0\\
12.55	0.3936\\
12.65	52.4472\\
12.75	43.8864\\
12.85	48.8064\\
12.95	1.8696\\
13.05	2.1648\\
13.15	0.0984\\
13.25	0\\
13.35	24.2064\\
13.45	0.6888\\
13.55	0\\
13.65	0\\
13.75	0\\
13.85	0.3936\\
13.95	11.316\\
14.05	0\\
14.15	1.2792\\
14.25	39.1632\\
14.35	12.5952\\
14.45	0.5904\\
14.55	2.8536\\
14.65	9.5448\\
14.75	0\\
14.85	0\\
14.95	0.2952\\
15.05	0.5904\\
15.15	1.2792\\
15.25	0\\
15.35	0\\
15.45	0\\
15.55	1.476\\
15.65	0.0984\\
15.75	0\\
15.85	0\\
15.95	0\\
16.05	6.6912\\
16.15	0\\
16.25	0.2952\\
16.35	0\\
16.45	0\\
16.55	0\\
16.65	0\\
16.75	0\\
16.85	0\\
16.95	0\\
17.05	0.984\\
17.15	0\\
17.25	0.1968\\
17.35	0.6888\\
17.45	0.7872\\
17.55	0\\
17.65	0\\
17.75	0\\
17.85	31.8816\\
17.95	6.4944\\
18.05	18.0072\\
18.15	27.7488\\
18.25	56.1864\\
18.35	38.0808\\
18.45	30.1104\\
18.55	86.4936\\
18.65	32.6688\\
18.75	6.396\\
18.85	74.8824\\
18.95	36.1128\\
19.05	78.6425196850412\\
19.15	0\\
};
\addlegendentry{Quasi deterministic}

\end{axis}
\end{tikzpicture}%
      \caption{Throughput over time with $R=4$ and $\gamma_{\rm th}=-\infty$~dB.}
      \label{fig:lroom_thr_qd_time}
  \end{subfigure}
  \hfill
  \begin{subfigure}[t]{0.48\textwidth}
      \centering
      \setlength\fwidth{0.8\columnwidth}
      \setlength\fheight{0.4\columnwidth}
      \input{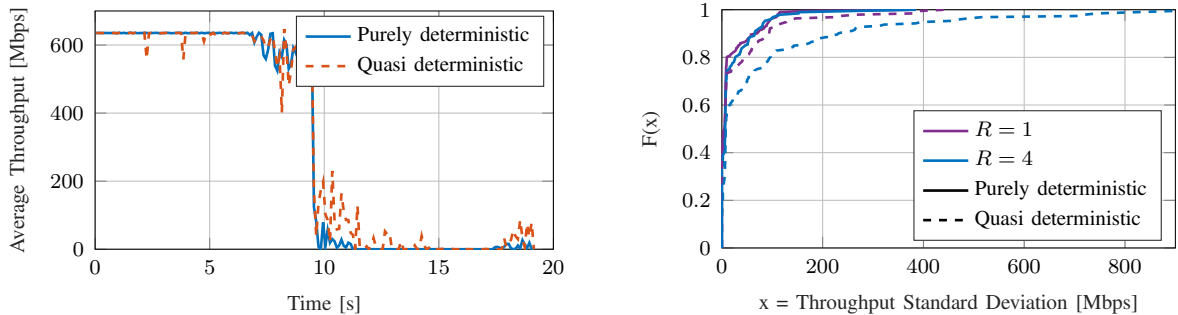}
      \caption{Throughput standard deviation for $\gamma_{\rm th} = -\infty$~dB.}
      \label{fig:lroom_thr_stdev_qd_udp800_ecdf}
  \end{subfigure}
  \hfill
    \setlength\belowcaptionskip{-.6cm}
  \caption{Throughput performance for the \textit{L-Room} scenario with a purely-deterministic and quasi-deterministic channel model, with a \gls{udp} CBR yielded traffic of 800~Mbps.}
  \label{fig:lroom_thr_qd}
\end{figure*}

A comparison between a purely-deterministic  and a quasi-deterministic channel with the \gls{qd} model described in \cref{sub:qd_model} is shown in \cref{fig:lroom_thr_qd}.
In general, from \cref{fig:lroom_thr_qd_time} it is possible to notice  that the added random rays from the \gls{qd} model tend to (i) increase the average received power and (ii) increase the frequency and amplitude of power fluctuations due to small scale fading, which is considered independent across subsequent time steps of 5 ms at a speed of 1.2~m/s.
These fluctuations can also affect the end-to-end performance, making it significantly less stable.

To further study these random fluctuations, the \gls{cdf} of the standard deviation of the throughput over 100~ms windows has been computed.
To obtain this metric, we first computed the average throughput over 5~ms sub-windows, i.e., the sampling period chosen for the ray-traced channel, and subsequently the standard deviation over 20 consecutive sub-windows.
This approach makes it possible to capture the deviation of the throughput over short time intervals, where it can be considered roughly constant.
Computing the standard deviation over the whole simulation, in fact, would yield a misleading metric, given the extreme differences over the almost 20~s long scenario.
\cref{fig:lroom_thr_stdev_qd_udp800_ecdf} shows how an increasing number of rays tends to increase the standard deviation of the throughput due to an increased small scale fading, especially when a quasi-deterministic model with random diffuse components is considered.
This effect should be taken into account when evaluating the performance of protocols for mmWave communications which adapt to the channel conditions, e.g., \gls{tcp}~\cite{zhang2019will}.

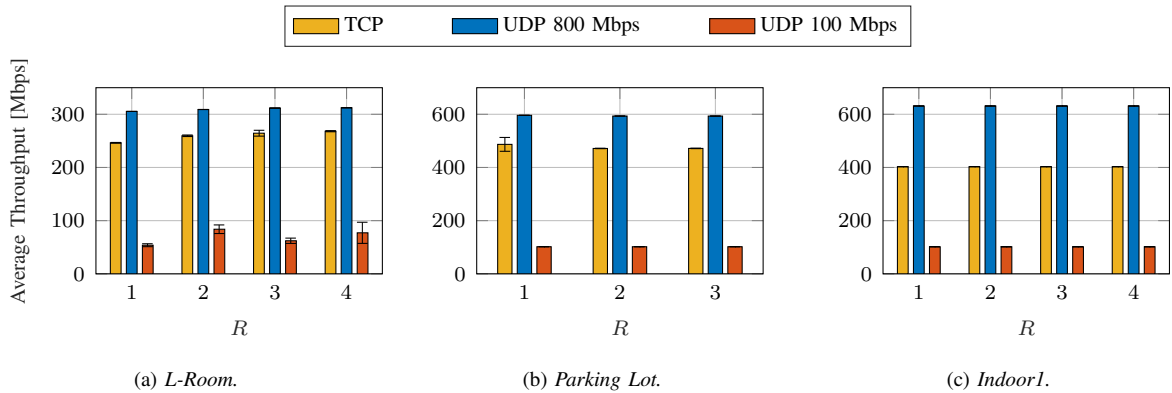
\begin{figure*}[tbp] 
  \centering
  \begin{subfigure}[t]{\textwidth}
      \centering
%
%
\definecolor{mycolor1}{rgb}{0.00000,0.44700,0.74100}%
\definecolor{mycolor2}{rgb}{0.85000,0.32500,0.09800}%
\definecolor{mycolor3}{rgb}{0.92900,0.69400,0.12500}%

\begin{tikzpicture}
\pgfplotsset{every tick label/.append style={font=\scriptsize}}

\begin{axis}[%
width=0,
height=0,
at={(0,0)},
scale only axis,
xmin=0,
xmax=0,
xtick={},
ymin=0,
ymax=0,
ytick={},
zmin=0,
zmax=0,
ztick={},
axis background/.style={fill=white},
legend style={legend cell align=center, align=center, draw=white!15!black, font=\scriptsize, at={(0, 0)}, anchor=center, /tikz/every even column/.append style={column sep=2em}},
legend columns=4,
]

\addplot[ybar, bar width=0.2, fill=mycolor3, draw=black, area legend] table[row sep=crcr] {%
0 0\\
};
\addplot[forget plot, color=white!15!black] table[row sep=crcr] {%
0 0\\
};
\addlegendentry{TCP}

\addplot[ybar, bar width=0.2, fill=mycolor1, draw=black, area legend] table[row sep=crcr] {%
0 0\\
};
\addplot[forget plot, color=white!15!black] table[row sep=crcr] {%
0 0\\
};
\addlegendentry{UDP 800~Mbps}

\addplot[ybar, bar width=0.2, fill=mycolor2, draw=black, area legend] table[row sep=crcr] {%
0 0\\
};
\addplot[forget plot, color=white!15!black] table[row sep=crcr] {%
0 0\\
};
\addlegendentry{UDP 100~Mbps}

\end{axis}
\end{tikzpicture}
  \end{subfigure}
  \\
  \hfill
  \begin{subfigure}[t]{0.3\textwidth}
      \centering
      \setlength\fwidth{0.8\columnwidth}
      \setlength\fheight{0.5\columnwidth}
%
%
\definecolor{mycolor1}{rgb}{0.00000,0.44700,0.74100}%
\definecolor{mycolor2}{rgb}{0.85000,0.32500,0.09800}%
\definecolor{mycolor3}{rgb}{0.92900,0.69400,0.12500}%
\begin{tikzpicture}

\pgfplotsset{every tick label/.append style={font=\scriptsize}}

\begin{axis}[%
width=0.961\fwidth,
height=\fheight,
at={(0\fwidth,0\fheight)},
scale only axis,
bar shift auto,
xmin=0.5,
xmax=4.5,
xtick={1,2,3,4},
xlabel style={font=\scriptsize\color{white!15!black}},
xlabel={$R$},
ymin=0,
ymax=350,
ylabel style={font=\scriptsize\color{white!15!black}},
ylabel={Average Throughput [Mbps]},
axis background/.style={fill=white},
xmajorgrids,
ymajorgrids,
legend style={legend cell align=left, align=left, draw=white!15!black, font=\scriptsize}
]

\addplot[ybar, bar width=0.15, fill=mycolor3, draw=black, area legend] table[row sep=crcr] {%
1 246.369779085051\\
2 259.603451687408\\
3 264.422240428048\\
4 268.1851554901\\
};
\addplot [color=black, draw=none, xshift=-0.21cm, forget plot]
 plot [error bars/.cd, y dir = both, y explicit]
 table[row sep=crcr, y error plus index=2, y error minus index=3]{%
1 246.369779085051  0.687707790179273 0.687707790179273\\
2 259.603451687408  1.45023903506841  1.45023903506841\\
3 264.422240428048  5.49896882270133  5.49896882270133\\
4 268.1851554901  1.05653815845405  1.05653815845405\\
};

\addplot[ybar, bar width=0.15, fill=mycolor1, draw=black, area legend] table[row sep=crcr] {%
1	305.35748644749\\
2	308.991302038\\
3	311.673501559097\\
4	312.291848622853\\
};
\addplot [color=black, draw=none, forget plot]
 plot [error bars/.cd, y dir = both, y explicit]
 table[row sep=crcr, y error plus index=2, y error minus index=3]{%
1 305.35748644749 0.0460428141512334  0.0460428141512334\\
2 308.991302038 0.0884741422805367  0.0884741422805367\\
3 311.673501559097  0.132373222400361 0.132373222400361\\
4 312.291848622853  0.081686071409368 0.081686071409368\\
};

\addplot[ybar, bar width=0.15, fill=mycolor2, draw=black, area legend] table[row sep=crcr] {%
1	53.826764218747\\
2	83.8689308745882\\
3	62.1729070259864\\
4	77.1206858204212\\
};
\addplot [color=black, draw=none, xshift=0.21cm, forget plot]
 plot [error bars/.cd, y dir = both, y explicit]
 table[row sep=crcr, y error plus index=2, y error minus index=3]{%
1 53.826764218747 2.76657263555373  2.76657263555373\\
2 83.8689308745882  8.09690004152083  8.09690004152083\\
3 62.1729070259864  4.99152476111355  4.99152476111355\\
4 77.1206858204212  19.8767623665522  19.8767623665522\\
};

\end{axis}
\end{tikzpicture}%
      \caption{\emph{L-Room.}}
      \label{fig:lroom_avg_thr_traffic_bars}
  \end{subfigure}
  \hfill
  \begin{subfigure}[t]{0.3\textwidth}
      \centering
      \setlength\fwidth{0.8\columnwidth}
      \setlength\fheight{0.5\columnwidth}
%
%
\definecolor{mycolor1}{rgb}{0.00000,0.44700,0.74100}%
\definecolor{mycolor2}{rgb}{0.85000,0.32500,0.09800}%
\definecolor{mycolor3}{rgb}{0.92900,0.69400,0.12500}%
\begin{tikzpicture}

\pgfplotsset{every tick label/.append style={font=\scriptsize}}

\begin{axis}[%
width=0.961\fwidth,
height=\fheight,
at={(0\fwidth,0\fheight)},
scale only axis,
bar shift auto,
xmin=0.5,
xmax=3.5,
xtick={1,2,3},
xlabel style={font=\scriptsize\color{white!15!black}},
xlabel={$R$},
ymin=0,
ymax=700,
axis background/.style={fill=white},
xmajorgrids,
ymajorgrids,
legend style={legend cell align=left, align=left, draw=white!15!black}
]

\addplot[ybar, bar width=0.15, fill=mycolor3, draw=black, area legend] table[row sep=crcr] {%
1 486.840724909162\\
2 471.463510479671\\
3 471.64734284148\\
};
\addplot [color=black, draw=none, xshift=-0.26cm, forget plot]
 plot [error bars/.cd, y dir = both, y explicit]
 table[row sep=crcr, y error plus index=2, y error minus index=3]{%
1 486.840724909162  26.0607799353553  26.0607799353553\\
2 471.463510479671  0.814778455915894 0.814778455915894\\
3 471.64734284148 0.879749606929958 0.879749606929958\\
};

\addplot[ybar, bar width=0.15, fill=mycolor1, draw=black, area legend] table[row sep=crcr] {%
1	596.139151904979\\
2	593.42528080402\\
3	593.343836710827\\
};
\addplot [color=black, draw=none, forget plot]
 plot [error bars/.cd, y dir = both, y explicit]
 table[row sep=crcr, y error plus index=2, y error minus index=3]{%
1 596.139151904979  0.36802077099214  0.36802077099214\\
2 593.42528080402 0.206411185994886 0.206411185994886\\
3 593.343836710827  0.198470125965617 0.198470125965617\\
};

\addplot[ybar, bar width=0.15, fill=mycolor2, draw=black, area legend] table[row sep=crcr] {%
1	101.985366856139\\
2	101.985366856139\\
3	101.985366856139\\
};
\addplot [color=black, draw=none, xshift=0.26cm, forget plot]
 plot [error bars/.cd, y dir = both, y explicit]
 table[row sep=crcr, y error plus index=2, y error minus index=3]{%
1 101.985366856139  6.22818168366676e-15  6.22818168366676e-15\\
2 101.985366856139  2.06565417708866e-14  2.06565417708866e-14\\
3 101.985366856139  1.64782198551098e-14  1.64782198551098e-14\\
};

\end{axis}
\end{tikzpicture}%
      \caption{\emph{Parking Lot.}}
      \label{fig:parkinglot_avg_thr_traffic_bars}
  \end{subfigure}
  \hfill
  \begin{subfigure}[t]{0.3\textwidth}
      \centering
      \setlength\fwidth{0.8\columnwidth}
      \setlength\fheight{0.5\columnwidth}
%
%
\definecolor{mycolor1}{rgb}{0.00000,0.44700,0.74100}%
\definecolor{mycolor2}{rgb}{0.85000,0.32500,0.09800}%
\definecolor{mycolor3}{rgb}{0.92900,0.69400,0.12500}%
\begin{tikzpicture}

\pgfplotsset{every tick label/.append style={font=\scriptsize}}

\begin{axis}[%
width=0.961\fwidth,
height=\fheight,
at={(0\fwidth,0\fheight)},
scale only axis,
bar shift auto,
xmin=0.5,
xmax=4.5,
xtick={1,2,3,4},
xlabel style={font=\scriptsize\color{white!15!black}},
xlabel={$R$},
ymin=0,
ymax=700,
axis background/.style={fill=white},
xmajorgrids,
ymajorgrids,
legend style={legend cell align=left, align=left, draw=white!15!black}
]

\addplot[ybar, bar width=0.15, fill=mycolor3, draw=black, area legend] table[row sep=crcr] {%
1 403.008677942185\\
2 403.008677942185\\
3 403.008677942185\\
4 403.008677942185\\
};
\addplot [color=black, draw=none, xshift = -0.21cm, forget plot]
 plot [error bars/.cd, y dir = both, y explicit]
 table[row sep=crcr, y error plus index=2, y error minus index=3]{%
1 403.008677942185  0 0\\
2 403.008677942185  0 0\\
3 403.008677942185  0 0\\
4 403.008677942185  0 0\\
};

\addplot[ybar, bar width=0.15, fill=mycolor1, draw=black, area legend] table[row sep=crcr] {%
1	631.2309377756\\
2 631.2309377756\\
3	631.2309377756\\
4	631.2309377756\\
};
\addplot [color=black, draw=none, forget plot]
 plot [error bars/.cd, y dir = both, y explicit]
 table[row sep=crcr, y error plus index=2, y error minus index=3]{%
1 631.2309377756  0 0\\
2 631.2309377756  0 0\\
3 631.2309377756  0 0\\
4 631.2309377756  0 0\\
};

\addplot[ybar, bar width=0.15, fill=mycolor2, draw=black, area legend] table[row sep=crcr] {%
1	101.84813810566\\
2	101.84813810566\\
3	101.84813810566\\
4	101.84813810566\\
};
\addplot [color=black, draw=none, xshift=0.21cm, forget plot]
 plot [error bars/.cd, y dir = both, y explicit]
 table[row sep=crcr, y error plus index=2, y error minus index=3]{%
1 101.84813810566 0 0\\
2 101.84813810566 0 0\\
3 101.84813810566 0 0\\
4 101.84813810566 0 0\\
};
\end{axis}
\end{tikzpicture}%
      \caption{\emph{Indoor1.}}
      \label{fig:indoor1_avg_thr_traffic_bars}
  \end{subfigure}
  \hfill
    \setlength\belowcaptionskip{-.6cm}
  \caption{Average throughput considering $\gamma_{\rm th} = -\infty$~dB.}
  \label{fig:avg_thr_traffic_bars}
\end{figure*}%

The average throughput for different configurations is shown in \cref{fig:avg_thr_traffic_bars}, including the 95\% \gls{ci}, often extremely narrow.
Both the \gls{los}-only scenarios, namely \textit{Parking Lot} and \textit{Indoor1}, show virtually no variations across different values of $R$ for all the three types of traffic considered.
Minor variations can only be observed for the \textit{L-Room} scenario, where the \gls{nlos} regime sets apart simulations with $R \leq 2$ from those with $R\geq 3$, not being able to exploit the last part of the path with lower interference and thus showing slightly lower performance.

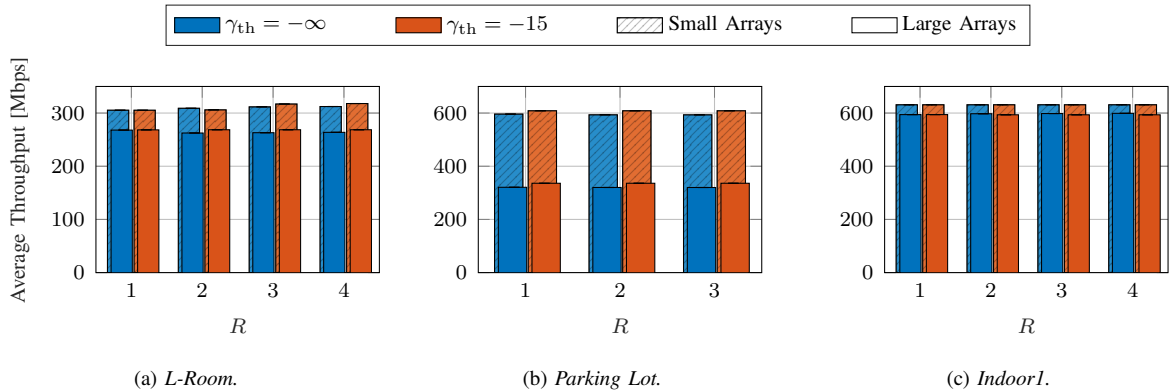
\begin{figure*}[tbp] 
  \centering
  \begin{subfigure}[t]{\textwidth}
      \centering
%
%
\definecolor{mycolor1}{rgb}{0.00000,0.44700,0.74100}%
\definecolor{mycolor2}{rgb}{0.85000,0.32500,0.09800}%
\definecolor{mycolor3}{rgb}{0.92900,0.69400,0.12500}%
\definecolor{mycolor4}{rgb}{0.49400,0.18400,0.55600}%

\begin{tikzpicture}
\pgfplotsset{every tick label/.append style={font=\scriptsize}}

\begin{axis}[%
width=0,
height=0,
at={(0,0)},
scale only axis,
xmin=0,
xmax=0,
xtick={},
ymin=0,
ymax=0,
ytick={},
zmin=0,
zmax=0,
ztick={},
axis background/.style={fill=white},
legend style={legend cell align=center, align=center, draw=white!15!black, font=\scriptsize, at={(0, 0)}, anchor=center, /tikz/every even column/.append style={column sep=2em}},
legend columns=4,
]

\addplot[ybar, bar width=0.4, fill=mycolor1, draw=black, area legend] table[row sep=crcr] {%
0 0\\
};
\addlegendentry{$\gamma_{\rm th} = -\infty$}

\addplot[ybar, bar width=0.4, fill=mycolor2, draw=black, area legend] table[row sep=crcr] {%
0 0\\
};
\addlegendentry{$\gamma_{\rm th} = -15$}

\addplot[ybar, bar width=0.4, fill=white, draw=black, area legend, postaction={pattern=north east lines, opacity=0.5}] table[row sep=crcr] {%
0 0\\
};
\addlegendentry{Small Arrays}

\addplot[ybar, bar width=0.4, fill=white, draw=black, area legend] table[row sep=crcr] {%
0 0\\
};
\addlegendentry{Large Arrays}

\end{axis}
\end{tikzpicture}%
  \end{subfigure}
  \\
  \hfill
  \begin{subfigure}[t]{0.3\textwidth}
      \centering
      \setlength\fwidth{0.8\columnwidth}
      \setlength\fheight{0.5\columnwidth}
%
%
\definecolor{mycolor1}{rgb}{0.00000,0.44700,0.74100}%
\definecolor{mycolor1light}{rgb}{0.1529,0.5529,0.7922}%
\definecolor{mycolor2}{rgb}{0.85000,0.32500,0.09800}%
\definecolor{mycolor2light}{rgb}{0.8824,0.4275,0.1961}%

\begin{tikzpicture}

\pgfplotsset{every tick label/.append style={font=\scriptsize}}

\begin{axis}[%
width=0.951\fwidth,
height=\fheight,
at={(0\fwidth,0\fheight)},
scale only axis,
bar shift auto,
xmin=0.5,
xmax=4.5,
xtick={1,2,3,4},
xlabel={$R$},
xlabel style={font=\scriptsize\color{white!15!black}},
ymin=0,
ymax=350,
ylabel style={font=\scriptsize\color{white!15!black}},
ylabel={Average Throughput [Mbps]},
axis background/.style={fill=white},
xmajorgrids,
ymajorgrids,
legend style={legend cell align=left, align=left, draw=white!15!black, font=\scriptsize}
]
\addplot[ybar, bar width=0.3, fill=mycolor1light, draw=black, area legend, postaction={pattern=north east lines, opacity=0.5}] table[row sep=crcr] {%
1	305.35748644749\\
2	308.991302038\\
3	311.673501559097\\
4	312.291848622853\\
};
\addplot [color=black, draw=none, xshift=-0.15cm, forget plot]
 plot [error bars/.cd, y dir = both, y explicit]
 table[row sep=crcr, y error plus index=2, y error minus index=3]{%
1 305.35748644749 0.0460428141512334  0.0460428141512334\\
2 308.991302038 0.0884741422805367  0.0884741422805367\\
3 311.673501559097  0.132373222400361 0.132373222400361\\
4 312.291848622853  0.081686071409368 0.081686071409368\\
};

\addplot[ybar, bar width=0.3, fill=mycolor2light, draw=black, area legend, postaction={pattern=north east lines, opacity=0.5}] table[row sep=crcr] {%
1	305.324318240173\\
2	306.034250540268\\
3	317.025362272727\\
4	317.902184545455\\
};
\addplot [color=black, draw=none, xshift=0.15cm, forget plot]
 plot [error bars/.cd, y dir = both, y explicit]
 table[row sep=crcr, y error plus index=2, y error minus index=3]{%
1 305.324318240173  0.0461805015566501  0.0461805015566501\\
2 306.034250540268  0.295886429607699 0.295886429607699\\
3 317.025362272727  0.0590312427845335  0.0590312427845335\\
4 317.902184545455  0.0704240264706906  0.0704240264706906\\
};

\end{axis}

\begin{axis}[%
width=0.951\fwidth,
height=\fheight,
at={(0\fwidth,0\fheight)},
scale only axis,
bar shift auto,
xmin=0.5,
xmax=4.5,
xtick={1,2,3,4},
hide x axis,
ymin=0,
ymax=350,
hide y axis,
legend style={legend cell align=left, align=left, draw=white!15!black}
]

\addplot[ybar, xshift=+0.05cm, bar width=0.3, fill=mycolor1, draw=black, area legend] table[row sep=crcr] {%
1 268.371160700522\\
2 262.885651692852\\
3 263.454041777907\\
4 264.13188886263\\
};
\addplot [color=black, draw=none, xshift=-0.10cm, forget plot]
 plot [error bars/.cd, y dir = both, y explicit]
 table[row sep=crcr, y error plus index=2, y error minus index=3]{%
1 268.371160700522  0.0404066478083949  0.0404066478083949\\
2 262.885651692852  0.117902146561363 0.117902146561363\\
3 263.454041777907  0.136566743953463 0.136566743953463\\
4 264.13188886263 0.0928713680715096  0.0928713680715096\\
};

\addplot[ybar, xshift=+0.05cm, bar width=0.3, fill=mycolor2, draw=black, area legend] table[row sep=crcr] {%
1 268.776011161097\\
2 269.014228184118\\
3 269.01908208301\\
4 269.01908208301\\
};
\addplot [color=black, draw=none, xshift=0.20cm, forget plot]
 plot [error bars/.cd, y dir = both, y explicit]
 table[row sep=crcr, y error plus index=2, y error minus index=3]{%
1 268.776011161097  0.0722349737608538  0.0722349737608538\\
2 269.014228184118  0.156843705397589 0.156843705397589\\
3 269.01908208301 0.159685574170307 0.159685574170307\\
4 269.01908208301 0.159685574170307 0.159685574170307\\
};

\end{axis}
\end{tikzpicture}%
      \caption{\emph{L-Room.}}
      \label{fig:lroom_avg_thr_ant_bars}
  \end{subfigure}
  \hfill
  \begin{subfigure}[t]{0.3\textwidth}
      \centering
      \setlength\fwidth{0.8\columnwidth}
      \setlength\fheight{0.5\columnwidth}
%
%
\definecolor{mycolor1}{rgb}{0.00000,0.44700,0.74100}%
\definecolor{mycolor1light}{rgb}{0.1529,0.5529,0.7922}%
\definecolor{mycolor2}{rgb}{0.85000,0.32500,0.09800}%
\definecolor{mycolor2light}{rgb}{0.8824,0.4275,0.1961}%
\begin{tikzpicture}

\pgfplotsset{every tick label/.append style={font=\scriptsize}}

\begin{axis}[%
width=0.951\fwidth,
height=\fheight,
at={(0\fwidth,0\fheight)},
scale only axis,
bar shift auto,
xmin=0.5,
xmax=3.5,
xtick={1,2,3},
xlabel style={font=\scriptsize\color{white!15!black}},
xlabel={$R$},
ymin=0,
ymax=700,
ylabel style={font=\scriptsize\color{white!15!black}},
axis background/.style={fill=white},
xmajorgrids,
ymajorgrids,
legend style={legend cell align=left, align=left, draw=white!15!black, font=\scriptsize}
]
\addplot[ybar, bar width=0.3, fill=mycolor1light, draw=black, area legend, postaction={pattern=north east lines, opacity=0.5}] table[row sep=crcr] {%
1	596.139151904979\\
2	593.42528080402\\
3	593.343836710827\\
};
\addplot [color=black, draw=none, xshift=-0.20cm,, forget plot]
 plot [error bars/.cd, y dir = both, y explicit]
 table[row sep=crcr, y error plus index=2, y error minus index=3]{%
1 596.139151904979  0.36802077099214  0.36802077099214\\
2 593.42528080402 0.206411185994886 0.206411185994886\\
3 593.343836710827  0.198470125965617 0.198470125965617\\
};

\addplot[ybar, bar width=0.3, fill=mycolor2light, draw=black, area legend, postaction={pattern=north east lines, opacity=0.5}] table[row sep=crcr] {%
1	608.64230812243\\
2	608.64230812243\\
3	608.64230812243\\
};
\addplot [color=black, draw=none, xshift=0.20cm,, forget plot]
 plot [error bars/.cd, y dir = both, y explicit]
 table[row sep=crcr, y error plus index=2, y error minus index=3]{%
1 608.64230812243 0.19871498896747  0.19871498896747\\
2 608.64230812243 0.19871498896747  0.19871498896747\\
3 608.64230812243 0.19871498896747  0.19871498896747\\
};

\end{axis}

\begin{axis}[%
width=0.951\fwidth,
height=\fheight,
at={(0\fwidth,0\fheight)},
scale only axis,
bar shift auto,
xmin=0.5,
xmax=3.5,
hide x axis,
ymin=0,
ymax=700,
hide y axis,
legend style={legend cell align=left, align=left, draw=white!15!black, font=\scriptsize}
]
\addplot[ybar, xshift=+0.05cm, bar width=0.3, fill=mycolor1, draw=black, area legend] table[row sep=crcr] {%
1 321.666946388312\\
2 320.952116990189\\
3 320.845442134851\\
};
\addplot [color=black, draw=none, xshift=-0.15cm, forget plot]
 plot [error bars/.cd, y dir = both, y explicit]
 table[row sep=crcr, y error plus index=2, y error minus index=3]{%
1 321.666946388312  0.102598759127528 0.102598759127528\\
2 320.952116990189  0.218036193354995 0.218036193354995\\
3 320.845442134851  0.14787469766584  0.14787469766584\\
};

\addplot[ybar, xshift=+0.05cm, bar width=0.3, fill=mycolor2, draw=black, area legend] table[row sep=crcr] {%
1 336.821266579716\\
2 336.821266579716\\
3 336.821266579716\\
};
\addplot [color=black, draw=none, xshift=0.25cm, forget plot]
 plot [error bars/.cd, y dir = both, y explicit]
 table[row sep=crcr, y error plus index=2, y error minus index=3]{%
1 336.821266579716  0.072515327611433 0.072515327611433\\
2 336.821266579716  0.072515327611433 0.072515327611433\\
3 336.821266579716  0.072515327611433 0.072515327611433\\
};

\end{axis}

\end{tikzpicture}%
      \caption{\emph{Parking Lot.}}
      \label{fig:parkinglot_avg_thr_ant_bars}
  \end{subfigure}
  \hfill
  \begin{subfigure}[t]{0.3\textwidth}
      \centering
      \setlength\fwidth{0.8\columnwidth}
      \setlength\fheight{0.5\columnwidth}
%
%
\definecolor{mycolor1}{rgb}{0.00000,0.44700,0.74100}%
\definecolor{mycolor1light}{rgb}{0.1529,0.5529,0.7922}%
\definecolor{mycolor2}{rgb}{0.85000,0.32500,0.09800}%
\definecolor{mycolor2light}{rgb}{0.8824,0.4275,0.1961}%
\begin{tikzpicture}

\pgfplotsset{every tick label/.append style={font=\scriptsize}}

\begin{axis}[%
width=0.951\fwidth,
height=\fheight,
at={(0\fwidth,0\fheight)},
scale only axis,
bar shift auto,
xmin=0.5,
xmax=4.5,
xtick={1,2,3,4},
xlabel style={font=\scriptsize\color{white!15!black}},
xlabel={$R$},
ymin=0,
ymax=700,
ylabel style={font=\color{white!15!black}},
axis background/.style={fill=white},
xmajorgrids,
ymajorgrids,
legend style={legend cell align=left, align=left, draw=white!15!black, font=\scriptsize}
]
\addplot[ybar, bar width=0.3, fill=mycolor1light, draw=black, area legend, postaction={pattern=north east lines, opacity=0.5}] table[row sep=crcr] {%
1	631.2309377756\\
2	631.2309377756\\
3	631.2309377756\\
4	631.2309377756\\
};
\addplot [color=black, draw=none, xshift=-0.15cm, forget plot]
 plot [error bars/.cd, y dir = both, y explicit]
 table[row sep=crcr, y error plus index=2, y error minus index=3]{%
1 631.2309377756  0 0\\
2 631.2309377756  0 0\\
3 631.2309377756  0 0\\
4 631.2309377756  0 0\\
};

\addplot[ybar, bar width=0.3, fill=mycolor2light, draw=black, area legend, postaction={pattern=north east lines, opacity=0.5}] table[row sep=crcr] {%
1	631.2309377756\\
2	631.2309377756\\
3	631.2309377756\\
4	631.2309377756\\
};
\addplot [color=black, draw=none, xshift=0.15cm, forget plot]
 plot [error bars/.cd, y dir = both, y explicit]
 table[row sep=crcr, y error plus index=2, y error minus index=3]{%
1 631.2309377756  0 0\\
2 631.2309377756  0 0\\
3 631.2309377756  0 0\\
4 631.2309377756  0 0\\
};
\end{axis}

\begin{axis}[%
width=0.951\fwidth,
height=\fheight,
at={(0\fwidth,0\fheight)},
scale only axis,
bar shift auto,
xmin=0.5,
xmax=4.5,
xtick={1,2,3,4},
hide x axis,
ymin=0,
ymax=700,
hide y axis,
legend style={legend cell align=left, align=left, draw=white!15!black}
]
\addplot[ybar, xshift=+0.05cm, bar width=0.3, fill=mycolor1, draw=black, area legend] table[row sep=crcr] {%
1 594.962925154336\\
2 598.274439510044\\
3 598.9914433317\\
4 599.746196883881\\
};\addplot [color=black, draw=none, xshift=-0.10cm, forget plot]
 plot [error bars/.cd, y dir = both, y explicit]
 table[row sep=crcr, y error plus index=2, y error minus index=3]{%
1 594.962925154336  0.0265190783934532  0.0265190783934532\\
2 598.274439510044  0.0133467284909489  0.0133467284909489\\
3 598.9914433317  0.014089040350831 0.014089040350831\\
4 599.746196883881  0.0246675485422865  0.0246675485422865\\
};

\addplot[ybar, xshift=+0.05cm, bar width=0.3, fill=mycolor2, draw=black, area legend] table[row sep=crcr] {%
1 595.005553904949\\
2 594.277461401274\\
3 594.277461401274\\
4 594.277461401274\\
};
\addplot [color=black, draw=none, xshift=0.2cm, forget plot]
 plot [error bars/.cd, y dir = both, y explicit]
 table[row sep=crcr, y error plus index=2, y error minus index=3]{%
1 595.005553904949  0.0297019653275436  0.0297019653275436\\
2 594.277461401274  0.0349197731586243  0.0349197731586243\\
3 594.277461401274  0.0349197731586243  0.0349197731586243\\
4 594.277461401274  0.0349197731586243  0.0349197731586243\\
};

\end{axis}

\end{tikzpicture}%
      \caption{\emph{Indoor1.}}
      \label{fig:indoor1_avg_thr_ant_bars}
  \end{subfigure}
  \hfill
    \setlength\belowcaptionskip{-.6cm}
  \caption{Average throughput for the 800~Mbps traffic considering different antenna architectures.}
  \label{fig:avg_thr_ant_bars}
\end{figure*}%

Similar results are shown in \cref{fig:avg_thr_ant_bars}, where two sets of antenna configurations are considered.
When smaller antenna arrays, and thus smaller antenna gains, are simulated, the average performance of the system decreases for all scenarios.
The largest performance hit is experienced by the \textit{Parking Lot} scenario, since the stronger path loss experienced as a result of the larger propagation distances involved in the outdoor scenario can be mitigated by the antenna gain.
The  performance drop observed in \cref{fig:avg_thr_ant_bars} can be also due to stronger interference.
Smaller arrays, in fact, are not able to create narrow beams, making TX\textsubscript{interf} interfere more strongly with RX\textsubscript{ref}.

\subsection{Computational Performance} 
\label{sub:performance}
The simulation techniques proposed in \cref{sec:mmwave_channel_simplifications} offer a trade-off between the simulation speedup given by the lower complexity, and a corresponding loss of accuracy.
\cref{sub:link_level_performance_results,sub:end_to_end_performance_results} analyzed in-depth the impact of the simplifications on the network metrics at two distinct levels.
Here, we compare the proposed simplifications from a computational complexity point of view, and then draw guidelines on the optimal combination of parameters that maximizes the accuracy.

For completeness, we need to distinguish the different contributions to the total runtime $T_{\text{tot}}$ required by a network simulation.
The first is the \gls{rt} runtime $T_{\text{\gls{rt}}}$, required by the  \gls{rt} to generate the \glspl{mpc} for the channel matrix.
The second contribution, $T_{\text{ns}}$, is due to the network simulator (either MATLAB or ns-3 in this work), which includes the computation of the channel matrix with the \gls{rt} data and what can be considered as simulation overhead.
As the same \gls{rt} channel trace can be used for an extensive simulation campaign, we define the overall runtime as $T_{\text{tot}} = T_{\text{\gls{rt}}} + 1000 \, T_{\text{ns}}$, i.e., the time required to simulate one \gls{rt} channel realization employed in 1000 network simulations, a typical if not conservative number when testing multiple parameters over multiple runs, e.g., for Monte Carlo analysis.

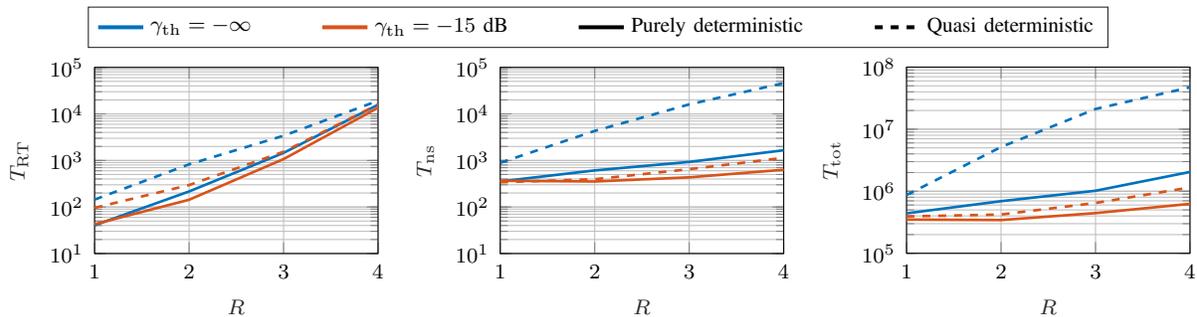
\begin{figure*}[tbp] 
  \centering
  \begin{subfigure}[t]{\textwidth}
      \centering
%
%
\definecolor{mycolor1}{rgb}{0.00000,0.44700,0.74100}%
\definecolor{mycolor2}{rgb}{0.85000,0.32500,0.09800}%
\definecolor{mycolor3}{rgb}{0.92900,0.69400,0.12500}%
\definecolor{mycolor4}{rgb}{0.49400,0.18400,0.55600}%

\begin{tikzpicture}
\pgfplotsset{every tick label/.append style={font=\scriptsize}}

\begin{axis}[%
width=0,
height=0,
at={(0,0)},
scale only axis,
xmin=0,
xmax=0,
xtick={},
ymin=0,
ymax=0,
ytick={},
zmin=0,
zmax=0,
ztick={},
axis background/.style={fill=white},
legend style={legend cell align=center, align=center, draw=white!15!black, font=\scriptsize, at={(0, 0)}, anchor=center, /tikz/every even column/.append style={column sep=2em}},
legend columns=4,
]
\addplot [color=mycolor1, line width=1.5pt]
  table[row sep=crcr]{%
0 0\\
};
\addlegendentry{$\gamma_{\rm th} = -\infty$}

\addplot [color=mycolor2, line width=1.5pt]
  table[row sep=crcr]{%
0 0\\
};
\addlegendentry{$\gamma_{\rm th} = -15$ dB}

\addplot [color=black, line width=1.5pt]
  table[row sep=crcr]{%
0 0\\
};
\addlegendentry{Purely deterministic}

\addplot [color=black, dashed, line width=1.5pt]
  table[row sep=crcr]{%
0 0\\
};
\addlegendentry{Quasi deterministic}

\end{axis}
\end{tikzpicture}%
  \end{subfigure}
  \\
  \hfill
  \begin{subfigure}[t]{0.3\textwidth}
      \centering
      \setlength\fwidth{0.8\columnwidth}
      \setlength\fheight{0.5\columnwidth}
%
%
\definecolor{mycolor1}{rgb}{0.00000,0.44700,0.74100}%
\definecolor{mycolor2}{rgb}{0.85000,0.32500,0.09800}%
\begin{tikzpicture}

\pgfplotsset{every tick label/.append style={font=\scriptsize}}

\begin{axis}[%
width=0.951\fwidth,
height=\fheight,
at={(0\fwidth,0\fheight)},
scale only axis,
xmin=1,
xmax=4,
xtick={1,2,3,4},
xlabel style={font=\scriptsize\color{white!15!black}},
xlabel={$R$},
ymode=log,
ymin=10,
ymax=100000,
ytick={10,100,1000,10000,100000,1000000,10000000,100000000},
yminorticks=true,
ylabel style={font=\scriptsize\color{white!15!black}},
ylabel={$T_{\rm RT}$},
axis background/.style={fill=white},
xmajorgrids,
ymajorgrids,
yminorgrids,
legend style={legend cell align=left, align=left, draw=white!15!black, font=\scriptsize}
]
\addplot [color=mycolor1, line width=1pt]
  table[row sep=crcr]{%
1	40.403648\\
2	215.781771\\
3	1453.282802\\
4	15745.789483\\
};

\addplot [color=mycolor2, line width=1pt]
  table[row sep=crcr]{%
1	42.753174\\
2	142.934651\\
3	1070.020165\\
4	13570.530595\\
};

\addplot [color=mycolor1, dashed, line width=1pt]
  table[row sep=crcr]{%
1	144.294291\\
2	825.992094\\
3	3397.045165\\
4	19162.520734\\
};

\addplot [color=mycolor2, dashed, line width=1pt]
  table[row sep=crcr]{%
1	95.837235\\
2	295.652972\\
3	1538.504787\\
4	15937.312243\\
};

\end{axis}
\end{tikzpicture}%
      \caption{Ray-Tracer runtime.}
      \label{fig:rt_runtime}
  \end{subfigure}
  \hfill
  \begin{subfigure}[t]{0.3\textwidth}
      \centering
      \setlength\fwidth{0.8\columnwidth}
      \setlength\fheight{0.5\columnwidth}
%
%
\definecolor{mycolor1}{rgb}{0.00000,0.44700,0.74100}%
\definecolor{mycolor2}{rgb}{0.85000,0.32500,0.09800}%
\begin{tikzpicture}

\pgfplotsset{every tick label/.append style={font=\scriptsize}}

\begin{axis}[%
width=0.951\fwidth,
height=\fheight,
at={(0\fwidth,0\fheight)},
scale only axis,
xmin=1,
xmax=4,
xtick={1,2,3,4},
xlabel style={font=\scriptsize\color{white!15!black}},
xlabel={$R$},
ylabel style={font=\scriptsize\color{white!15!black}},
ylabel={$T_{\rm ns}$},
ymode=log,
ymin=10,
ymax=100000,
ytick={10,100,1000,10000,100000,1000000,10000000,100000000},
yminorticks=true,
axis background/.style={fill=white},
xmajorgrids,
ymajorgrids,
yminorgrids,
legend style={legend cell align=left, align=left, draw=white!15!black, font=\scriptsize}
]
\addplot [color=mycolor1, line width=1pt]
  table[row sep=crcr]{%
1	354.051642036438\\
2	610.657903194427\\
3	929.993145465851\\
4	1655.16849894524\\
};

\addplot [color=mycolor2, line width=1pt]
  table[row sep=crcr]{%
1	365.728077459335\\
2	353.412846803665\\
3	433.530441856384\\
4	630.993070840836\\
};

\addplot [color=mycolor1, dashed, line width=1pt]
  table[row sep=crcr]{%
1	889.481681442261\\
2	4311.30041165352\\
3	16031.4926957607\\
4	45845.405634737\\
};

\addplot [color=mycolor2, dashed, line width=1pt]
  table[row sep=crcr]{%
1	340.061680793762\\
2	395.201657056808\\
3	649.102880859375\\
4	1139.08778209686\\
};

\end{axis}
\end{tikzpicture}%
      \caption{\Acrfull{ns3} runtime.}
      \label{fig:ns3_runtime}
  \end{subfigure}
  \hfill
  \begin{subfigure}[t]{0.3\textwidth}
      \centering 
      \setlength\fwidth{0.8\columnwidth}
      \setlength\fheight{0.5\columnwidth}
%
%
\definecolor{mycolor1}{rgb}{0.00000,0.44700,0.74100}%
\definecolor{mycolor2}{rgb}{0.85000,0.32500,0.09800}%
\begin{tikzpicture}

\pgfplotsset{every tick label/.append style={font=\scriptsize}}

\begin{axis}[%
width=0.951\fwidth,
height=\fheight,
at={(0\fwidth,0\fheight)},
scale only axis,
xmin=1,
xmax=4,
xtick={1,2,3,4},
xticklabels={{$1$},{$2$},{$3$},{$4$}},
xlabel style={font=\scriptsize\color{white!15!black}},
xlabel={$R$},
ymode=log,
ymin=100000,
ymax=100000000,
ytick={10,100,1000,10000,100000,1000000,10000000,100000000},
yminorticks=true,
ylabel style={font=\scriptsize\color{white!15!black}},
ylabel={$T_{\rm tot}$},
axis background/.style={fill=white},
xmajorgrids,
ymajorgrids,
yminorgrids,
legend style={legend cell align=left, align=left, draw=white!15!black, font=\scriptsize}
]
\addplot [color=mycolor1, line width=1pt]
  table[row sep=crcr]{%
1	440066.118999105\\
2	691207.87199622\\
3	1017950.80301036\\
4	2044775.3185777\\
};

\addplot [color=mycolor2, line width=1pt]
  table[row sep=crcr]{%
1	349105.233862095\\
2	343018.116087539\\
3	444077.607121024\\
4	625738.691748793\\
};

\addplot [color=mycolor1, dashed, line width=1pt]
  table[row sep=crcr]{%
1	868647.518663864\\
2	5159889.35731502\\
3	21153418.7968072\\
4	47258237.0172416\\
};

\addplot [color=mycolor2, dashed, line width=1pt]
  table[row sep=crcr]{%
1	392153.746485259\\
2	421354.081259506\\
3	646665.60382417\\
4	1148127.33471157\\
};

\end{axis}
\end{tikzpicture}%
      \caption{Total runtime.}
      \label{fig:campaign_runtime}
  \end{subfigure}
  \hfill
    \setlength\belowcaptionskip{-.6cm}
  \caption{Simulation runtime vs.
$R$ and $\gamma_{\rm th}$.
The total runtime (\cref{fig:campaign_runtime}) accounts for an \gls{rt} simulation (\cref{fig:rt_runtime}) and $1000$ ns-3 simulator runs (\cref{fig:ns3_runtime}).
A purely-deterministic channel and a quasi-deterministic channel are considered.}
  \label{fig:runtime}
\end{figure*}%

\cref{fig:runtime} shows the \gls{rt}, the ns-3, and the corresponding total runtime, for the \textit{L-Room} scenario.
The figure  compares the impact on the computational complexity of the simplification introduced by the reduction of the maximum order of reflection $R$.
We consider the two extreme values of $\gamma_{\rm th}$, i.e., $-$15~dB and $-\infty$~dB, and we compare a purely-deterministic channel with a quasi-deterministic channel that includes the random diffuse components introduced by the QD model.
First, it can be observed that  $R$ has the greatest impact on the runtime.
In particular, the \gls{rt} runtime $T_{\rm RT}$ increases by more than 2 orders of magnitude when increasing $R$, and the ns-3 runtime $T_{\rm ns}$ experiences a similar effect.
The impact of the \gls{qd} model is clearly visible in \cref{fig:ns3_runtime}, and, consequently, in \cref{fig:campaign_runtime}.
Nevertheless, note that in this case increasing the \gls{mpc} thresholding $\gamma_{\rm th}$ to $-$15~dB can effectively reduce the gap between the runtime with and without \gls{qd} model, for every reflection order.

Additionally, to summarize the conclusions from \cref{sub:link_level_performance_results,sub:end_to_end_performance_results} quantitatively, we express the difference of the network performance between the simplified models and the baseline\footnote{Recall that the reference baseline is obtained without simplifications, i.e., with the maximum available reflection order for the scenario ($R=4$ for \emph{Indoor1} and \emph{L-Room} and $R=3$ for \emph{Parking Lot}), and with $\gamma_{\rm th} = -\infty$~dB.} considering the \gls{nrmse}, computed as~\cite{testolina2020simplified}
\begin{equation}
  \text{NRMSE} =  \frac{\text{RMSE}}{\sigma_{\hat{x}}}
  = \frac{\sqrt{\frac{1}{N} \sum_{n=1}^N \qty[\qty(x_n-\hat{x_n})^2]}}{\sigma_{\hat{x}}},
\end{equation}
where $x$ is the metric with the configuration of interest, $\hat{x}$ is the baseline metric, and $\sigma_{\hat{x}}$ represents the standard deviation of the baseline metric.
This metric evaluates the distance between each baseline-simplified pair of a given simulated metric in the time domain.
Thus, it can detect punctual variations such as sharp drops or spikes, which may be relevant for communication protocols.
As in \cite{testolina2020simplified}, we compare it with a speedup metric, defined as the factor by which the overall simulation $T_{\rm tot}$ runtime is reduced compared to the baseline.
For the remainder of this section, we will consider 0.05 as the maximum acceptable value for the \gls{nrmse}, meaning that we deem acceptable an RMSE equal to 5\% of the standard deviation of the considered metric.


\begin{figure*}[tbp] 
  \begin{subfigure}[t]{\textwidth}
    \centering
%
%
\definecolor{mycolor1}{rgb}{0.00000,0.44700,0.74100}%
\definecolor{mycolor2}{rgb}{0.85000,0.32500,0.09800}%
\definecolor{mycolor3}{rgb}{0.92900,0.69400,0.12500}%
\definecolor{mycolor4}{rgb}{0.49400,0.18400,0.55600}%

\begin{tikzpicture}
\pgfplotsset{every tick label/.append style={font=\scriptsize}}

\begin{axis}[%
width=0,
height=0,
at={(0,0)},
scale only axis,
xmin=0,
xmax=0,
xtick={},
ymin=0,
ymax=0,
ytick={},
zmin=0,
zmax=0,
ztick={},
axis background/.style={fill=white},
legend style={legend cell align=center, align=center, draw=white!15!black, font=\scriptsize, at={(0, 0)}, anchor=center, /tikz/every even column/.append style={column sep=2em}},
legend columns=4,
]
\addplot[only marks, mark=*, mark options={}, mark size=2.5pt, draw=black, fill=black] table[row sep=crcr]{%
x y\\
0 0\\
};
\addlegendentry{$R=1$}

\addplot[only marks, mark=square*, mark options={}, mark size=2.5pt, draw=black, fill=black] table[row sep=crcr]{%
x y\\
0 0\\
};
\addlegendentry{$R=2$}

\addplot[only marks, mark=diamond*, mark options={}, mark size=2.5pt, draw=black, fill=black] table[row sep=crcr]{%
x y\\
0 0\\
};
\addlegendentry{$R=3$}

\addplot[only marks, mark=triangle*, mark options={rotate=180}, mark size=2.5pt, draw=black, fill=black] table[row sep=crcr]{%
x y\\
0 0\\
};
\addlegendentry{$R=4$}

\addplot[only marks, mark=*, mark options={}, mark size=2.5pt, draw=mycolor1, fill=mycolor1] table[row sep=crcr]{%
x y\\
0 0\\
};
\addlegendentry{$\gamma_{\rm th}=-\infty$}

\addplot[only marks, mark=*, mark options={}, mark size=2.5pt, draw=mycolor2, fill=mycolor2] table[row sep=crcr]{%
x y\\
0 0\\
};
\addlegendentry{$\gamma_{\rm th}=-40$}

\addplot[only marks, mark=*, mark options={}, mark size=2.5pt, draw=mycolor3, fill=mycolor3] table[row sep=crcr]{%
x y\\
0 0\\
};
\addlegendentry{$\gamma_{\rm th}=-25$}

\addplot[only marks, mark=*, mark options={}, mark size=2.5pt, draw=mycolor4, fill=mycolor4] table[row sep=crcr]{%
x y\\
0 0\\
};
\addlegendentry{$\gamma_{\rm th}=-15$}

\end{axis}
\end{tikzpicture}%
  \end{subfigure}
  \\
  \hfill
  \begin{subfigure}[t]{0.32\textwidth}
      \centering
      \setlength\fwidth{0.8\columnwidth}
      \setlength\fheight{0.35\columnwidth}
%
%
\definecolor{mycolor1}{rgb}{0.00000,0.44700,0.74100}%
\definecolor{mycolor2}{rgb}{0.85000,0.32500,0.09800}%
\definecolor{mycolor3}{rgb}{0.92900,0.69400,0.12500}%
\definecolor{mycolor4}{rgb}{0.49400,0.18400,0.55600}%
\begin{tikzpicture}

\pgfplotsset{every tick label/.append style={font=\scriptsize}}

\begin{axis}[%
width=0.961\fwidth,
height=\fheight,
at={(0\fwidth,0\fheight)},
scale only axis,
xmin=1,
xmax=4.5,
xlabel={Speedup factor},
xlabel style={font=\scriptsize\color{white!15!black}},
ymin=0,
ymax=0.07,
scaled y ticks = false,
y tick label style={/pgf/number format/.cd,
                    fixed,
                    },
extra y ticks={0.05},
extra y tick labels={},
extra y tick style={grid style={black, dashed}},
ylabel={NRMSE},
ylabel style={font=\scriptsize\color{white!15!black}},
axis background/.style={fill=white},
xmajorgrids,
ymajorgrids,
legend style={legend cell align=left, align=left, draw=white!15!black},
ylabel shift=-5pt,
]
\addplot[only marks, mark=*, mark options={}, mark size=2.5pt, draw=mycolor1, fill=mycolor1] table[row sep=crcr]{%
x	y\\
2.14777421571913	0.0482458607656119\\
};
\addplot[only marks, mark=square*, mark options={}, mark size=2.5pt, draw=mycolor1, fill=mycolor1] table[row sep=crcr]{%
x	y\\
2.69971724653508	0.00726010706725523\\
};
\addplot[only marks, mark=diamond*, mark options={}, mark size=2.5pt, draw=mycolor1, fill=mycolor1] table[row sep=crcr]{%
x	y\\
1.377129680226	0.00250852909833157\\
};
\addplot[only marks, mark=triangle*, mark options={rotate=180}, mark size=2.5pt, draw=mycolor1, fill=mycolor1] table[row sep=crcr]{%
x	y\\
1	0\\
};
\addplot[only marks, mark=*, mark options={}, mark size=2.5pt, draw=mycolor2, fill=mycolor2] table[row sep=crcr]{%
x	y\\
2.41142208511198	0.0482458607656119\\
};
\addplot[only marks, mark=square*, mark options={}, mark size=2.5pt, draw=mycolor2, fill=mycolor2] table[row sep=crcr]{%
x	y\\
2.81041940327027	0.00724675928033252\\
};
\addplot[only marks, mark=diamond*, mark options={}, mark size=2.5pt, draw=mycolor2, fill=mycolor2] table[row sep=crcr]{%
x	y\\
1.74820915458627	0.00408947705674702\\
};
\addplot[only marks, mark=triangle*, mark options={rotate=180}, mark size=2.5pt, draw=mycolor2, fill=mycolor2] table[row sep=crcr]{%
x	y\\
1.56847050154326	0.00406713260843107\\
};
\addplot[only marks, mark=*, mark options={}, mark size=2.5pt, draw=mycolor3, fill=mycolor3] table[row sep=crcr]{%
x	y\\
2.16702564814304	0.0482344458834449\\
};
\addplot[only marks, mark=square*, mark options={}, mark size=2.5pt, draw=mycolor3, fill=mycolor3] table[row sep=crcr]{%
x	y\\
3.09138761977511	0.0250827663201449\\
};
\addplot[only marks, mark=diamond*, mark options={}, mark size=2.5pt, draw=mycolor3, fill=mycolor3] table[row sep=crcr]{%
x	y\\
3.14663415392064	0.0250828277469519\\
};
\addplot[only marks, mark=triangle*, mark options={rotate=180}, mark size=2.5pt, draw=mycolor3, fill=mycolor3] table[row sep=crcr]{%
x	y\\
2.74839831594686	0.0250828277469519\\
};
\addplot[only marks, mark=*, mark options={}, mark size=2.5pt, draw=mycolor4, fill=mycolor4] table[row sep=crcr]{%
x	y\\
2.33561808810392	0.0636449863233284\\
};
\addplot[only marks, mark=square*, mark options={}, mark size=2.5pt, draw=mycolor4, fill=mycolor4] table[row sep=crcr]{%
x	y\\
4.10036754446053	0.0636460571720891\\
};
\addplot[only marks, mark=diamond*, mark options={}, mark size=2.5pt, draw=mycolor4, fill=mycolor4] table[row sep=crcr]{%
x	y\\
3.73605817997949	0.0636460571720891\\
};
\addplot[only marks, mark=triangle*, mark options={rotate=180}, mark size=2.5pt, draw=mycolor4, fill=mycolor4] table[row sep=crcr]{%
x	y\\
3.81538773828162	0.0636460571720891\\
};

\end{axis}
\end{tikzpicture}%
      \caption{\textit{Indoor1.}}
      \label{fig:scatter_sinr_indoor1}
  \end{subfigure}
  \hfill
  \begin{subfigure}[t]{0.32\textwidth}
      \centering
      \setlength\fwidth{0.8\columnwidth}
      \setlength\fheight{0.35\columnwidth}
%
%
\definecolor{mycolor1}{rgb}{0.00000,0.44700,0.74100}%
\definecolor{mycolor2}{rgb}{0.85000,0.32500,0.09800}%
\definecolor{mycolor3}{rgb}{0.92900,0.69400,0.12500}%
\definecolor{mycolor4}{rgb}{0.49400,0.18400,0.55600}%
\begin{tikzpicture}

\pgfplotsset{every tick label/.append style={font=\scriptsize}}

\begin{axis}[%
width=0.961\fwidth,
height=\fheight,
at={(0\fwidth,0\fheight)},
scale only axis,
xmin=1,
xmax=6,
xlabel={Speedup factor},
xlabel style={font=\scriptsize\color{white!15!black}},
ymin=0,
ymax=0.15,
scaled y ticks = false,
y tick label style={/pgf/number format/.cd,
                    fixed,
                    },
extra y ticks={0.05},
extra y tick labels={},
extra y tick style={grid style={black, dashed}},
ylabel={NRMSE},
ylabel style={font=\scriptsize\color{white!15!black}},
axis background/.style={fill=white},
xmajorgrids,
ymajorgrids,
legend style={legend cell align=left, align=left, draw=white!15!black},
ylabel shift=-5pt,
]
\addplot[only marks, mark=*, mark options={}, mark size=2.5pt, draw=mycolor1, fill=mycolor1] table[row sep=crcr]{%
x	y\\
5.00956007802015	0.0575024304573171\\
};
\addplot[only marks, mark=square*, mark options={}, mark size=2.5pt, draw=mycolor1, fill=mycolor1] table[row sep=crcr]{%
x	y\\
3.32254262025651	0.0283395651571489\\
};
\addplot[only marks, mark=diamond*, mark options={}, mark size=2.5pt, draw=mycolor1, fill=mycolor1] table[row sep=crcr]{%
x	y\\
2.04418577584242	0.0112138647536373\\
};
\addplot[only marks, mark=triangle*, mark options={rotate=180}, mark size=2.5pt, draw=mycolor1, fill=mycolor1] table[row sep=crcr]{%
x	y\\
1	0\\
};
\addplot[only marks, mark=*, mark options={}, mark size=2.5pt, draw=mycolor2, fill=mycolor2] table[row sep=crcr]{%
x	y\\
5.09968417653581	0.0575024304573171\\
};
\addplot[only marks, mark=square*, mark options={}, mark size=2.5pt, draw=mycolor2, fill=mycolor2] table[row sep=crcr]{%
x	y\\
3.61708093989835	0.0283437949104894\\
};
\addplot[only marks, mark=diamond*, mark options={}, mark size=2.5pt, draw=mycolor2, fill=mycolor2] table[row sep=crcr]{%
x	y\\
2.29176366022729	0.0135368155632886\\
};
\addplot[only marks, mark=triangle*, mark options={rotate=180}, mark size=2.5pt, draw=mycolor2, fill=mycolor2] table[row sep=crcr]{%
x	y\\
1.35530303615285	0.00827466216679956\\
};
\addplot[only marks, mark=*, mark options={}, mark size=2.5pt, draw=mycolor3, fill=mycolor3] table[row sep=crcr]{%
x	y\\
5.25229825839127	0.0574853919932469\\
};
\addplot[only marks, mark=square*, mark options={}, mark size=2.5pt, draw=mycolor3, fill=mycolor3] table[row sep=crcr]{%
x	y\\
3.31274897851959	0.044685241303026\\
};
\addplot[only marks, mark=diamond*, mark options={}, mark size=2.5pt, draw=mycolor3, fill=mycolor3] table[row sep=crcr]{%
x	y\\
3.01326883273326	0.0378322798963866\\
};
\addplot[only marks, mark=triangle*, mark options={rotate=180}, mark size=2.5pt, draw=mycolor3, fill=mycolor3] table[row sep=crcr]{%
x	y\\
2.09401167009292	0.0372674071055149\\
};
\addplot[only marks, mark=*, mark options={}, mark size=2.5pt, draw=mycolor4, fill=mycolor4] table[row sep=crcr]{%
x	y\\
4.93945893701995	0.124826271446493\\
};
\addplot[only marks, mark=square*, mark options={}, mark size=2.5pt, draw=mycolor4, fill=mycolor4] table[row sep=crcr]{%
x	y\\
5.35541640040585	0.120401770358716\\
};
\addplot[only marks, mark=diamond*, mark options={}, mark size=2.5pt, draw=mycolor4, fill=mycolor4] table[row sep=crcr]{%
x	y\\
4.476106297967	0.119363434262936\\
};
\addplot[only marks, mark=triangle*, mark options={rotate=180}, mark size=2.5pt, draw=mycolor4, fill=mycolor4] table[row sep=crcr]{%
x	y\\
2.42485888697905	0.119446975096224\\
};

\end{axis}
\end{tikzpicture}%
      \caption{\textit{L-Room.}}
      \label{fig:scatter_sinr_lroom}
  \end{subfigure}
  \hfill
  \begin{subfigure}[t]{0.32\textwidth}
      \centering
      \setlength\fwidth{0.8\columnwidth}
      \setlength\fheight{0.35\columnwidth}
%
%
\definecolor{mycolor1}{rgb}{0.00000,0.44700,0.74100}%
\definecolor{mycolor2}{rgb}{0.85000,0.32500,0.09800}%
\definecolor{mycolor3}{rgb}{0.92900,0.69400,0.12500}%
\definecolor{mycolor4}{rgb}{0.49400,0.18400,0.55600}%
\begin{tikzpicture}

\pgfplotsset{every tick label/.append style={font=\scriptsize}}

\begin{axis}[%
width=0.961\fwidth,
height=\fheight,
at={(0\fwidth,0\fheight)},
scale only axis,
xmin=1,
xmax=15,
xlabel={Speedup factor},
xlabel style={font=\scriptsize\color{white!15!black}},
ymin=0,
ymax=0.8,
scaled y ticks = false,
y tick label style={/pgf/number format/.cd,
                    fixed,
                    },
extra y ticks={0.05},
extra y tick labels={},
extra y tick style={grid style={black, dashed}},
ylabel={NRMSE},
ylabel style={font=\scriptsize\color{white!15!black}},
axis background/.style={fill=white},
xmajorgrids,
ymajorgrids,
legend style={legend cell align=left, align=left, draw=white!15!black},
ylabel shift=-5pt,
]
\addplot[only marks, mark=*, mark options={}, mark size=2.5pt, draw=mycolor1, fill=mycolor1] table[row sep=crcr]{%
x	y\\
10.3582297738184	0.0886029234039624\\
};
\addplot[only marks, mark=square*, mark options={}, mark size=2.5pt, draw=mycolor1, fill=mycolor1] table[row sep=crcr]{%
x	y\\
6.35169945553091	0.0151234740614676\\
};
\addplot[only marks, mark=diamond*, mark options={}, mark size=2.5pt, draw=mycolor1, fill=mycolor1] table[row sep=crcr]{%
x	y\\
1	0\\
};
\addplot[only marks, mark=*, mark options={}, mark size=2.5pt, draw=mycolor2, fill=mycolor2] table[row sep=crcr]{%
x	y\\
8.24434556214731	0.0886011773929203\\
};
\addplot[only marks, mark=square*, mark options={}, mark size=2.5pt, draw=mycolor2, fill=mycolor2] table[row sep=crcr]{%
x	y\\
6.7695043670877	0.0341950244542623\\
};
\addplot[only marks, mark=diamond*, mark options={}, mark size=2.5pt, draw=mycolor2, fill=mycolor2] table[row sep=crcr]{%
x	y\\
1.02633952718758	0.0341967554242368\\
};
\addplot[only marks, mark=*, mark options={}, mark size=2.5pt, draw=mycolor3, fill=mycolor3] table[row sep=crcr]{%
x	y\\
10.7329876239027	0.0981340540478461\\
};
\addplot[only marks, mark=square*, mark options={}, mark size=2.5pt, draw=mycolor3, fill=mycolor3] table[row sep=crcr]{%
x	y\\
8.77968125007818	0.0981340540478461\\
};
\addplot[only marks, mark=diamond*, mark options={}, mark size=2.5pt, draw=mycolor3, fill=mycolor3] table[row sep=crcr]{%
x	y\\
1.09038808174497	0.0981340540478461\\
};
\addplot[only marks, mark=*, mark options={}, mark size=2.5pt, draw=mycolor4, fill=mycolor4] table[row sep=crcr]{%
x	y\\
9.71230457229708	0.719832011726014\\
};
\addplot[only marks, mark=square*, mark options={}, mark size=2.5pt, draw=mycolor4, fill=mycolor4] table[row sep=crcr]{%
x	y\\
9.23932927153579	0.719832011726014\\
};
\addplot[only marks, mark=diamond*, mark options={}, mark size=2.5pt, draw=mycolor4, fill=mycolor4] table[row sep=crcr]{%
x	y\\
1.13256502755412	0.719832011726014\\
};

\end{axis}
\end{tikzpicture}%
      \caption{\textit{Parking Lot.}}
      \label{fig:scatter_sinr_parkinglot}
  \end{subfigure}
  \hfill
  \setlength\belowcaptionskip{-.6cm}
  \caption{Trade-off between the \gls{sinr} performance and the speedup obtained with the different simplification parameters for the three scenarios.
  The dashed black line at NRMSE=0.05 represents the maximum acceptable value for the NRMSE.
  As in \cref{sub:link_level_performance_results}, results and runtimes for the link-level MATLAB simulator have been considered.
  }
  \label{fig:scatter_sinr}
\end{figure*}
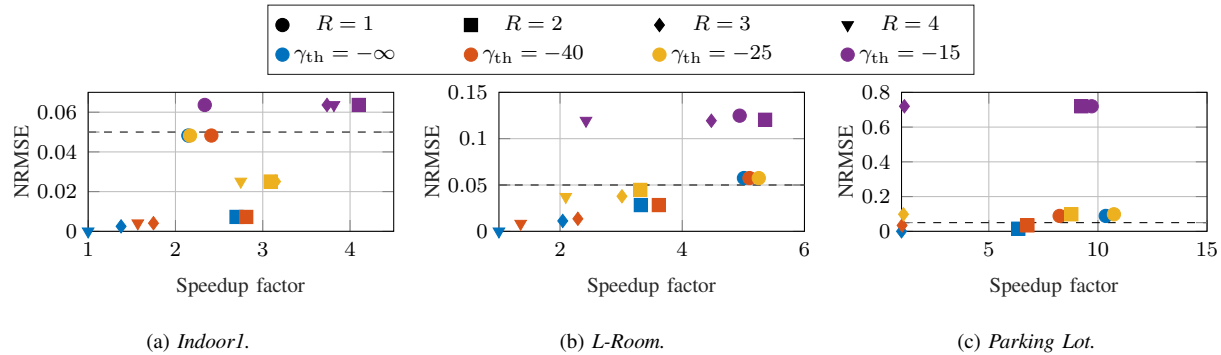%

\paragraph*{Link-Level Performance}
The  variations of the \gls{sinr} behavior due to the simplifications are shown in \cref{fig:scatter_sinr}.
In general it is possible to notice that markers with the same shape tend to increase with increasing steepness as the relative threshold increases, thus showing diminishing returns for the largest value of $\gamma_{\rm th}$.

For the \emph{Indoor1} scenario (\cref{fig:scatter_sinr_indoor1}), significant deviation from the baseline occurs with $R=1$, and with $\gamma_{\rm th}=-15$~dB.
Nevertheless, for all the considered cases, the \gls{nrmse} is smaller than 0.07, confirming what was anticipated in \cref{sub:link_level_performance_results}, i.e., that only minor changes take place even with the most aggressive simplifications.
Within the maximum accepted NRMSE it is possible to speed up the simulator up to a factor of 3.14 with an \gls{nrmse} as small as 0.025 ($R=3$, $\gamma_{\rm th}=-25$ dB), although $R=2$, $\gamma_{\rm th}=-25$ comes in close with a similar \gls{nrmse} and a speedup of 3.09.

Good performance is also obtained in the \textit{L-Room} scenario (\cref{fig:scatter_sinr_lroom}).
In this case, choosing $\gamma_{\rm th}<-15$~dB and $R>1$ makes it possible for the \gls{nrmse} to remain below 0.05, but an overall speedup of a factor of 3.3 is obtained with $R=2$ and $\gamma_{\rm th}=-40$~dB.
In this case, using $R=1$ even with $\gamma_{\rm th}=-25$~dB might still be acceptable, with an NRMSE of only 0.057 but a speedup factor equal to 5.3.

On the contrary, \cref{fig:scatter_sinr_parkinglot} witnesses a much more severe accuracy loss for the \textit{Parking Lot} scenario, as already highlighted in \cref{sub:link_level_performance_results}.
Specifically, the \gls{nrmse} increases almost tenfold with respect to the previous case with $\gamma_{\rm th}=-15$ dB.
This is due to the higher absorption coefficient of the materials that are present in the outdoor environment, that, together with the increased path lengths, lead to a significant decrease in the average \gls{mpc} power.
Thus, a low $\gamma_{\rm th}$ severely degrades the overall received power by removing a large percentage (even up to 90\%) of such \glspl{mpc}, whose overall contribution is far from negligible.
In this case, the best configuration would be $R=2$ and $\gamma_{\rm th}=-40$~dB, achieving a speedup of 6.8 with an \gls{nrmse} of 0.034.

\begin{figure*}[tbp] 
  \begin{subfigure}[t]{\textwidth}
    \centering
%
%
\definecolor{mycolor1}{rgb}{0.00000,0.44700,0.74100}%
\definecolor{mycolor2}{rgb}{0.85000,0.32500,0.09800}%
\definecolor{mycolor3}{rgb}{0.92900,0.69400,0.12500}%
\definecolor{mycolor4}{rgb}{0.49400,0.18400,0.55600}%

\begin{tikzpicture}
\pgfplotsset{every tick label/.append style={font=\scriptsize}}

\begin{axis}[%
width=0,
height=0,
at={(0,0)},
scale only axis,
xmin=0,
xmax=0,
xtick={},
ymin=0,
ymax=0,
ytick={},
zmin=0,
zmax=0,
ztick={},
axis background/.style={fill=white},
legend style={legend cell align=center, align=center, draw=white!15!black, font=\scriptsize, at={(0, 0)}, anchor=center, /tikz/every even column/.append style={column sep=2em}},
legend columns=4,
]
\addplot[only marks, mark=*, mark options={}, mark size=2.5pt, draw=black, fill=black] table[row sep=crcr]{%
x y\\
0 0\\
};
\addlegendentry{$R=1$}

\addplot[only marks, mark=square*, mark options={}, mark size=2.5pt, draw=black, fill=black] table[row sep=crcr]{%
x y\\
0 0\\
};
\addlegendentry{$R=2$}

\addplot[only marks, mark=diamond*, mark options={}, mark size=2.5pt, draw=black, fill=black] table[row sep=crcr]{%
x y\\
0 0\\
};
\addlegendentry{$R=3$}

\addplot[only marks, mark=triangle*, mark options={rotate=180}, mark size=2.5pt, draw=black, fill=black] table[row sep=crcr]{%
x y\\
0 0\\
};
\addlegendentry{$R=4$}

\addplot[only marks, mark=*, mark options={}, mark size=2.5pt, draw=mycolor1, fill=mycolor1] table[row sep=crcr]{%
x y\\
0 0\\
};
\addlegendentry{$\gamma_{\rm th}=-\infty$}

\addplot[only marks, mark=*, mark options={}, mark size=2.5pt, draw=mycolor2, fill=mycolor2] table[row sep=crcr]{%
x y\\
0 0\\
};
\addlegendentry{$\gamma_{\rm th}=-40$}

\addplot[only marks, mark=*, mark options={}, mark size=2.5pt, draw=mycolor3, fill=mycolor3] table[row sep=crcr]{%
x y\\
0 0\\
};
\addlegendentry{$\gamma_{\rm th}=-25$}

\addplot[only marks, mark=*, mark options={}, mark size=2.5pt, draw=mycolor4, fill=mycolor4] table[row sep=crcr]{%
x y\\
0 0\\
};
\addlegendentry{$\gamma_{\rm th}=-15$}

\end{axis}
\end{tikzpicture}%
  \end{subfigure}
  \\
  \hfill
  \begin{subfigure}[t]{0.32\textwidth}
      \centering
      \setlength\fwidth{0.8\columnwidth}
      \setlength\fheight{0.35\columnwidth}
%
%
\definecolor{mycolor1}{rgb}{0.00000,0.44700,0.74100}%
\definecolor{mycolor2}{rgb}{0.85000,0.32500,0.09800}%
\definecolor{mycolor3}{rgb}{0.92900,0.69400,0.12500}%
\definecolor{mycolor4}{rgb}{0.49400,0.18400,0.55600}%
\begin{tikzpicture}
\pgfplotsset{every tick label/.append style={font=\scriptsize}}

\begin{axis}[%
width=0.961\fwidth,
height=\fheight,
at={(0\fwidth,0\fheight)},
scale only axis,
xmin=1,
xmax=4,
xlabel={Speedup factor},
xlabel style={font=\scriptsize\color{white!15!black}},
ymin=0,
ymax=0.05,
scaled y ticks = false,
y tick label style={/pgf/number format/.cd,
                    fixed,
                    },
extra y ticks={0.05},
extra y tick labels={},
extra y tick style={grid style={black, dashed}},
ylabel={NRMSE},
ylabel style={font=\scriptsize\color{white!15!black}},
axis background/.style={fill=white},
xmajorgrids,
ymajorgrids,
legend style={legend cell align=left, align=left, draw=white!15!black}
]
\addplot[only marks, mark=*, mark options={}, mark size=2.5pt, draw=mycolor1, fill=mycolor1] table[row sep=crcr]{%
x	y\\
3.65944956846694	0\\
};
\addplot[only marks, mark=square*, mark options={}, mark size=2.5pt, draw=mycolor1, fill=mycolor1] table[row sep=crcr]{%
x	y\\
2.43822570657934	0\\
};
\addplot[only marks, mark=diamond*, mark options={}, mark size=2.5pt, draw=mycolor1, fill=mycolor1] table[row sep=crcr]{%
x	y\\
1.59815576998518	0\\
};
\addplot[only marks, mark=triangle*, mark options={rotate=180}, mark size=2.5pt, draw=mycolor1, fill=mycolor1] table[row sep=crcr]{%
x	y\\
1	0\\
};
\addplot[only marks, mark=*, mark options={}, mark size=2.5pt, draw=mycolor2, fill=mycolor2] table[row sep=crcr]{%
x	y\\
3.57754805753355	0\\
};
\addplot[only marks, mark=square*, mark options={}, mark size=2.5pt, draw=mycolor2, fill=mycolor2] table[row sep=crcr]{%
x	y\\
2.65726135739738	0\\
};
\addplot[only marks, mark=diamond*, mark options={}, mark size=2.5pt, draw=mycolor2, fill=mycolor2] table[row sep=crcr]{%
x	y\\
1.74865977108953	0\\
};
\addplot[only marks, mark=triangle*, mark options={rotate=180}, mark size=2.5pt, draw=mycolor2, fill=mycolor2] table[row sep=crcr]{%
x	y\\
1.55170659049487	0\\
};
\addplot[only marks, mark=*, mark options={}, mark size=2.5pt, draw=mycolor3, fill=mycolor3] table[row sep=crcr]{%
x	y\\
3.48941184794853	0\\
};
\addplot[only marks, mark=square*, mark options={}, mark size=2.5pt, draw=mycolor3, fill=mycolor3] table[row sep=crcr]{%
x	y\\
2.78428435487577	0\\
};
\addplot[only marks, mark=diamond*, mark options={}, mark size=2.5pt, draw=mycolor3, fill=mycolor3] table[row sep=crcr]{%
x	y\\
2.86662732395355	0\\
};
\addplot[only marks, mark=triangle*, mark options={rotate=180}, mark size=2.5pt, draw=mycolor3, fill=mycolor3] table[row sep=crcr]{%
x	y\\
2.86383684359972	0\\
};
\addplot[only marks, mark=*, mark options={}, mark size=2.5pt, draw=mycolor4, fill=mycolor4] table[row sep=crcr]{%
x	y\\
3.88620056351731	0\\
};
\addplot[only marks, mark=square*, mark options={}, mark size=2.5pt, draw=mycolor4, fill=mycolor4] table[row sep=crcr]{%
x	y\\
3.58912062306577	0\\
};
\addplot[only marks, mark=diamond*, mark options={}, mark size=2.5pt, draw=mycolor4, fill=mycolor4] table[row sep=crcr]{%
x	y\\
3.87281307201182	0\\
};
\addplot[only marks, mark=triangle*, mark options={rotate=180}, mark size=2.5pt, draw=mycolor4, fill=mycolor4] table[row sep=crcr]{%
x	y\\
3.78762541424776	0\\
};

\end{axis}
\end{tikzpicture}%
      \caption{\textit{Indoor1.}}
      \label{fig:scatter_thr_indoor1}
  \end{subfigure}
  \hfill
  \begin{subfigure}[t]{0.32\textwidth}
      \centering
      \setlength\fwidth{0.8\columnwidth}
      \setlength\fheight{0.35\columnwidth}
%
%
\definecolor{mycolor1}{rgb}{0.00000,0.44700,0.74100}%
\definecolor{mycolor2}{rgb}{0.85000,0.32500,0.09800}%
\definecolor{mycolor3}{rgb}{0.92900,0.69400,0.12500}%
\definecolor{mycolor4}{rgb}{0.49400,0.18400,0.55600}%
\begin{tikzpicture}

\pgfplotsset{every tick label/.append style={font=\scriptsize}}

\begin{axis}[%
width=0.961\fwidth,
height=\fheight,
at={(0\fwidth,0\fheight)},
scale only axis,
xmin=1,
xmax=5,
xlabel={Speedup factor},
xlabel style={font=\scriptsize\color{white!15!black}},
ymin=0,
ymax=0.16,
scaled y ticks = false,
y tick label style={/pgf/number format/.cd,
                    fixed,
                    },
extra y ticks={0.05},
extra y tick labels={},
extra y tick style={grid style={black, dashed}},
ylabel style={font=\scriptsize\color{white!15!black}},
axis background/.style={fill=white},
xmajorgrids,
ymajorgrids,
legend style={legend cell align=left, align=left, draw=white!15!black}
]
\addplot[only marks, mark=*, mark options={}, mark size=2.5pt, draw=mycolor1, fill=mycolor1] table[row sep=crcr]{%
x	y\\
4.71886987802412	0.0537608503568379\\
};
\addplot[only marks, mark=square*, mark options={}, mark size=2.5pt, draw=mycolor1, fill=mycolor1] table[row sep=crcr]{%
x	y\\
2.73528608213464	0.0346069422764507\\
};
\addplot[only marks, mark=diamond*, mark options={}, mark size=2.5pt, draw=mycolor1, fill=mycolor1] table[row sep=crcr]{%
x	y\\
1.79389199176546	0.0280490848897467\\
};
\addplot[only marks, mark=triangle*, mark options={rotate=180}, mark size=2.5pt, draw=mycolor1, fill=mycolor1] table[row sep=crcr]{%
x	y\\
1	0\\
};
\addplot[only marks, mark=*, mark options={}, mark size=2.5pt, draw=mycolor2, fill=mycolor2] table[row sep=crcr]{%
x	y\\
4.39403613671621	0.0537608503568379\\
};
\addplot[only marks, mark=square*, mark options={}, mark size=2.5pt, draw=mycolor2, fill=mycolor2] table[row sep=crcr]{%
x	y\\
3.11764392349607	0.0346069422764507\\
};
\addplot[only marks, mark=diamond*, mark options={}, mark size=2.5pt, draw=mycolor2, fill=mycolor2] table[row sep=crcr]{%
x	y\\
1.78295045747831	0.0243102896712767\\
};
\addplot[only marks, mark=triangle*, mark options={rotate=180}, mark size=2.5pt, draw=mycolor2, fill=mycolor2] table[row sep=crcr]{%
x	y\\
1.3656354119253	0.0277020111699796\\
};
\addplot[only marks, mark=*, mark options={}, mark size=2.5pt, draw=mycolor3, fill=mycolor3] table[row sep=crcr]{%
x	y\\
4.32827628186468	0.0537608503568379\\
};
\addplot[only marks, mark=square*, mark options={}, mark size=2.5pt, draw=mycolor3, fill=mycolor3] table[row sep=crcr]{%
x	y\\
3.58049744111927	0.0332382753835736\\
};
\addplot[only marks, mark=diamond*, mark options={}, mark size=2.5pt, draw=mycolor3, fill=mycolor3] table[row sep=crcr]{%
x	y\\
2.57780190082245	0.127550398385135\\
};
\addplot[only marks, mark=triangle*, mark options={rotate=180}, mark size=2.5pt, draw=mycolor3, fill=mycolor3] table[row sep=crcr]{%
x	y\\
1.88882012364103	0.121999275751062\\
};
\addplot[only marks, mark=*, mark options={}, mark size=2.5pt, draw=mycolor4, fill=mycolor4] table[row sep=crcr]{%
x	y\\
4.56819994512694	0.0537631531653711\\
};
\addplot[only marks, mark=square*, mark options={}, mark size=2.5pt, draw=mycolor4, fill=mycolor4] table[row sep=crcr]{%
x	y\\
4.72602733733684	0.0544945946292001\\
};
\addplot[only marks, mark=diamond*, mark options={}, mark size=2.5pt, draw=mycolor4, fill=mycolor4] table[row sep=crcr]{%
x	y\\
3.84471355749737	0.152958575844679\\
};
\addplot[only marks, mark=triangle*, mark options={rotate=180}, mark size=2.5pt, draw=mycolor4, fill=mycolor4] table[row sep=crcr]{%
x	y\\
2.59231871720043	0.155948605171316\\
};

\end{axis}
\end{tikzpicture}%
      \caption{\textit{L-Room.}}
      \label{fig:scatter_thr_lroom}
  \end{subfigure}
  \hfill
  \begin{subfigure}[t]{0.32\textwidth}
      \centering
      \setlength\fwidth{0.8\columnwidth}
      \setlength\fheight{0.35\columnwidth}
%
%
\definecolor{mycolor1}{rgb}{0.00000,0.44700,0.74100}%
\definecolor{mycolor2}{rgb}{0.85000,0.32500,0.09800}%
\definecolor{mycolor3}{rgb}{0.92900,0.69400,0.12500}%
\definecolor{mycolor4}{rgb}{0.49400,0.18400,0.55600}%
\begin{tikzpicture}

\pgfplotsset{every tick label/.append style={font=\scriptsize}}

\begin{axis}[%
width=0.961\fwidth,
height=\fheight,
at={(0\fwidth,0\fheight)},
scale only axis,
xmin=1,
xmax=3.5,
xlabel={Speedup factor},
xlabel style={font=\scriptsize\color{white!15!black}},
ymin=0,
ymax=0.5,
scaled y ticks = false,
y tick label style={/pgf/number format/.cd,
                    fixed,
                    },
extra y ticks={0.05},
extra y tick labels={},
extra y tick style={grid style={black, dashed}},
ylabel style={font=\scriptsize\color{white!15!black}},
axis background/.style={fill=white},
xmajorgrids,
ymajorgrids,
legend style={legend cell align=left, align=left, draw=white!15!black}
]
\addplot[only marks, mark=*, mark options={}, mark size=2.5pt, draw=mycolor1, fill=mycolor1] table[row sep=crcr]{%
x	y\\
2.5608346181777	0.163970789664203\\
};
\addplot[only marks, mark=square*, mark options={}, mark size=2.5pt, draw=mycolor1, fill=mycolor1] table[row sep=crcr]{%
x	y\\
1.50193329043406	0.0291566737120494\\
};
\addplot[only marks, mark=diamond*, mark options={}, mark size=2.5pt, draw=mycolor1, fill=mycolor1] table[row sep=crcr]{%
x	y\\
1	0\\
};
\addplot[only marks, mark=*, mark options={}, mark size=2.5pt, draw=mycolor2, fill=mycolor2] table[row sep=crcr]{%
x	y\\
2.56877093084264	0.164163252219731\\
};
\addplot[only marks, mark=square*, mark options={}, mark size=2.5pt, draw=mycolor2, fill=mycolor2] table[row sep=crcr]{%
x	y\\
1.89543286475749	0.0480336813453542\\
};
\addplot[only marks, mark=diamond*, mark options={}, mark size=2.5pt, draw=mycolor2, fill=mycolor2] table[row sep=crcr]{%
x	y\\
1.54503274210074	0.0480282302088579\\
};
\addplot[only marks, mark=*, mark options={}, mark size=2.5pt, draw=mycolor3, fill=mycolor3] table[row sep=crcr]{%
x	y\\
2.63193895838597	0.166885670995383\\
};
\addplot[only marks, mark=square*, mark options={}, mark size=2.5pt, draw=mycolor3, fill=mycolor3] table[row sep=crcr]{%
x	y\\
2.28974964347462	0.166885670995383\\
};
\addplot[only marks, mark=diamond*, mark options={}, mark size=2.5pt, draw=mycolor3, fill=mycolor3] table[row sep=crcr]{%
x	y\\
2.11751466791023	0.166885670995383\\
};
\addplot[only marks, mark=*, mark options={}, mark size=2.5pt, draw=mycolor4, fill=mycolor4] table[row sep=crcr]{%
x	y\\
3.35316258174224	0.417692857176549\\
};
\addplot[only marks, mark=square*, mark options={}, mark size=2.5pt, draw=mycolor4, fill=mycolor4] table[row sep=crcr]{%
x	y\\
3.19687056385543	0.417692857176549\\
};
\addplot[only marks, mark=diamond*, mark options={}, mark size=2.5pt, draw=mycolor4, fill=mycolor4] table[row sep=crcr]{%
x	y\\
2.51063268043686	0.417692857176549\\
};

\end{axis}
\end{tikzpicture}%
      \caption{\textit{Parking Lot.}}
      \label{fig:scatter_thr_parkinglot}
  \end{subfigure}
  \hfill
    \setlength\belowcaptionskip{-.6cm}
  \caption{Trade-off between the throughput performance and the speedup obtained with the different simplification parameters for the three scenarios.
As in \cref{sub:end_to_end_performance_results}, for the throughput ns-3 has been considered.}
  \label{fig:scatter_thr}
\end{figure*}
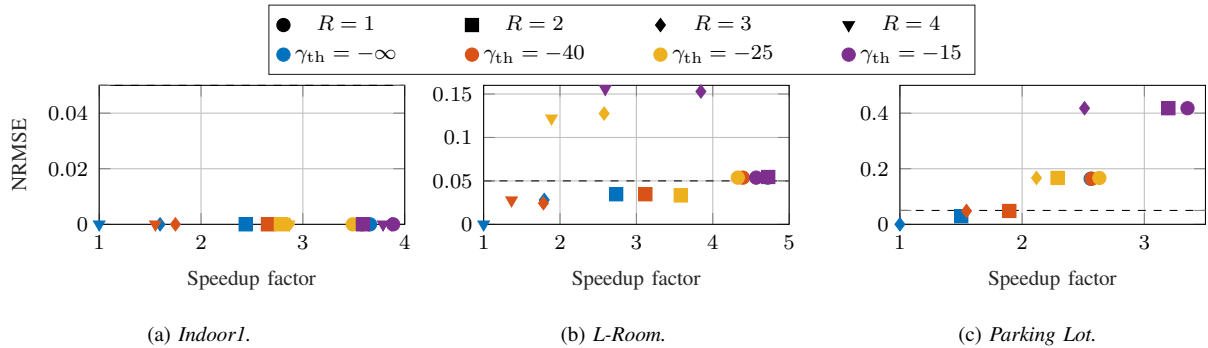%

\paragraph*{End-to-End Performance}
\cref{fig:scatter_thr} reports the \gls{nrmse} of the throughput vs. the speedup for end-to-end simulations.
Similarly to what happened for the link-level performance, $\gamma_{\rm th}$ shows diminishing returns especially for both the \textit{L-Room} and the \textit{Parking Lot} scenarios.

\cref{fig:scatter_thr_indoor1} shows that virtually no variations occur when introducing simplifications in the \emph{Indoor1} scenario.
This suggests that setting $R=1$ and $\gamma_{\rm th}=-15$~dB can speed up the simulation by a factor of almost 4 with negligible accuracy loss with respect to the baseline configuration.

Good results are obtained also for the \textit{L-Room} scenario in \cref{fig:scatter_thr_lroom}, and with $R=2$ and $\gamma_{\rm th}=-25$~dB it is possible to gain a 3.6 speedup factor with an \gls{nrmse} of 0.033, or even a 4.7 speedup if an \gls{nrmse} of 0.054 is still accepted using $R=1$.

Similarly to the link-level case, the end-to-end simulations for the \textit{Parking Lot} scenario in \cref{fig:scatter_thr_parkinglot} are much more severely affected by the simplifications.
With both $\gamma_{\rm th}\geq-25$~dB and $R=1$, the \gls{nrmse} increases beyond acceptable levels.
Nevertheless, setting $R=2$ and $\gamma_{\rm th}=-40$~dB offers an \gls{nrmse} as low as 0.048 with a 1.9 speedup factor.

\subsection{Design Guidelines} 
\label{sub:design_guidelines}

Without focusing on the exact numbers, that are specific to the scenarios and the simulation tools employed in this work, it is still possible to draw some general guidelines for an efficient \gls{rt} simulation.
\begin{itemize}
  \item \textit{Environment.} The simulation scenario plays a key role in the simplification choice.
  Specifically, when considering indoor \gls{los} simulations, the secondary rays can be neglected with good approximation and very significant time savings.
  When considering also \gls{nlos} conditions in indoor scenarios, instead, less flexibility should be expected, although it is still possible to consistently reduce the runtime with a minor accuracy loss.
  Finally, outdoor scenarios should be treated carefully, as aggressive simplifications can have detrimental effects on the fidelity of the simulations.
  Nevertheless, a working point can be identified which offers a significant speedup.
  \item \textit{Simplification Strategy.} Although their effect can vary substantially depending on the considered scenario, some general considerations can be drawn regarding the simplification techniques.
  The link-level metrics, such as the \gls{sinr}, benefit, in terms of runtime, more from a reduction of the maximum reflection order rather than from an increase of the \gls{mpc} threshold.
  For both link-level and end-to-end metrics, an aggressive thresholding policy leads to a performance degradation that is not justified by a corresponding speedup improvement.
  Finally, end-to-end results require a balanced use of both simplification techniques to achieve an optimal working point.
  In general, it can be noticed that $R=2$ and a relative threshold $\gamma_{\rm th}=[-40,-25]$~dB consistently yield a good balance of accuracy and speedup in almost all cases, and for this reason it is the suggested configuration.
\end{itemize}




\section{Conclusions and Future Work} 
\label{sec:conclusions_and_future_work}

Accurate and scalable simulations are the basis for the design and evaluation of future 5G networks operating at \gls{mmwave} frequencies.
The usage of large antenna arrays, however, requires the modeling of the spatial dimension of the channel, for example through \glspl{rt}, which can model the propagation of the different multipath components of a mmWave signal based on the geometry of the scenario.
In this paper, we discussed two possible simplification techniques to decrease the computational complexity of \gls{rt}-based channel and network simulations, to improve the scalability of these techniques.
We showed that, while the proposed strategies decrease the number of multipath components to be actually computed, they have a different impact on the physical layer and end-to-end performance with different scenarios, applications, and antenna array configurations.
We believe that the insights that resulted from the extensive profiling and performance evaluation can guide researchers in designing accurate, yet scalable, simulations of mmWave networks.

As future work, we plan to extend the profiling and comparison considering end-to-end measurements in actual mmWave deployments, reproducing the scenario in the \gls{rt}.
Moreover, we also plan to study the impact of the simplifications on different beamforming architectures (e.g., hybrid beamforming).

\section*{Acknowledgments}
We want to thank our collaborators at NIST for their support and for providing high-quality data.
In particular, we want to thank (in alphabetical order) Anuraag Bodi, Camillo Gentile, Nada Golmie, Chiehping Lai, Tanguy Ropitault, Neeraj Varshney, Jian Wang.


\end{document}